\documentclass[twocolumn]{aastex62}

\received{2019 July 15}
\revised{2019 August 12}
\accepted{2019 August 15}
\submitjournal{ApJ}

\shortauthors{D\'ek\'any et al.}
\begin{document}

\title{Into the Darkness: Classical and Type II Cepheids in the Zona Galactica Incognita}

\correspondingauthor{Istv\'an D\'ek\'any}
\email{dekany@uni-heidelberg.de}

\author[0000-0002-0786-7307]{Istv\'an D\'ek\'any}
\affil{Astronomisches Rechen-Institut, Zentrum f\"ur Astronomie der Universit\"at Heidelberg,
M\"onchhofstr. 12-14, 69120 Heidelberg, Germany}

%\author{et al.}

\author{Gergely Hajdu}
%\altaffiliation{Millennium Institute of Astrophysics, Santiago, Chile}
\affiliation{Astronomisches Rechen-Institut, Zentrum f\"ur Astronomie der Universit\"at Heidelberg,
M\"onchhofstr. 12-14, 69120 Heidelberg, Germany}
\affiliation{Instituto de Astrof\'isica, Facultad de F\'isica, Pontificia Universidad Cat\'olica de Chile, Av. Vicu\~na Mackenna 4860, 782-0436 Macul, Santiago, Chile}
\affiliation{Millennium Institute of Astrophysics, Santiago, Chile}

\author{Eva K. Grebel}
\affiliation{Astronomisches Rechen-Institut, Zentrum f\"ur Astronomie der Universit\"at Heidelberg,
M\"onchhofstr. 12-14, 69120 Heidelberg, Germany}

%\author{Daniel Majaess}
%\affiliation{Mount Saint Vincent University, Halifax, Nova Scotia, Canada}
%\affiliation{Saint Mary's University, Halifax, Nova Scotia, Canada}

\author{M\'arcio Catelan}
%\altaffiliation{Millennium Institute of Astrophysics, Santiago, Chile}
\affiliation{Instituto de Astrof\'isica, Facultad de F\'isica, Pontificia Universidad Cat\'olica de Chile, Av. Vicu\~na Mackenna 4860, 782-0436 Macul, Santiago, Chile}
\affiliation{Millennium Institute of Astrophysics, Santiago, Chile}

%\affiliation{Millennium Institute of Astrophysics, Santiago, Chile}



\begin{abstract}

The far side of the Milky Way's disk is one of the most concealed parts of the known Universe due to extremely high interstellar extinction and point source density toward low Galactic latitudes. Large time-domain photometric surveys operating in the near-infrared hold great potential for the exploration of these vast uncharted areas of our Galaxy. We conducted a census of distant classical and type II Cepheids along the southern Galactic mid-plane using near-infrared photometry from the VISTA Variables in the V\'ia L\'actea survey. We performed a machine-learned classification of the Cepheids based on their infrared light curves using a convolutional neural network. We have discovered 640 distant classical Cepheids with up to $\sim$40 magnitudes of visual extinction, and over 500 type II Cepheids, most of them located in the inner bulge. Intrinsic color indices of individual Cepheids were predicted from sparse photometric data using a neural network, allowing their use as accurate reddening tracers. They revealed a steep, spatially varying near-infrared extinction curve toward the inner bulge. Type II Cepheids in the Galactic bulge were also employed to measure robust mean selective-to-absolute extinction ratios. They trace a centrally concentrated spatial distribution of the old bulge population with a slight elongation, consistent with earlier results from RR~Lyrae stars. Likewise, the classical Cepheids were utilized to trace the Galactic warp and various substructures of the Galactic disk, and to uncover significant vertical and radial age gradients of the thin disk population at the far side of the Milky Way.

\end{abstract}

\keywords{Delta Cepheid variable stars, Population II Cepheid variable stars, Catalogs, Surveys, Galactic bulge, Milky Way disk, Interstellar extinction}


\section{Introduction} \label{sec:intro}

\noindent Six decades have passed since the first 21-cm radio surveys of Galactic neutral hydrogen gas started \citep[e.g.,][]{1957BAN....13..247S,1957BAN....13..201W} with the promise of mapping the detailed structure of the Milky Way's disk. But in spite of the tremendous progress since then, large swathes of our Galaxy at its far side have remained unexplored. 

The structure of the Galactic disk has been traced with two different approaches: (i) by the velocity mapping of interstellar gas, i.e., by measuring density peaks of neutral and ionized hydrogen gas (HI and HII; see, e.g., \citealt{2009ARA&A..47...27K,2014A&A...569A.125H,2017PASP..129i4102K}, and references therein) and by mapping giant molecular clouds (GMCs) using CO lines or masers \citep[e.g.,][and references therein]{2016SciA....2E0878X,2018A&A...616L..15X}; and (ii) by using young stars (e.g., classical Cepheids, OB stars) and young open clusters as tracers \citep[e.g.,][]{2009MNRAS.398..263M,2019NatAs...3..320C,2019Sci...365..478S}.

Large-scale maps of the perturbed surface density of HI using the 21-cm hyperfine transition lack general consensus, and studies using different  techniques to transform radial velocity distributions to face-on maps of HI density suffer from tension. Observational data have been interpreted with various four-arm logarithmic spiral models with both anomalously large pitch angles and signs of a global non-axisymmetric configuration of the spiral structure \citep{2006Sci...312.1773L}, and also axisymmetric models with their global configuration being in qualitative agreement with maps based on other tracers \citep[e.g.,][]{2017PASP..129i4102K}.

Current maps from the velocity mapping of HII regions and GMCs are based on large compilations of kinematic data \cite[e.g.,][]{2003A&A...397..133R,2009A&A...499..473H}, and favor models with 3--€"4 arms with poorly constrained pitch angles \citep[see][and references therein]{2015MNRAS.450.4277V}. Polynomial spiral models were proposed by \citet{2014A&A...569A.125H} to ease the tension between different gas tracers.

In addition to the major limitation of the velocity mapping methods being blind toward the Galactic center and anticenter, they also suffer from a near--far ambiguity toward the I-st and IV-th Galactic quadrants, and are tied to assumptions on the bulk kinematic properties of the Galaxy, such as the standard rotation curve and the local standard of rest€. Uncertainties in these parameters introduce biases in the distance measurements, and systematic velocity offsets of young objects from the standard Galactic rotation as suggested by the spiral density-wave theory \citep{1969ApJ...155..721L,2006Sci...311...54X,2015MNRAS.454..626H} imply additional controversy in correlating gas cloud velocities with distances.

Most of the above issues are mitigated by accurate direct distance measurements using parallaxes of young stars associated with gas clouds and radio interferometric parallaxes of masers \citep[e.g.,][]{2009ApJ...700..137R,2012ApJ...751..157F}, but such data are available only for a few objects at the near side of the disk, with only a recent venture to the far side \citep{2017Sci...358..227S}.

Young stellar objects, such as OB stars, young open clusters, and classical Cepheid variable stars provide alternative means to trace the large-scale spatial structures of the Galactic disk. The latter are particularly apt young population tracers due to their accurate period-luminosity (PL) relations in the near-infrared (near-IR) photometric bands \citep[e.g.,][]{2015AJ....149..117M}, and the relationship between their pulsation periods and ages \citep{2005ApJ...621..966B,2016A&A...591A...8A}. Until very recently, disk Cepheids were known only within a $\sim$6~kpc radius around the Sun due to observational challenges posed by high interstellar extinction and source density endemic to low Galactic latitudes. Hence they could only be used to trace local spiral arm features \citep[e.g.,][]{2009MNRAS.398..263M}, similarly to OB stars \citep{2019MNRAS.487.1400C}.

Lately, large time-domain photometric surveys boosted the number of known disk Cepheids. In particular, the discovery of thousands of disk Cepheids by the Wide-field Infrared Survey Explorer \cite[WISE,][]{2018ApJS..237...28C} and the Optical Gravitational Lensing Experiment \citep[OGLE,][]{2018AcA....68..315U} provided a breakthrough in their census. The resulting Cepheid catalogs allowed it to map the Galactic warp \citep{2019Sci...365..478S,2019NatAs...3..320C} and large substructures at the near side of the disk \citep{2019Sci...365..478S}. Further significant contributions to the census of classical Cepheids were recently provided by the All-Sky Automated Survey for Supernovae \citep[ASAS-SN,][]{2018MNRAS.477.3145J}, and Gaia \citep{2019A&A...625A..14R}. In spite of the numerous recent discoveries, a vast section at the far side of the disk in the I-st and the IV-th Galactic quadrants, dubbed as the {\em ``Zona Galactica Incognita''} by \cite{2017AstRv..13..113V}, remained almost devoid of known Cepheids due to the extremely high attenuation by interstellar dust.

Although current photometric surveys operating at infrared wavelengths enable us to detect Cepheids at the far side of the disk, well beyond any optical survey's horizon imposed by interstellar dust, our understanding of the extinction's wavelength dependence (i.e., the ``extinction curve'') is still a fundamental limiting factor for the usage of stellar tracers in the disk. Trivially, the accurate knowledge of the (mean) selective-to-absolute extinction ratio (and its uncertainty) over an area of study is of critical importance for consistent distance estimates of highly attenuated objects. A good illustration of this is the current debate concerning the existence of classical Cepheids within the bulge volume. In a previous study, we reported the discovery of numerous classical Cepheids and speculated that they trace a young, thin stellar disk spanning across the inner Galaxy \citep{2015ApJ...812L..29D}. However, using a largely common stellar sample, \citet{2016MNRAS.462..414M} argued that the inner Galaxy is free from classical Cepheids (except for the nuclear bulge, \citealt{2011Natur.477..188M,2013MNRAS.429..385M}). The two studies arrived to opposing conclusions about the physical nature of the inner Milky Way mostly due to a $\sim$10$\%$ difference in the $A(K_s)/E(H-K_s)$ extinction ratio, adopting its value from two different studies based the same technique and obtained for the same area by the same group \citep{2006ApJ...638..839N,2009ApJ...696.1407N}.

Further complications arise from systematic uncertainties in the intrinsic magnitudes of various extinction tracers, the propagation of photometric zero-point errors into the extinction law through the color excesses, biases in the extinction law due to the differences in the spectral energy distribution of different tracers in conjunction with the use of broadband filters, etc. --- a comprehensive discussion of all these complications is provided by \citet{2019ApJ...877..116W}. A very important open question about the near-IR extinction is its spatial variation, as current studies of this matter are in significant tension, and descriptions of the reddening curve range from universal near-IR extinction \citep[e.g.,][]{2014ApJ...788L..12W,2016A&A...593A.124M,2016ApJ...821...78S} to highly variable extinction curves \citep[e.g.,][]{2009ApJ...699.1209F,2009ApJ...707..510Z}. Furthermore, if the near-IR extinction law does vary spatially, it is important to assess the typical angular scales of such variation, as to whether a ``mean'' extinction curve can be adopted over an extended area without introducing large-scale biases in distance estimates.

The exploitation of near-IR time-domain surveys for the census of distant Cepheids also poses the technical challenge of light curve classification. While this is usually straightforward in case of optical data, and even its automation has become a routine task, near-IR light curves of pulsating stars lack the abundance of features found in their optical counterparts \citep[see, e.g.,][]{2016A&A...595A..82E}. Combined with noisy photometry and sub-optimal sampling, this can lead to high classification ambiguity even for the most skillful domain expert. The distinction of classical Cepheids from type II Cepheids, i.e., old, low-mass, He-burning pulsating stars \citep{2015pust.book.....C} is particularly challenging because despite their very different physical parameters and evolutionary status, their near-IR light curves are quite similar. To make things more complicated, high-quality near-IR time-series photometry of variable stars with firm classifications are generally scarce, hence attempts to develop machine-learned classification models are complicated by modest-sized training sets.

In this study, we leverage the near-IR photometric database of the VISTA Variables in the V\'ia L\'actea ESO Public survey \citep[VVV,][]{2010NewA...15..433M} to conduct a deep census of distant Cepheids in the {\em Zona Galactica Incognita}. In Sect.~\ref{sec:obs}, we discuss the data acquisition, photometric calibration and variability search, then present a machine-learned light curve classifier for Cepheids in Sect.~\ref{sec:classif}, which we deploy on VVV data to discover over a thousand new, distant Cepheid variables. We employ the Cepheids as extinction tracers in Sect.~\ref{sec:extinction}, to establish robust mean near-IR extinction coefficients and probe the spatial variation of the near-IR reddening law in the direction of the bulge. We analyze the spatial distributions of type II and classical Cepheids in Sect.~\ref{sec:distribution}, using them as population tracers of the inner bulge and the far side of the Galactic disk, respectively. We summarize our findings in Sect.~\ref{sec:conclusions}.

\section{Observations, calibration, variability analysis}\label{sec:obs}

\subsection{Data}\label{subsec:data}

\noindent Our study is based on photometric time-series of low Galactic latitude regions, acquired by the VVV survey in the near-IR $JHK_s$ passbands of the VISTA system. Our target area is the combination of those of our earlier studies \citep{2015ApJ...812L..29D,2018ApJ...857...54D}, namely the VVV's entire $\sim$4$^\circ$-wide disk footprint along the Galactic equator, and a $\sim$3$^\circ$-wide adjacent area toward the inner bulge. These regions cover a total of $\sim$286 square degrees in the longitudinal range of $-65^\circ \lesssim l \lesssim +10^\circ$, and consist of the VVV fields d001--d152, b313--b332 and b335--b354, as defined by \citet{2010NewA...15..433M}, excluding the extremely crowded nuclear bulge region toward fields b333 and b334.

Each VVV field was observed at 50--100 epochs in the $K_s$ band with a non-uniform, space-varying cadence and a total baseline of $\sim$5 years; and at 1--10 epochs in the $J$ and $H$ bands. Limiting apparent magnitudes are highly position-dependent and vary with interstellar extinction and source density, ranging from $\sim$15.5 to $\sim$18.5~mag in the $K_s$ band  and from $\sim$16.5 to $\sim$20~mag in the $J$ band \citep[for more details, see][]{2012A&A...537A.107S}.

Our analysis is based on the standard data products of the VISTA Data Flow System \citep[VDFS,][]{2004SPIE.5493..401E}, provided by the Cambridge Astronomy Survey Unit (CASU). Details of the image processing and aperture photometry are discussed by \citet{2004SPIE.5493..411I}. We used the photometric measurements made on detector frame stacks called {\em pawprints}. At each observational epoch, sequences of 6 pawprints are acquired within a $\sim$3-minute interval with positional offsets in order to fill the gaps between the detector's 16 chips. The CASU photometric source tables of individual pawprints were positionally cross-matched using our earlier procedure discussed by \citet{2018ApJ...857...54D}, providing unified catalogs for each field for a total of $\sim$$3\cdot10^8$ point sources.

\subsection{Photometric calibration}\label{subsec:zpcal}

\noindent CASU provide an absolute photometric calibration for VVV data following the method of \citet{2018MNRAS.474.5459G}. In their approach, photometric zero-points (ZPs) are determined separately for each individual pawprint using local secondary standard stars from the 2MASS survey \citep{2006AJ....131.1163S}, and robust universal transformation formulae between the two photometric systems. Intra-detector (i.e., chip-wise) sensitivity variations are measured and corrected for on a monthly basis.

As a result of a detailed investigation of the calibration accuracy of VVV as provided by CASU, we concluded that the approach by \citet{2018MNRAS.474.5459G} can lead to significant residual variations in the photometric ZPs in the $J,H,K_s$ bands for our target area, seriously affecting scientific conclusions based on the data. Our findings are presented in full detail by \citet{zpcalib}, and here we only give a brief summary of the issues.

Inaccuracies in the CASU ZPs have been traced back to two main root causes: (i) significant intra-detector variations on time scales much shorter than one month, and (ii) the presence of numerous objects that appear as unresolved blends in 2MASS, but are well-resolved by VVV. Both contribute to time-varying photometric ZP offsets between measurements taken for the same object at the same epoch by different chips, as well as systematic biases in the ZPs varying on longer time-scales. The combined effect of the two can be rather destructive, not only biasing mean stellar magnitudes, but also distorting the light curves, thus affecting their classification.

\begin{figure*}
%\plottwo{fig1a.pdf}{fig1b.pdf}
\gridline{
          \fig{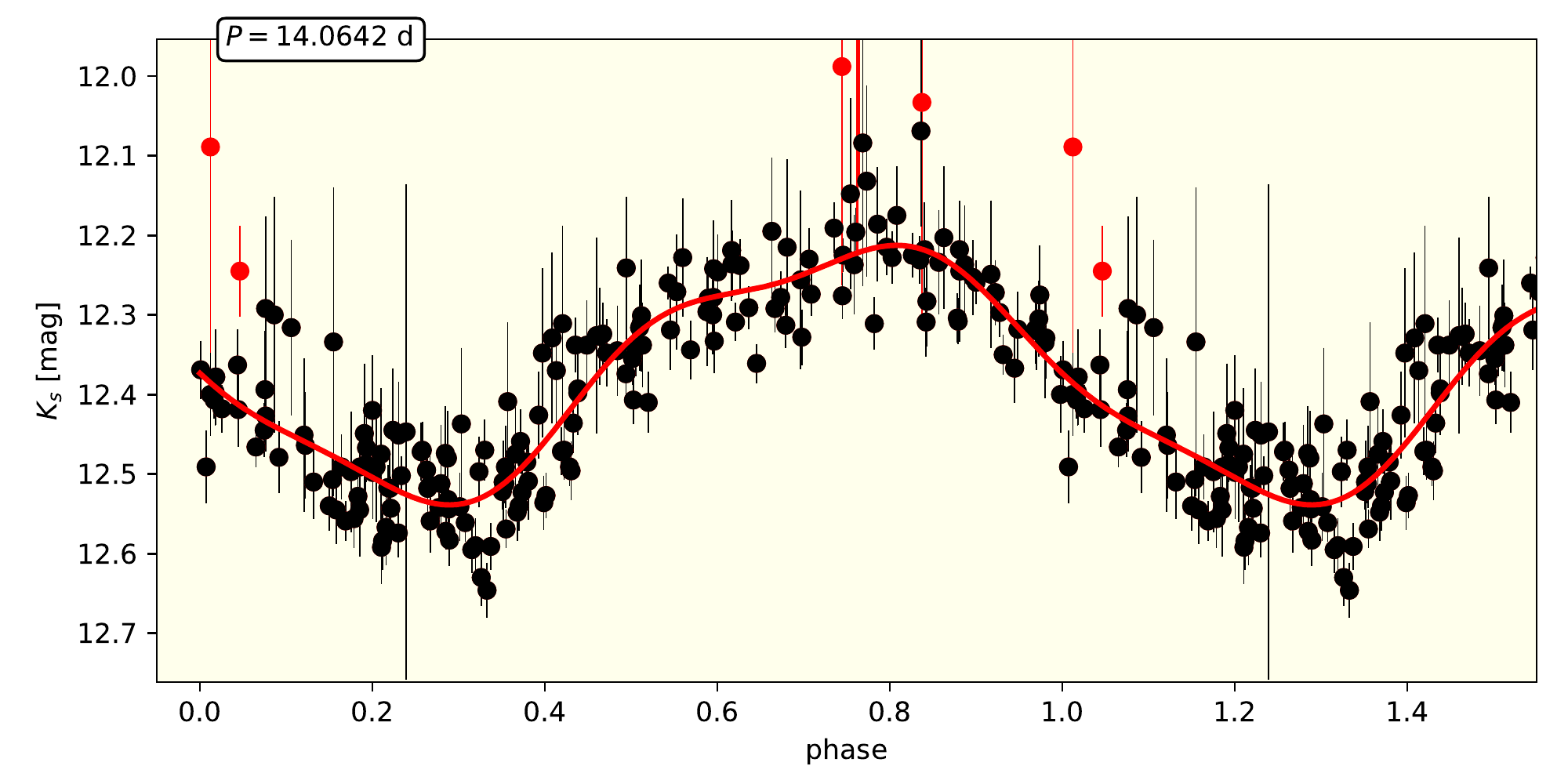}{0.48\textwidth}{}
          \fig{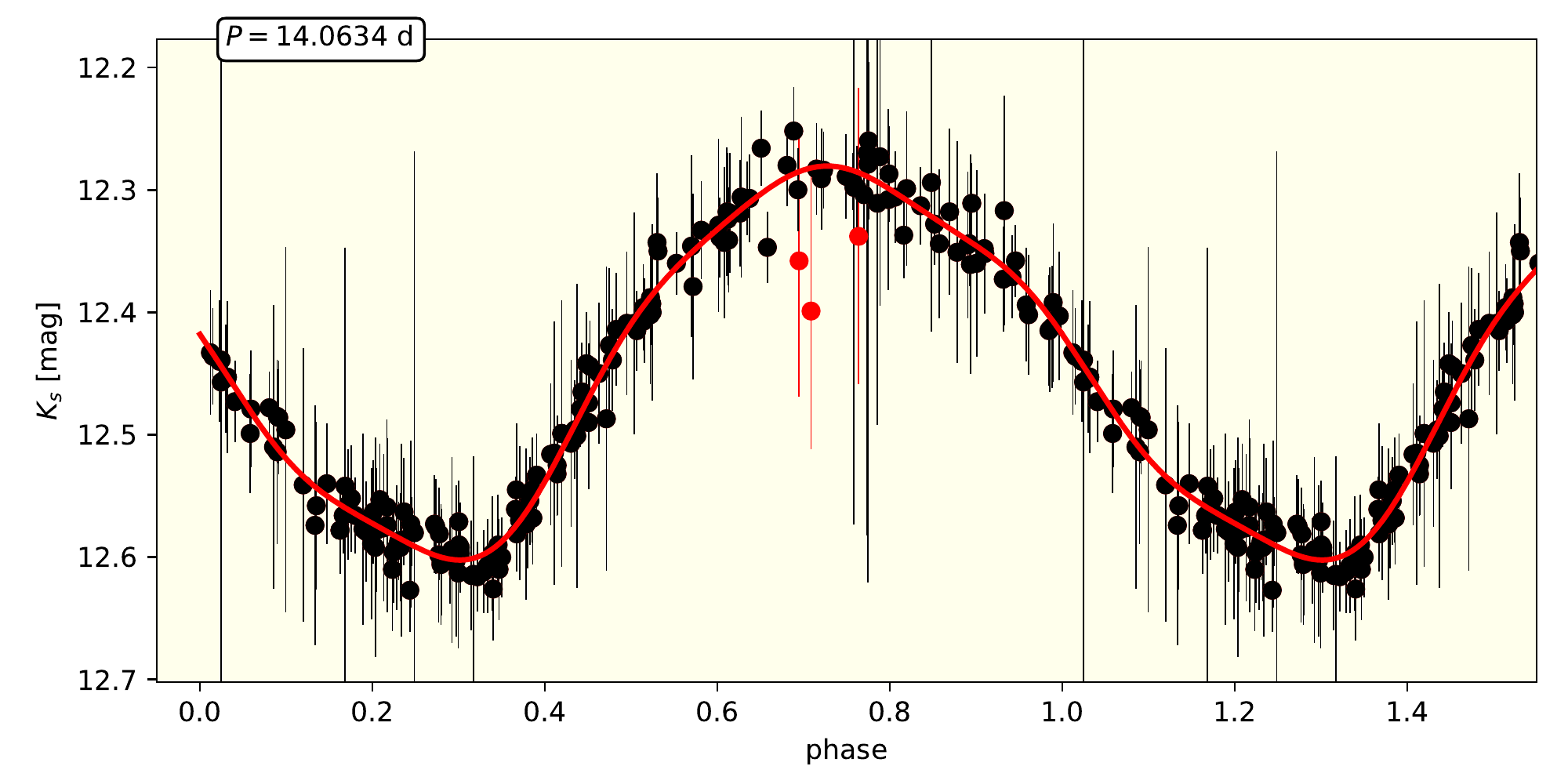}{0.48\textwidth}{}
          }
\caption{
$K_s$-band light curves of the same Cepheid toward the central bulge, phase-folded with the best-fitting period shown in the upper left corner. The left panel shows photometry calibrated by CASU, the right panel displays recalibrated photometry according to \citet{zpcalib}.\label{fig:zpcal}
}
\end{figure*}

In order to correct for these ZP anomalies, we recalibrated the CASU photometry, following the method of \citet{zpcalib}. In this approach, ZPs were determined separately for each of VISTA's 16 detector chips within each pawprint, and the calibration's volatility to blended objects was eliminated by using a fine-tuned positional cross-matching between 2MASS and VISTA sources, followed by a robust regression. The effect of the recalibration is demonstrated on a Cepheid light curve in Fig.~\ref{fig:zpcal}. The original CASU calibration resulted in a distorted light curve and biased mean apparent magnitude, while the recalibrated light curve revealed the true photometric potential of the VVV survey.

\subsection{Variability search}\label{subsec:varsearch}

\noindent Sources with putative light variation were selected following the same procedure as in our earlier RR~Lyrae census \citep{2018ApJ...857...54D}, which is discussed in the aforementioned study in detail. In brief, we employed variability indices that take advantage of the correlated sampling of the VVV light curves in measuring the ratio of the point-to-point and global scatters of the photometry. We used this method to pre-select light curves, thus reducing the sample size before period search. Typically, $10^4$ objects were selected in each field, based on their significance levels estimated from Monte Carlo simulations.

We used the Generalized Lomb-Scargle Periodogram method \citep{2009A&A...496..577Z} to search for periodic signals in the [3.8~day, 40~day] interval using the procedure described by \citet{2018ApJ...857...54D}. The lower limit was chosen to avoid confusion between the different pulsation modes of classical Cepheids \citep[see, e.g.,][]{2015AJ....149..117M} and to constrain the analysis to fundamental-mode pulsators, while the upper limit is imposed by the photometric sampling. By applying the additional selection criterion of $(K_{s,{\rm max.}}-K_{s,{\rm min.}}-\sigma)>0.15$ on each light curve, where $\sigma$ is the weighted standard deviation of the measurements, we narrowed down our sample of Cepheid candidates to $\sim$$4\cdot10^4$ objects. In the next step of the analysis, we applied a machine-learned light curve classifier on these candidates, which we present in the next Section.

\section{Light curve representation and classification}\label{sec:classif}

\noindent Supervised machine learning became a routine approach for light curve classification due to the requirement of automated procedures posed by the large volumes of data from synoptic surveys. Much of the progress has been done for data in optical passbands \citep[see, e.g.,][and references therein]{2016A&A...587A..18K}, while time-series classification in other wavelengths remained a challenge. The classification of $K_s$-band light curves of pulsating variable stars is inherently more difficult because they have generally smaller amplitudes compared to their optical counterparts, and their subtler features are more easily washed out by observational noise. In addition, accurately classified objects with high-quality $K_s$ light curves are scarce, despite the vast amount of data from surveys like VVV, resulting in modest training sets. Due to these challenges, a machine-learned classifier for Cepheids in the near-IR has been lacking.

A common approach in astronomical light curve classification is to derive a large variety of descriptive statistics on the photometric time-series \citep[e.g.,][]{2011ApJ...733...10R}, and use them together with the parameters of the light curve's model representation as descriptive variables, i.e., input {\em features} of the classification problem. Usually, these features are fed into a traditional classification model such as a random forest classifier, a support vector machine, etc., designed to work well on structured data \citep[see, e.g.,][]{2007A&A...475.1159D,2011ApJ...733...10R,2012ApJS..203...32R}. Although this approach has been proven successful in general, it has some possible drawbacks. In certain cases, the light curve shape might not be efficiently captured by such features, and if the information about a characteristic detail is carried by multiple correlated features, certain classes might populate complex manifolds in the resulting feature space, which in turn can hinder the learning process in the absence of a sizable training set.

\begin{figure}[t]
\includegraphics[width=0.48\textwidth]{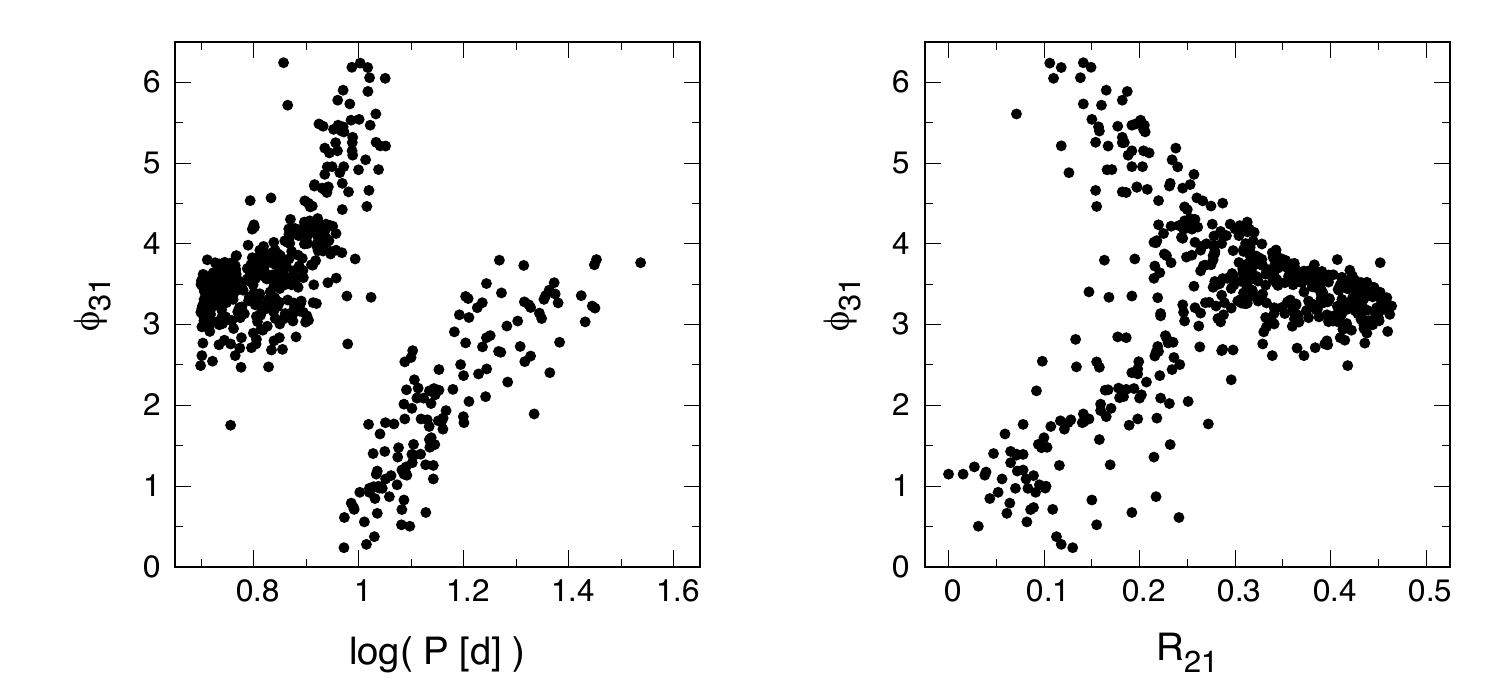}
\caption{
Fourier parameters of the $I$-band light curves of fundamental-mode classical Cepheids in the Large Magellanic Cloud, observed by the OGLE survey \citep{2015AcA....65..297S}.
\label{fig:fourier_par_dcep}}
\end{figure}

We illustrate this problem with the Hertzsprung progression (\citealt{1926BAN.....3..115H}, see also \citealt{2015pust.book.....C}) observed in classical Cepheids in the $\sim$[5,14]~day period range, whereby a resonance between two pulsation modes causes a bump in the light curve, and the pulsation phase in which it occurs decreases with increasing period. This bump is a characteristic light curve detail that can be used to distinguish some classical Cepheids from other types of variable stars (e.g., type II Cepheids), therefore a classifier should be capable to learn about its occurrence. In the customary Fourier representation of a periodic light curve, the position and the size of the bump is described by the first few $R_{ij}=A_i/A_j$ and $\phi_{ij}=\phi_i-i\phi_j$ parameters, where $A_i$ and $\phi_i$ are the amplitude and phase of the $i$-th term in the fitted Fourier-series, respectively. However, it would be difficult for a model to learn the occurrence of the Hertzsprung bump in Cepheid light curves based on the $A_{ij}$ and $\phi_{ij}$ features because they populate a complicated manifold in the feature space, as clearly shown by their various marginal distributions displayed in Fig.~\ref{fig:fourier_par_dcep}.

In order to circumvent such challenges, we adopted a different approach of light curve classification, in which the shape of the light variation is directly perceived by the model, similarly to the human brain when visually inspecting phase diagrams.
%In this analogy, we require that the model can directly `see' details in the light curve such as (Hertzsprung) bumps, plateaus, rising branches, etc., and the interdependency between them.
In other words, rather than a standard classification problem on structured data, we approach light curve classification more like a computer vision problem on one-dimensional `images', i.e., {\em sequence data}. We achieve this by using a convolutional neural network (CNN) as our classification model.

In the following subsections, we outline the functionality and advantages of CNNs, describe the input features of the classifier, and discuss the procedure of model selection, training, and evaluation in the standard supervised machine-learning framework.

\subsection{Convolutional neural networks}\label{subsec:cnn}

\noindent Over the last decade, CNNs \citep{cnnproc} have become very widespread in supervised machine learning, mainly due to their substantial contribution to the rapid advancement of computer vision applications such as image classification and object detection \citep[e.g.,][]{NIPS2012_4824,2014arXiv1409.1556S}. At the same time, CNNs have been also successfully employed in one-dimensional (sequence) data in diverse fields ranging from natural language processing \citep[e.g.,][]{2016arXiv161208083D} to medical diagnosis \citep[e.g.,][]{2017arXiv170701836R}. Recently, the effectiveness of CNNs for astronomical time-series classification has been demonstrated for transiting exoplanets and supernovae \citep[see][and references therein]{2018MNRAS.474..478P,2018AJ....155...94S,2019arXiv190100461B}. In the following, we provide a concise outline of the components of CNNs, and their basic functionality and standard architectures, in comparison with classical, fully connected neural networks.

A classical neural network consists of numerous interconnected units called neurons distributed in layers, where the $i$th neuron of layer $l$ performs the following mathematical operation on its input vector $\mathbf{x}$:

\begin{equation}
a_i^{[l+1]} = g(\mathbf{w_i^{[l]} x}+b_i^{[l]}),
\end{equation}

\noindent where the elements of the weight vector $\mathbf{w_i}$ and the bias term $b_i$ are free parameters of the model, $g$ is a nonlinear {\em activation function}, and its output $a_i$ is called an activation \citep[see, eg.,][]{hastie}. In the classical multilayer perceptron (MLP) architecture, the data are propagated through $L$ hidden layers of neurons, and the neurons of neighboring layers are fully interconnected, meaning that the vector $\mathbf{a^{[l]}}$ containing the activations of all neurons of layer $l$ serves as the input of all neurons in layer $l+1$. The final layer performs the prediction, and its activation function (and the corresponding loss function) depends on the type of the problem to be solved (i.e., regression or classification). An MLP is capable to model very complicated, highly nonlinear functions, which makes it a very versatile tool for machine learning problems on high-dimensional structured data. However, due to their fully connected architecture, they require a large number of parameters to learn spatial correlations in sequence (or image) data. CNNs, on the other hand, are best suited for the latter data types by design, their sensitivity to spatial correlations being hardwired into their model architecture.

The fundamental component of a CNN is a convolutional layer. The main part of a model consists of $L$ subsequent convolutional layers where the output of layer $l$ forms the input of layer $l+1$, and the input of the first layer is the (normalized) data. In layer $l$, the following operator is applied on the input:

\begin{equation}\label{eq:conv}
g(~* \mathcal{F}_i^{[l]} + b_i^{[l]} ),
\end{equation}

\noindent where $\mathcal{F}_i^{[l]}$ is the $i$-th filter of the layer with size $f^{[l]}$, $b_i$ is a bias term, `$*$' is the cross-correlation operator, and $g$ is a non-linear function. In more practical terms, the filter $\mathcal{F}$ is stepped over the input sequence with strides of $s^{[l]}$, and at each position, its elements are multiplied with the underlying values of the input sequence, and summed up to provide the corresponding element of the output sequence. The size (i.e., the number of elements) $f^{[l]}$ of the filters is usually in the range of [3,9] and is odd by convention. Generally, the input sequence of a layer and its filters have $n_{\rm c}^{[l]}\geq1$ number of channels, and their cross-correlation results are added up, resulting in a single-channel output per filter. Consequently, a convolutional layer with $n_f^{[l]}$ filters produces an output sequence with $n_f^{[l]}$ channels, thus $n_f^{[l]}=n_c^{[l+1]}$. 

The 2-dimensional version of the above procedure is analogous to the `convolution' operation in image processing, whereby the image is cross-correlated with specifically designed filters in order to detect low-level features, such as edges using the Sobel-filter \citep{sobel}. The substantial difference is that the elements of each filter $\mathcal{F}_i^{[l]}$ and the corresponding bias terms in a CNN are not hand-designed, but are free parameters that are learned by the model. Since the output of a convolution layer serves as the input of the next one, filters of deeper layers learn to detect more complex features. By applying the nonlinear function $g$ on the output sequence of each filter, we allow the model to learn complex interdependencies between the various intermediate features.

The advantage of a CNN with respect to an MLP is that its parameters are shared by distant elements in the input sequence, thus a CNN has much fewer parameters, hence requires smaller training sets and is less prone to overfitting. For the same reason, CNNs are also highly insensitive to translations of the input sequences, whereas the latter would highly affect the performance of MLPs.

As the input is propagated through multiple convolution layers, the length of the sequence shrinks due to the cross-correlations, unless the input is padded with a sufficient $p^{[l]}$ number of zero elements, resulting in an output sequence of length $n^{[l]} = \lfloor{ (n^{[l-1]}+2p^{[l]}-f^{[l]}) / s^{[l]} +1}\rfloor$. Padding also avoids information loss at the beginning and the end of a sequence. After a convolutional layer, the lengths of the output sequences can be reduced by applying an optional pooling (i.e., binning) layer. 
%A pooling layer essentially performs a binning of the sequence by taking the maximum (or the mean) value using a specified bin size and stride length (with optional padding) in each channel. 
The number of channels generally increases while the sequence is propagated through the model as usually an increasing number of filters are applied in consecutive layers. 

%\begin{equation}
%n^{[l]} = \Bigg\lfloor{ \frac{n^{[l-1]}+2p^{[l]}-f^{[l]}}{s^{[l]}} +1}\Bigg\rfloor,
%\end{equation}

A series of convolutional and pooling layers can be considered as an encoder that transforms the input data sequence into a into a complex, high-dimensional feature space.  Following this, the final multi-channel output sequence can be vectorized (``flattened'') and propagated into a classical fully connected neural network in order to learn even more complex interdependencies in the data \citep[e.g.,][]{NIPS2012_4824}. Alternatively, it can be reduced into a single feature vector via global pooling (i.e., taking the per-channel maximum or mean of the last output, see \citealt{2013arXiv1312.4400L}). In either case, the output from the last neural layer is directed into the final layer which performs a softmax regression to predict the class probabilities $\hat y_i\in[0,1]$, and its units have the form:

\begin{equation}
\hat y_j = \frac{\exp(\mathbf{w_j a}+b_j)}{\sum_{k=1}^{N}\exp(\mathbf{w_k a}+b_k) }.\label{eq:softmax}
\end{equation}

In Eq.~\ref{eq:softmax}, the $\mathbf{w_i}$ weight vectors and $b_i$ bias terms are free parameters, $\mathbf{a}$ is the final feature vector, $N$ is the number of classes, and the predicted class is simply the one with the highest probability. The optimal model parameters are found by minimizing the categorical cross-entropy cost function:

\begin{equation}
J = \frac{1}{M} \sum_{i=1}^{M} \sum_{j=1}^{N} -y_{ij} \log{\hat y_{ij}}~,
\end{equation}

\noindent where $M$ is the number of training examples and $y_{ij}$ is the ground-truth class vector of the training example $i$, i.e., it takes a value of 1 if the object is of class $j$ and 0 otherwise.

In both MLPs and CNNs, the partial derivatives of $J$ with respect to the model parameters can be explicitly expressed, thus $J$ can be minimized numerically using a gradient descent based optimization algorithm.

A CNN has several {\em hyperparameters} that are kept fixed during the optimization of the model parameters. Some hyperparameters determine the model complexity, such as $L$, $f^{[l]}$, $s^{[l]}$, $p^{[l]}$, the exact functional form of $g$, the number of fully connected layers if employed; others govern the optimization process of choice. They are optimized via a standard $k$-fold cross-validation procedure, whereby the training data are randomly split into $k$ disjunct sets, the model is optimized $k$ times for a fixed combination of hyperparameters, each time using the union of $k-1$ datasets for training, and the held out set for performance evaluation using some metric. The hyperparameters that optimize the performance metric of choice describe the optimal model architecture. The tuning of the hyperparameters is a largely experimental trial-and-error process relying on insight from earlier applications of CNNs, without aiming to either achieve or prove that the chosen model provides the {\em global} maximum of the performance metric among {\em all} theoretically possible architectures for the given training data.

\subsection{Light curve representation}\label{subsec:lcfit}

Since our objects of interest are monoperiodic Cepheids, we could simply use phase-folded light curves as input sequences for the CNN if the phases of the observations were identical for all objects. Since this is naturally not the case, we first compute a normalized model representation of each light curve, and evaluate them on a common phase grid. The resulting `synthetic' magnitudes are then used as input sequences for the classifier, i.e., the input features that describe the {\em shape} of the light variation; while the information on its {\em scale} is carried by two additional features, namely the period and the peak-to-valley amplitude.

We model the light-curves with a truncated Fourier-series of the form:

\begin{equation}
m(t) = \sum^M_{i=0} a_i \sin \left( i \frac{2 \pi (t-t_0)}{P} + \phi_i \right),
\end{equation}\label{eq:lcmodel}

\noindent where $P$ is the period, $t$ is the time, the amplitudes $a_i$ and phases $\phi_i$ are free parameters, and the number of Fourier terms $M$ is a hyperparameter. We perform a robust nonlinear fit employing the Huber loss function \citep{huberloss}, which uses the standard squared loss for points within a $\delta$ deviation and linear absolute deviation beyond it, resulting in decreased volatility to outliers. We used values of $0.03\leq\delta\leq0.05$ based on empirical tests. Initial values of the period were computed by the GLS method.  The optimal period, amplitudes, and phases are found by the Trust Region Reflective optimization algorithm \citep{trf} implemented in {\tt scipy}, employed in conjunction with an iterative outlier rejection. We determined the optimal value of $M$ via 10-fold cross-validation by maximizing the coefficient of determination ($R^2$ score) using the implementation in the {\tt scikit-learn} package. In the cross-validation, we stratify the folds according to the pulsation phase in order to avoid randomly introducing phase gaps. In the case of the VVV data, the above procedure was performed for all apertures, and the optimal aperture was selected to be the one that yielded the lowest Huber cost.

We phase-align the light curves by matching the phase of the first Fourier term $\phi_1$ of all objects, and define it to be the zero phase. The accuracy of this alignment step is not crucial for the classification, thanks to the translation invariance of CNNs, but it avoids important light curve features to fall close the sequence edges, i.e., phases 0 and 1. Subsequently, the fitted light curves are strandardized to zero mean and unit variance, which aids the convergence of the optimization algorithm of the classifier. Finally, the fitted model is evaluated on an equidistant grid of 38 phase values between 0 and 1, and the resulting sequences serve as the input sequence for the classifier.

\subsection{Training set}\label{sec:trset}

In order to establish a dataset for training the classifier, we surveyed the literature for high-quality $K_s$-band light curves of accurately classified fundamental-mode classical and type II Cepheids. In general, candidates for our training set were objects with both high signal-to-noise optical time-series photometry with good phase coverage (allowing their unambiguous classification and accurate period determination) and precise $K_s$-band follow-up photometry. At the time of this writing, such objects are not available in large numbers.

We compiled the training data for classical Cepheids from \citet{1984ApJS...54..547W}, \citet{1992AAS...93...93L}, \citet{1997PASP..109..645B} and \citet{2011ApJS..193...12M} for objects in the Galactic field, from \citet{2004AJ....128.2239P} for Cepheids in the LMC, and from the NIR survey of the SMC by \citet{2018MNRAS.481.4206I}. For our type II Cepheid training set, we relied entirely on the VVV photometry of objects identified toward the Galactic bulge by the OGLE-IV survey \citep{2017AcA....67..297S} and we also collected the VVV data of classical Cepheids in the same catalog.

We processed the light curves of the training set candidates according to Sect.~\ref{subsec:lcfit} but with fixed pulsation period when an accurate value from the literature was available. This was followed by a rigorous quality control by visual inspection, rejecting objects with noisy or sparse photometry and manually tuning the regression's hyperparameters when necessary. As a result, we obtained 140 classical and 356 type II Cepheids for the training set from literature data.

Due to the very limited number of classical Cepheids available for training, we selected {\em bona fide} objects from our Cepheid candidates toward the bulge area following our classification method discussed in our earlier study \citep{2015ApJ...812L..29D}. This tentative classification method is based on the (in)consistency between a star's distance and extinction under the assumptions that it is either a classical or type II Cepheid. We compute their $E(J-K_s),E(H-K_s)$ reddening values using the objects' mean magnitudes and PL relations (see Sect.~\ref{sec:extinction} for details). We also obtain the $E'(J-K_s)$ values from the extinction map of \citet{2012A&A...543A..13G} at the positions of the objects, which measures the cumulative reddening up to the mean locus of red clump stars in the Galactic bulge. We compute the corresponding $A_K$ and $A_K'$ absolute extinction values and their errors assuming the selective-to-absolute extinction ratios of \citet{2016A&A...593A.124M}.

\begin{figure}
\plotone{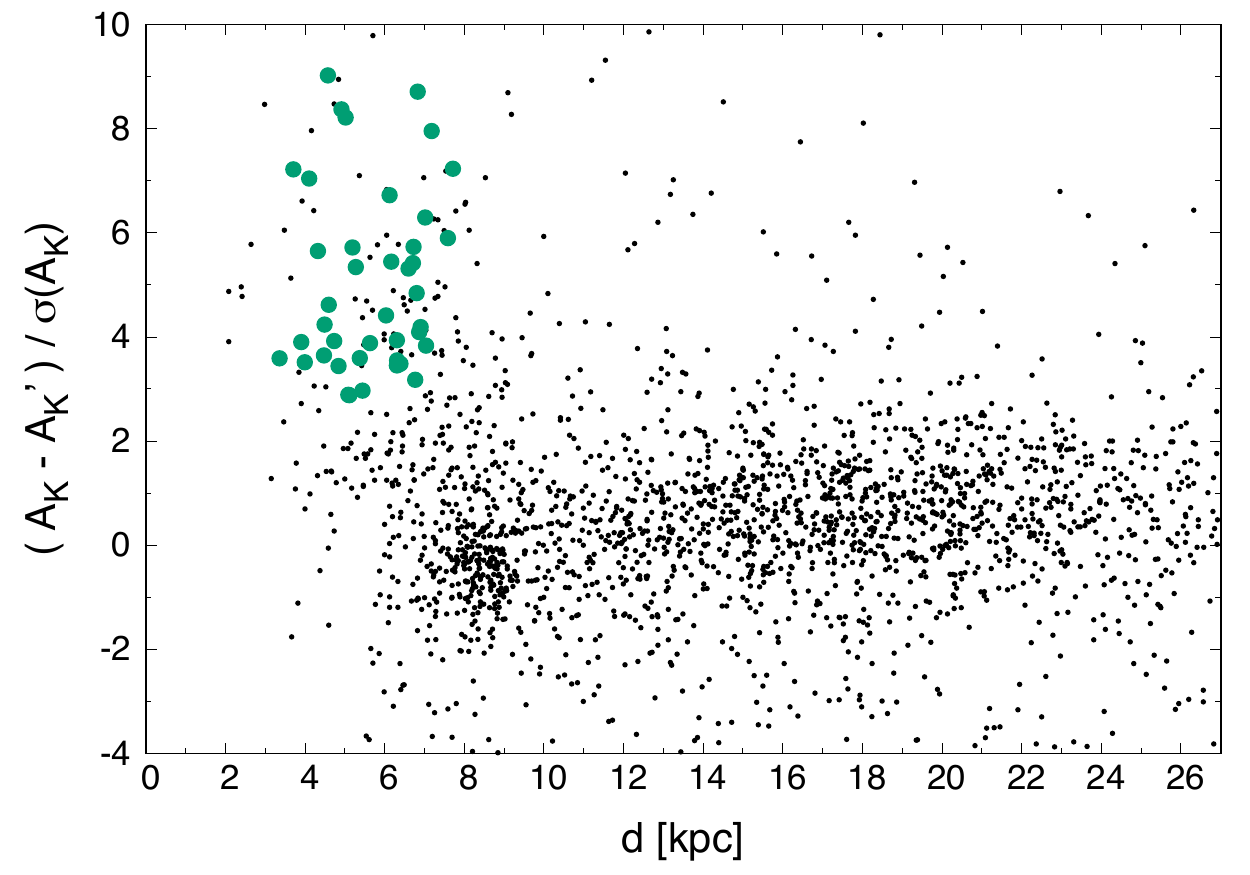}
\caption{
Weighted standard deviations of the absolute $K_s$-band extinction computed from the VVV photometry (see text) with respect to the corresponding extinction values from \citet{2012A&A...543A..13G} as a function of Heliocentric distance for our bulge Cepheid candidates (black points), under the assumption that all objects are type II Cepheids. Green points show the {\em bona fide} classical Cepheids in our sample selected for the classifier's training set.
\label{fig:trsel}}
\end{figure}

Figure~\ref{fig:trsel} shows the weighted standard deviation:
$$\delta = \frac{A_K-A_K'}{\sqrt{\sigma^2(A_K)+\sigma^2(A_K')}},$$
i.e. the difference between the two extinction values normalized with their total error as a function of Heliocentric distance $d$ for the sample of Cepheid candidates in the VVV bulge area, assuming that all of them are type II Cepheids. Since type II Cepheids are highly concentrated in the bulge volume, most of the true type II Cepheids in our sample are expected to cluster around $\delta=0$, $d\approx8$~kpc, i.e., where Fig.~\ref{fig:trsel} indeed shows an overdensity. The farther an object is from this overdensity, the less likely it is to be a real type II bulge Cepheid because of the inconsistency between its distance and extinction. Consequently, objects at the far right side of this plot are likely not Cepheids of any type, while those toward the upper-left corner of the plot are likely classical Cepheids beyond the bulge (due to the notion that their extinctions and distances become consistent under the classical Cepheid assumption).

We selected the objects in the $\delta \geq 3$, $d<7.5$~kpc range and performed a thorough visual inspection. We found 48 objects with high-quality light curves that could be visually classified as classical Cepheids with high confidence, and included these objects in our training set, raising the number of training examples for this class to 188. We note that our selection is rather insensitive to our assumption of the extinction law within the range of values reported in the literature, and we obtain the same selection using the extinction ratios derived later in Sect.~\ref{sec:extinction}.

\begin{figure}[t]
\plotone{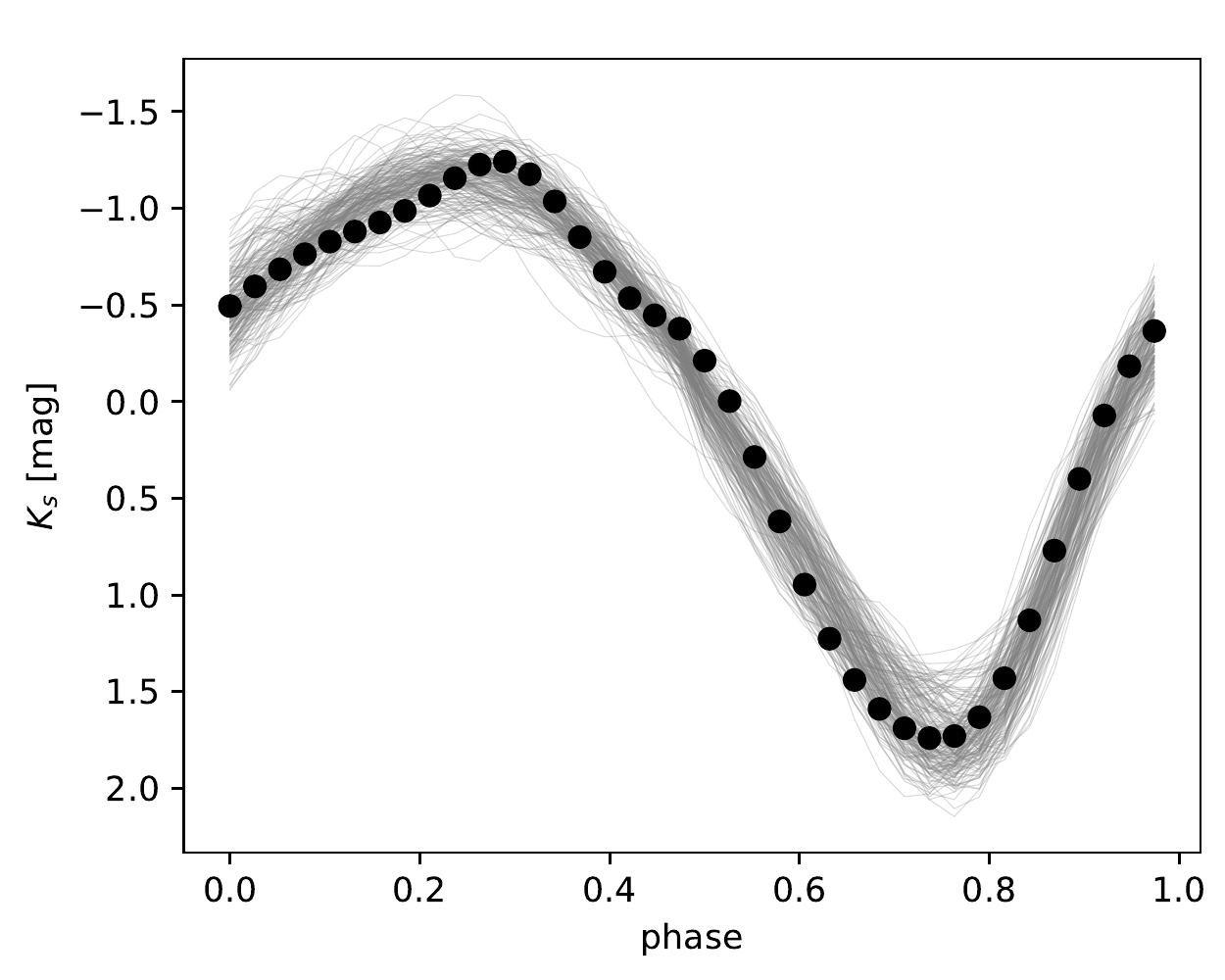}
\caption{
Phase-folded, phase-aligned and standardized light curve models of all classical Cepheids in our training set (gray curves). Black points denote the light curve model evaluated in 38 equidistant phase points for one training example, which serves as an input sequence of our CNN classifier (see Sect.~\ref{subsec:cnn}).
\label{fig:trset}}
\end{figure}

The identifiers, periods, amplitudes, and references of the classical Cepheids (DCEP) and type II Cepheids (T2CEP) in our training set are listed in Table~\ref{tab:trset}. Figure~\ref{fig:trset} shows the phase-aligned, strandardized model representations of the light curves of classical Cepheids in our training set.

%\startlongtable
\begin{deluxetable*}{lDccll}
\tablecaption{Classifier training set of classical and type II Cepheids\tablenotemark{a} \label{tab:trset}}
\tablehead{
\colhead{ID} & \multicolumn2c{period~[d]} & \colhead{amplitude~[mag]} & type & \colhead{ref. to period} & \colhead{ref. to light curve}
}
\decimals
%\colnumbers
%\decimalcolnumbers
\startdata
AQ~Pup      &    30.104     & 0.468  & DCEP & \citet{1992AAS...93...93L} & \citet{1992AAS...93...93L}     \\
BB~Sgr      &    6.63739   & 0.173  & DCEP  & \citet{1992AAS...93...93L} & \citet{1992AAS...93...93L}    \\
BETA~Dor  &    9.84247  & 0.206  & DCEP  &  \citet{1992AAS...93...93L} & \citet{1992AAS...93...93L}   \\
BF~Oph       &    4.06775  & 0.180  & DCEP &  \citet{1992AAS...93...93L} & \citet{1992AAS...93...93L}     \\
BG~Lac       &    5.33191  & 0.192  & DCEP &  \citet{1997PASP..109..645B} & \citet{1997PASP..109..645B} \\
BM~Per      &    22.9509   & 0.468  & DCEP &  \citet{2011ApJS..193...12M} & \citet{2011ApJS..193...12M}   \\
BN~Pup       &   13.6731   & 0.419  & DCEP &  \citet{1992AAS...93...93L} & \citet{1992AAS...93...93L}      \\
CN~Cep       &   9.50167   & 0.156  & DCEP &  \citet{1992AAS...93...93L} & \citet{1992AAS...93...93L}      \\
\enddata
%\tablenotetext{a}{The full version of this table is available in the online edition.}
\tablecomments{This table is available in its entirety in machine-readable form.}
\end{deluxetable*}

We collected training data for non-Cepheid variable stars in the studied period range from the VVV survey. We performed a variability analysis of the $K_s$-band photometric data of tiles b292--b296 according to Sect.~\ref{subsec:varsearch}. These fields lie toward Baade's window \citep[see][]{2010NewA...15..433M}, outside of the target area of our Cepheid search; and we chose them because they boast a high number of photometric epochs, and they lie in crowded regions, thus the data distribution is very similar to that of our target area. Moreover, the public OGLE-IV catalog of Cepheids is highly complete in these fields, enabling us to have a clean sample by excluding all known Cepheids. We collected a total of 498 light curves with high signal-to-noise ratio of non-Cepheid, but otherwise unclassified variable stars for our training set.

\subsection{Model selection and optimization}\label{subsec:modeloptim}

\noindent We experimented with a sizable variety of small CNN architectures using the components discussed in Sect.~\ref{subsec:cnn} to find the best-performing classifier. Since only a modest training set is at our disposal, we also had to apply regularization to combat overfitting. We reached an optimal bias-variance tradeoff (i.e., the golden middle between under- and overfitting the data) by tuning the model complexity (through the hyperparameters) and the regularization parameters, and evaluating the model performance by standard 10-fold cross-validation.

\begin{figure}[t]
%\plotone{cnn1.pdf}
\centering
\includegraphics[width=0.5\textwidth]{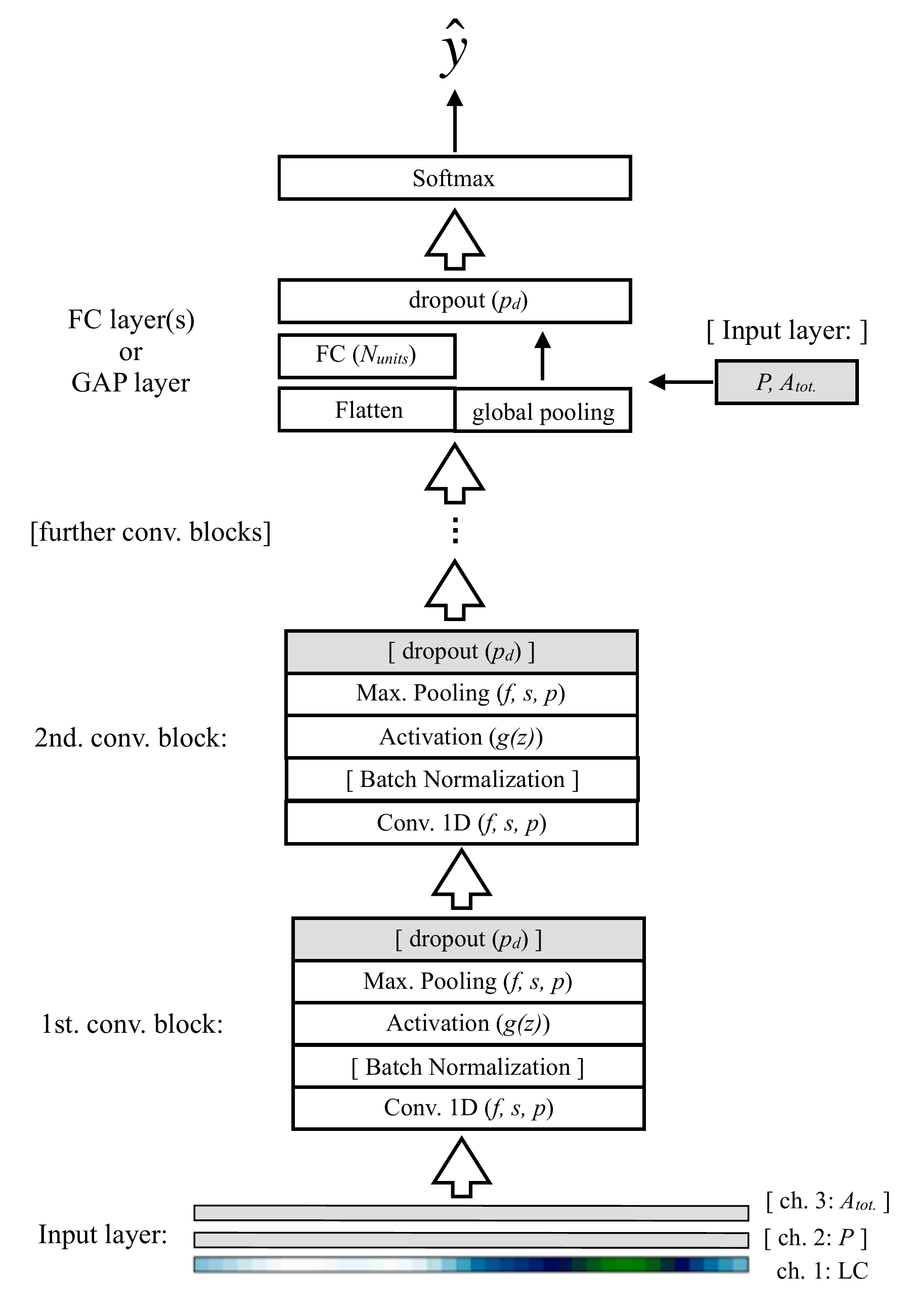}
\caption{
Summary of the CNN architectures evaluated in our model selection procedure, with data propagating from bottom to top (see text for details). Each rectangle represents a building block of the model, and their tunable hyperparameters are listed in italics. Optional model components are shown inside squared brackets, alternative model components are shown side by side. The first channel in the input layer is a standardized synthetic light curve (LC), visualized as a one-dimensional image.
\label{fig:cnn}}
\end{figure}

Figure~\ref{fig:cnn} summarizes the family of CNN architectures in our model selection process.
% We tried and evaluated a variety of small CNN architectures assembled from the basic elements discussed in Sect.~\ref{subsec:cnn}.
Following the input layer, i.e., the standardized input sequence, the first part of the CNNs consists of a small number (2--4) of convolution blocks. Each block starts with a convolutional layer, with its input formed by the output of the previous block (or from the input layer in case of the first block), followed by an activation layer (see Eq.~\ref{eq:conv}), and a pooling layer. The activation is optionally preceded by an additional layer performing batch normalization \citep{bnpaper}.
%whereby the output of the previous layer is normalized by subtracting its mean and dividing by its standard deviation.
Modern optimization algorithms use only a subset of the training data (a ``mini-batch'') in every iteration, thus the data in consecutive iterations might suffer from covariate shift, which slows down the optimization. Batch normalization eliminates this effect, and also has a slight regularization effect on the network. 

The output of the last convolutional block is either flattened (vectorized) and fed into a classical, fully connected (FC) neural network with 1--2 hidden layers, or propagated into a global pooling layer. The output of this is fed into the final softmax unit which predicts the class of the input light curve.
% by outputting 3 probabilities, one for each of the 3 classes.

We add regularization to our models using the Dropout method by \citet{dropoutpaper}. This technique is based on the idea of preventing a neural network's units from excessive co-adaptation to the data by randomly dropping units from the network during training with some probability $P_d$, which is a tunable hyperparameter. Once trained, the full network is applied to the target data by appropriately scaling down its parameters. We apply dropout before the final softmax layer and (optionally) after the convolution blocks.

We discussed how our sequence data are propagated forward through the network, but our input data consist of two additional features, namely the period and the amplitude. There are two different ways of including such ancillary data in a CNN: (i) by using a second input layer after the convolution blocks, and concatenating them with the outcoming feature vector or (ii) adding them along with the sequence data in the first input layer using additional channels. In the latter, the two additional channels consist of the standardized values of $P$ and $A$, each repeated in its respective channel to match the sequence data in length. The latter method requires a larger number of parameters in the CNN than the former, but enforces the entire network to co-adapt to all the input data, which might lead to a better solution.

We implemented the CNN models in the TensorFlow \citep{2016arXiv160304467A} and Keras \citep{keras} programming frameworks.
%\footnote{The source codes are publicly available via GitHub.}
The models were trained using the Adam optimization method \citep{2014arXiv1412.6980K} using a mini-batch size of 256 and a learning rate of 0.005, and leaving the rest of its hyperparameters at their default values. We iterated the training process through several thousand training epochs until convergence was reached on the training set. We evaluated the performance of each model using the standard classification accuracy as our performance metric:

\begin{equation}\label{eq:accuracy}
\mathcal{A} = \frac{N_{tp} + N_{tn}}{N_{tp} + N_{tn} + N_{fp} + N_{fn}},
\end{equation}

\noindent where $N_{tp}, N_{tn}, N_{fp}, N_{fn}$ are the number of true positives, true negatives, false positives, and false negatives in the validation set, respectively.

\begin{figure*}
\plottwo{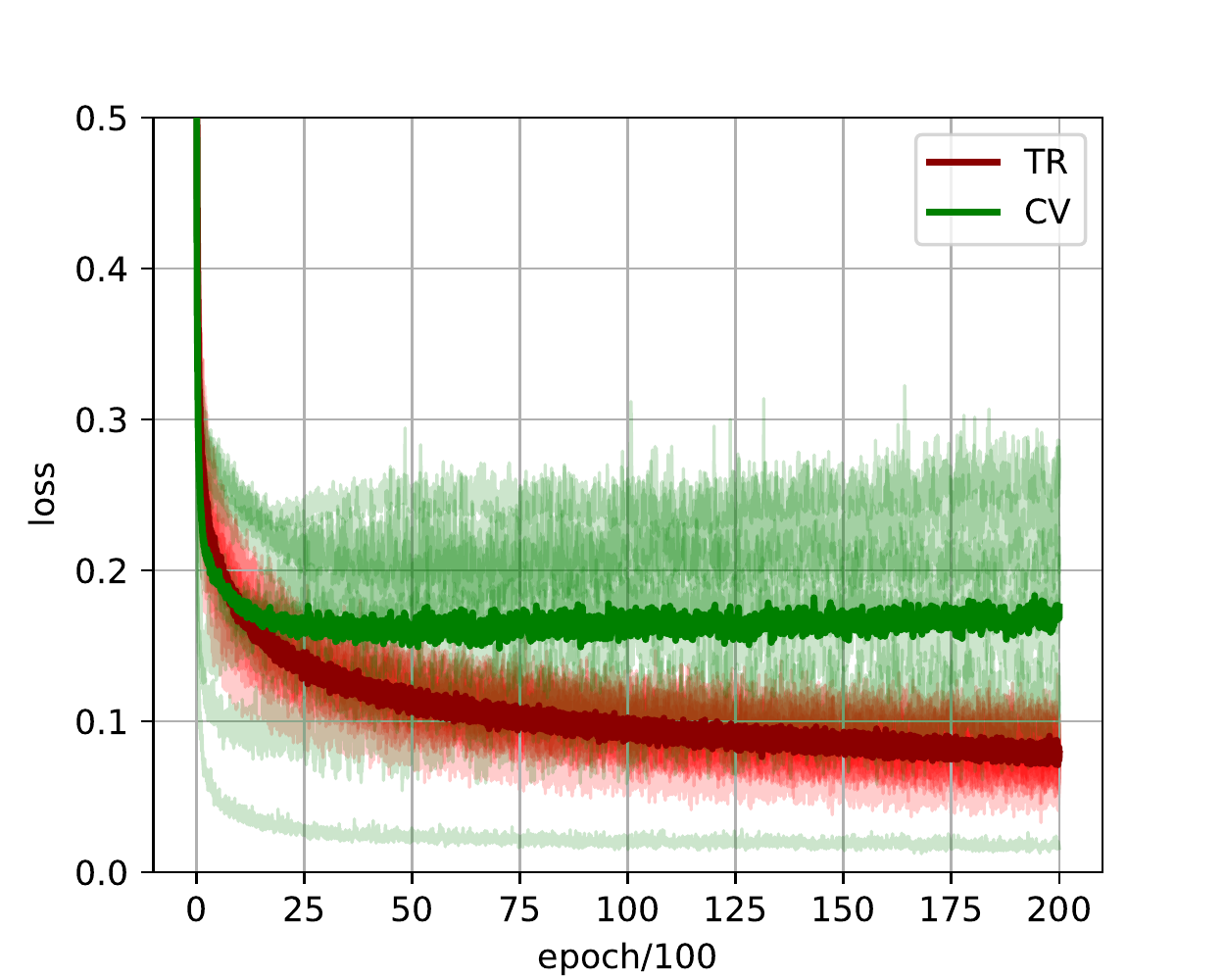}{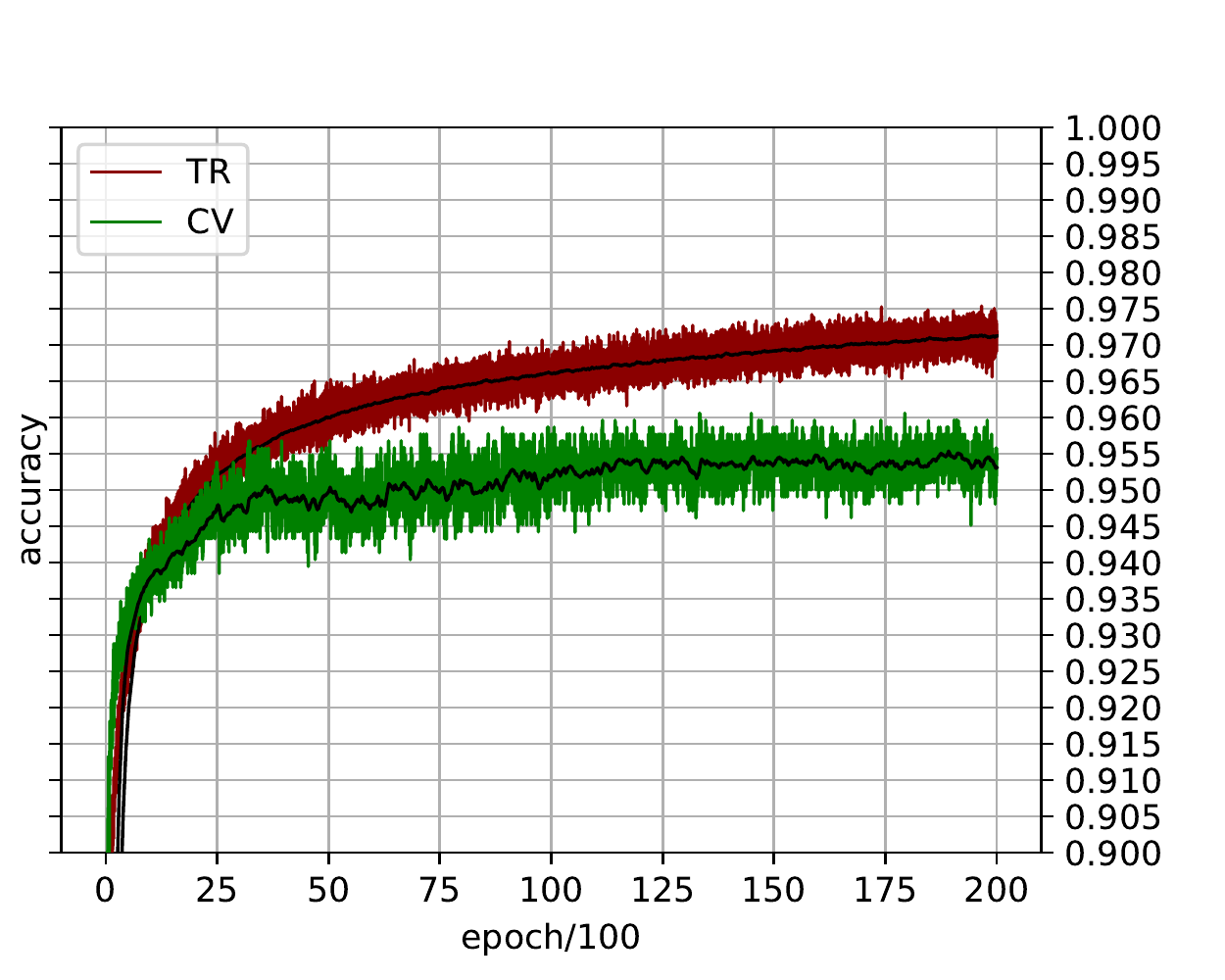}
\caption{
Left: training (TR, red) and cross-validation (CV, green) loss as a function of training epochs. Semi-transparent thin curves show the loss per fold, thick curves show their mean. Right: classification accuracy of the training and validation sets as a function of training epoch (with same notation as above). The black curves show exponentially weighted moving averages.
\label{fig:loss_acc}}
\end{figure*}

The architecture of the best-performing model is summarized by Table~\ref{tab:bestcnn}. The model has a 3-channel input layer including the normalized light curve, as well as $P$ and $A$. The convolutional part consists of two blocks including batch normalization and pooling, followed by a global maximum pooling (GMP) layer and a softmax layer. We also use two Dropout layers, one after the first convolutional block, the second after the GMP layer. For the non-linear function in the activation layers, we used the Exponential Linear Unit (ELU) introduced by \citet{2015arXiv151107289C}. The model has a total of 4179 parameters. Figure~\ref{fig:loss_acc} shows the evolution of the loss and the mean accuracy during the training of our best CNN. The values of the cross-validation set converge to a constant value close to the asymptotes of the training set, showing a good bias-variance tradeoff.

%\startlongtable
\begin{deluxetable*}{lllr}
\tablecaption{Properties of our convolutional neural network classifier\label{tab:bestcnn}}
\tablehead{
\colhead{Layer} & \colhead{hyperparameters} & \colhead{output shape} & \colhead{num. of param.}
}
%\decimals
\startdata
Input & none & (38,3) & 0 \\
Convolution & 24 filters, $f=3$, $s=2$, $p=1$ & (19,24) & 240 \\
Batch Norm. & none & (19,24) & 96 \\
Activation & $g=ELU$ & (19,24) & 0 \\
Max. Pooling & $f=2$, $s=2$, $p=0$ & (9,24) & 0 \\
Dropout & $P_d=0.14$ & (9,24) & 0 \\
Convolution & 48 filters, $f=3$, $s=2$, $p=1$ & (5,48) & 3504 \\
Batch Norm. & none & (5,48) & 192 \\
Activation & $g=ELU$ & (5,48) & 0 \\
Global Max. Pooling & none & (1,48) & 0 \\
Dropout & $P_d=0.5$ & (1,48) & 0 \\
Softmax & none & (1,3) & 147 \\
\enddata
\tablecomments{Direction of data propagation is from top to bottom. Hyperparameters $f$, $s$, $p$, $P_d$, and $g$ denote filter size, stride length, padding size, dropout probability, and activation function, respectively. Our notation for a layer's output shape is {\em (sequence length, number of channels).}}
\end{deluxetable*}

\subsection{Performance, data mismatch}\label{subsect:performance}

\noindent The best approach to obtain an unbiased estimate of the classification performance is applying the model on an explicit test set, i.e., part of the labeled data set used neither for training, nor for cross-validation, commonly amounting to some $20\%$ of all labeled data. Unfortunately, in our case this approach is undesirable due to the modest amount of the labeled data available (see Sect.~\ref{sec:trset}): an explicit test set would either withdraw too much information from the training and cross-validation, thus limiting its accuracy, or it would be too small to provide a reasonable estimate on the performance. Hence we use the cross-validation set to estimate our final CNN's efficiency, which is also a common approach in the machine learning literature, noting that it might provide slightly upward biased estimates.

\begin{figure}
\plotone{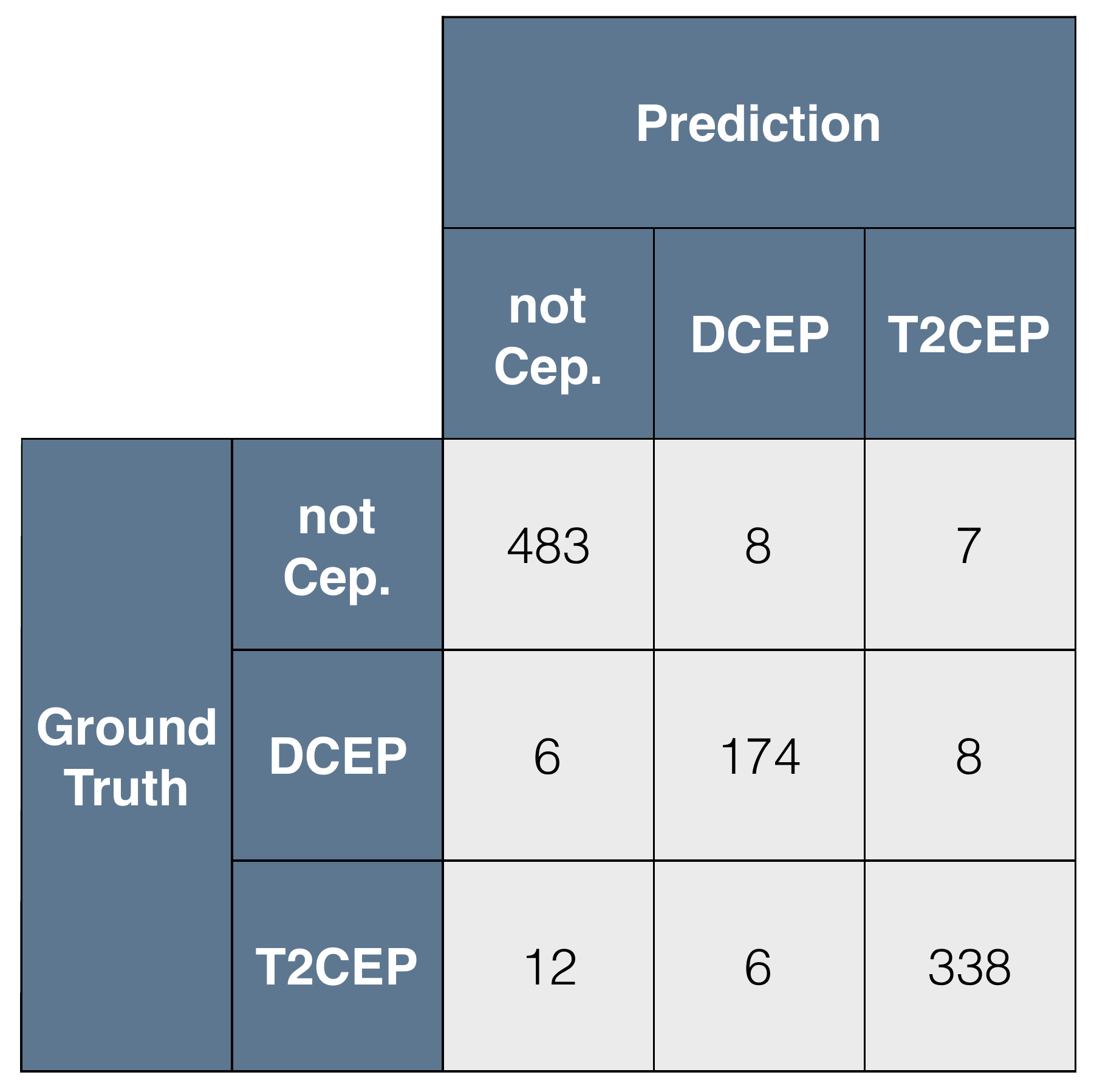}
\caption{
Confusion matrix of our CNN classifier measured by 10-fold cross-validation.
\label{fig:confmat}}
\end{figure}

Figure~\ref{fig:confmat} shows the confusion matrix measured on the validation set, comparing the prediction to the ground truth for each class. The accuracy (Eq.~\ref{eq:accuracy}) is the ratio of the sum of the diagonal elements and the sum of all elements of this matrix. We also computed various performance estimates for the two binarized classification problems of: (i) classical Cepheids {\em vs} everything else, and (ii) type II Cepheids {\em vs} everything else. We use the the following metrics: precision $\mathcal{P}$ (i.e., purity), recall $\mathcal{R}$ (i.e., completeness), false alarm probability $FAP$ and $F_1$ score, defined as:

\begin{eqnarray}\label{eq:metrics}
\mathcal{P} &=& N_{tp}/(N_{tp}+N_{fp}) \\
\mathcal{R} &=& N_{tp}/(N_{tp}+N_{fn}) \\
FAP &=& N_{fp}/(N_{fp}+N_{tn}) \\
F_1 &=& 2\mathcal{P}\mathcal{R}/(\mathcal{P}+\mathcal{R}).
\end{eqnarray}

\noindent The resulting values for each performance metric are displayed in Table~\ref{tab:perf}.

\begin{deluxetable}{ccc}
\tablecaption{Performance metrics for our CNN classifier measured by 10-fold cross-validation.\label{tab:perf}}
\tablehead{
\colhead{Metric} & \multicolumn2c{Value}
}
%\decimals
\startdata
$\mathcal{A}$ & \multicolumn2c{0.95} \\
\hline
& DCEP & T2CEP \\
\hline
$\mathcal{A}$ & 0.97 & 0.97 \\
$\mathcal{P}$ & 0.93 & 0.96 \\
$\mathcal{R}$ & 0.93 & 0.95 \\
$F_1$ & 0.93 & 0.95 \\
$FAP$ & 0.02 & 0.02 \\
\enddata
\end{deluxetable}

The classification performance on the target (i.e., unlabeled) data set might differ from the performance estimated on the labeled data set (regardless of whether we use the cross-validation set or an explicit test set for the estimation) in case their data distributions are significantly different. This is commonly referred to as the data mismatch problem. A simple example for this is if our labeled data set used for training and cross-validation consists of light curves with high signal-to-noise ratio, whereas our target set is composed of noisy data (e.g., faint objects). A performance measure obtained via cross-validation will probably overestimate the classifier's true performance on noisy data. Moreover, the model that yields optimal performance on the high-quality data set might not even be the optimal model for the classification of the noisy data.

A common approach to combat the data mismatch problem is to perform the model selection and evaluation by using the high-quality data for the training set, and use a cross-validation set whose distribution matches that of the target data. However, in case of a large photometric survey such as the VVV, the data distribution in the light curves is highly variable. It is not only a continuous function of the object's flux (i.e., apparent magnitude), but also a very complicated function of the position of the object due to spatially varying levels of crowding, the sampling (which also depends on the apparent magnitude), and even the location on the detector; rendering the aforementioned approach for handling data mismatch unrealistically complicated. Consequently, this approach is not adopted (and the problem of data mismatch is generally disregarded) by the astronomical literature.

However, we still intend to estimate the effect of the data mismatch affecting our classifier by using a more straightforward data synthesis approach. We collected a large amount of light curves from highly crowded regions of the VVV survey for sources that do not show any significant intrinsic variability. These light curves, consisting of characteristic noise of the VVV photometry, were added to the signals of randomly selected examples from our training set. This way we created $\sim$33,000 synthetic light curves with known classes. We applied our CNN classifier on this test data set and compared its predictions to the ground truth to estimate its performance.

\begin{figure*}
\plottwo{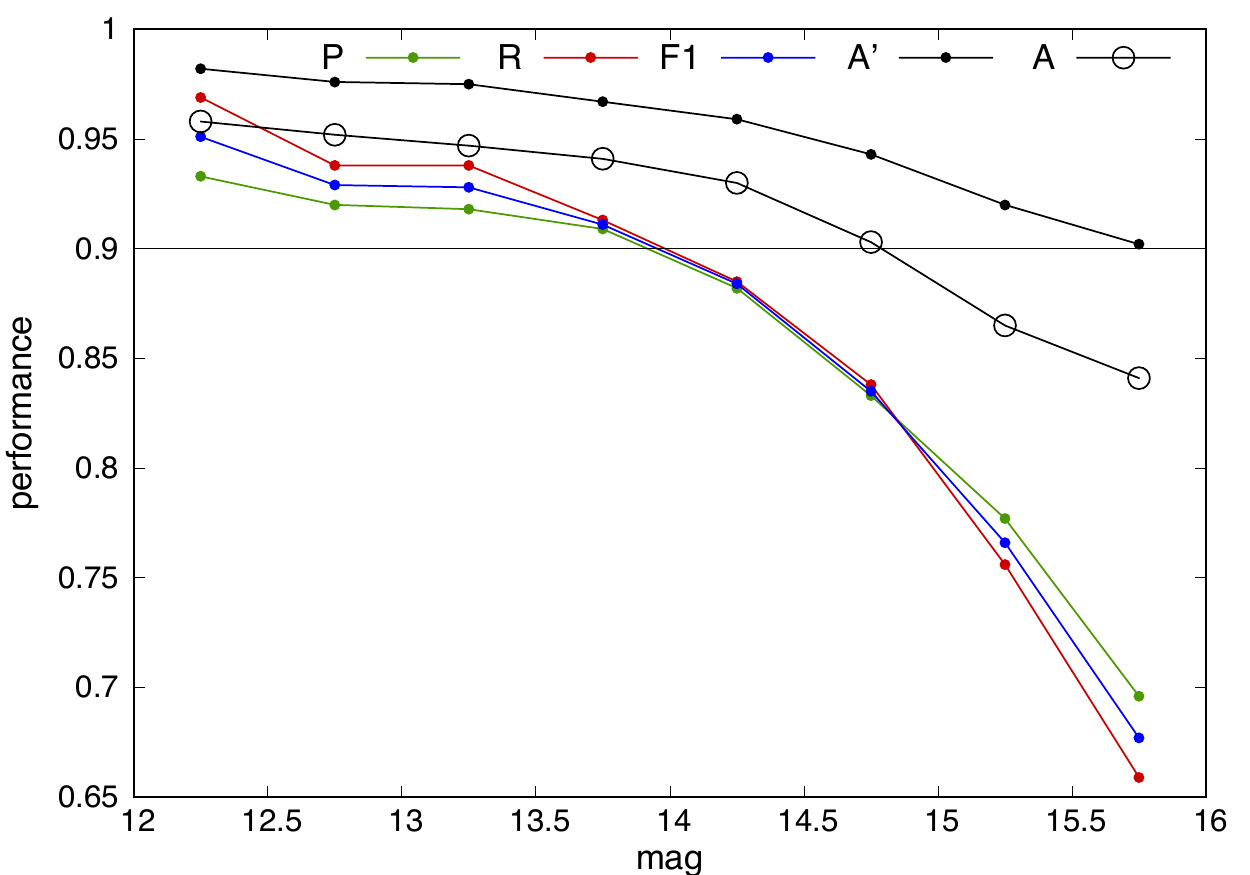}{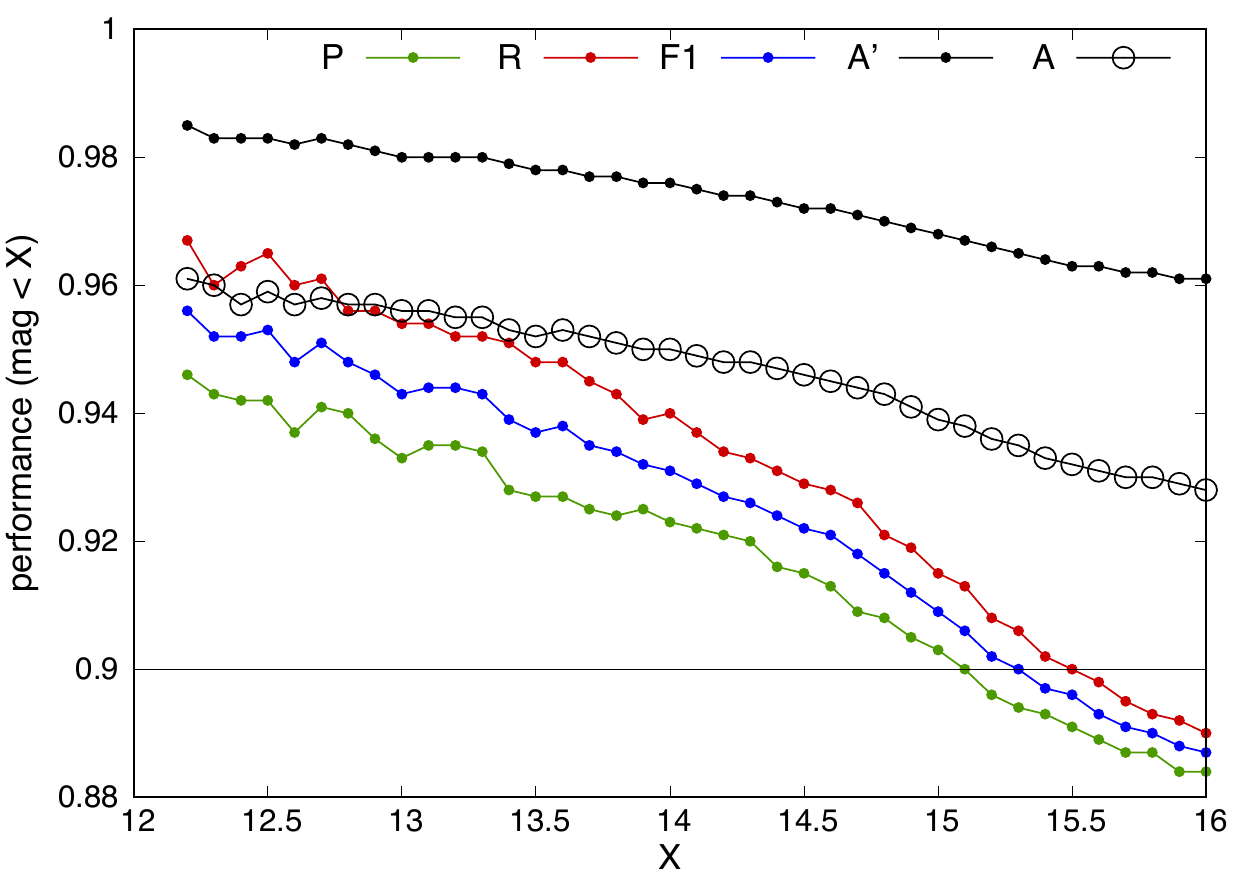}
\caption{
Left panel: performance estimates as a function of mean apparent $K_s$ magnitude, using a bin size of $0.5$~mag. Right panel: as in left panel but showing cumulative distributions. The figure keys have the following notation: A: overall accuracy; P, R, F$_1$, A': binary precision, recall, F$_1$ score, and accuracy for classical Cepheids, respectively.
\label{fig:perf_mag}}
\end{figure*}

Figure~\ref{fig:perf_mag} shows the distributions of various performance measures as a function of mean apparent $K_s$ magnitude. We note that these estimates might have a slight positive bias, since we relied on the same labeled data set that we used for the training and model selection. However, the distributions capture the effect of the objects' brightness (highly correlated with their signal-to-noise ratio, $S/N$) on the classifier's efficiency. We can observe that the performance stays above $0.9$ for stars brighter than $\sim$14~mag in all metrics, and it falls rapidly beyond $\sim$14.5~mag. The cumulative distribution shows, however, that both precision and recall stay above $0.9$ for a sample of objects brighter than 15~mag (the distribution of apparent magnitudes matches that of the VVV survey).

\subsection{Deployment and visual inspection}

\noindent We applied the CNN classifier described in Sect.~\ref{sec:classif} on our dataset of Cepheid candidates selected according to Sect.~\ref{subsec:varsearch}. In order to keep the expected overall performance reasonably high, we rejected all candidates with $S/N<60$ and phase coverage $\Phi<0.85$ from further analysis. Cepheids with noisier light curves are generally indistinguishable from other types of variable stars because the noise washes out characteristic details from their phase diagrams; and lower phase coverage causes most of our light curve fits to diverge, biasing the input of the CNN. By applying the cut on $S/N$, we omit a varying  fraction of stars increasing with apparent magnitude, with approximately $\sim$7$\%$ for $K_s<13$, $\sim$13$\%$ for $13<K_s<14$, $\sim$30$\%$ for $14<K_s<15$, and $\sim$54$\%$ of stars fainter than 15~mag.

The light curves classified as DCEP or T2CEP were visually inspected in order to increase the sample's purity, and objects with obvious misclassification were excluded from further analysis. The light curves rejected in this way fall into two main categories: obvious photometric blemishes and misclassified eclipsing binaries. The former arise from the fact that our training set does not cover all peculiar light curves of arbitrary shape, therefore our model has to extrapolate the parameter space covered by the training set, which can result in a small number of false positives, but these can be extremely easily found and rejected. The latter is due the confusion of Cepheids with contact or semi-detached binaries with one of their minima being under-sampled. Such cases occur in the disk area where the time distribution of the photometric measurements is often strongly clustered due to the lack of an appropriate sampling strategy.

\begin{figure*}
%\plottwo{0720706060_dff.pdf}{0720706060_dff_2p.pdf}
\gridline{\fig{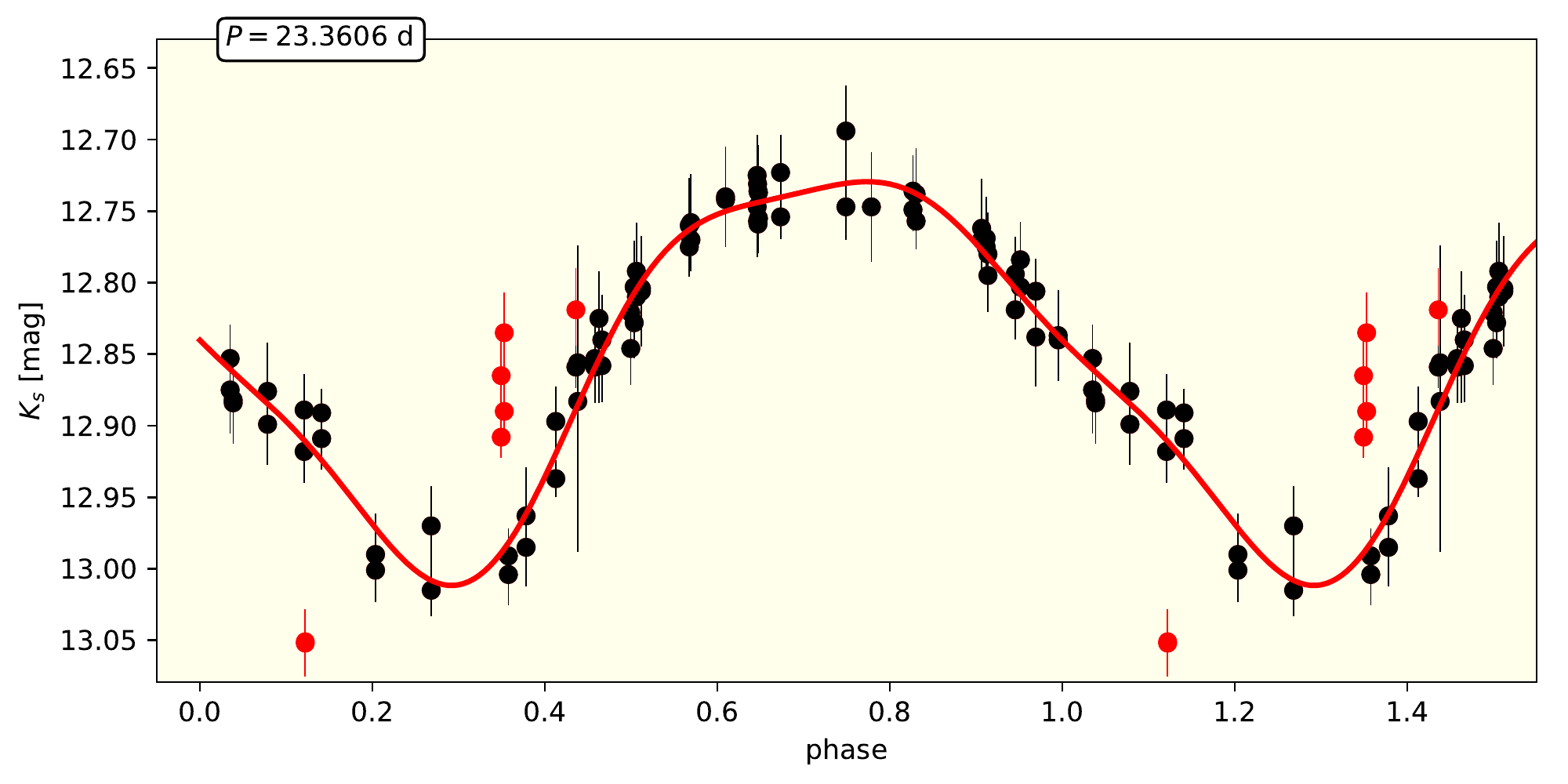}{0.5\textwidth}{}
\fig{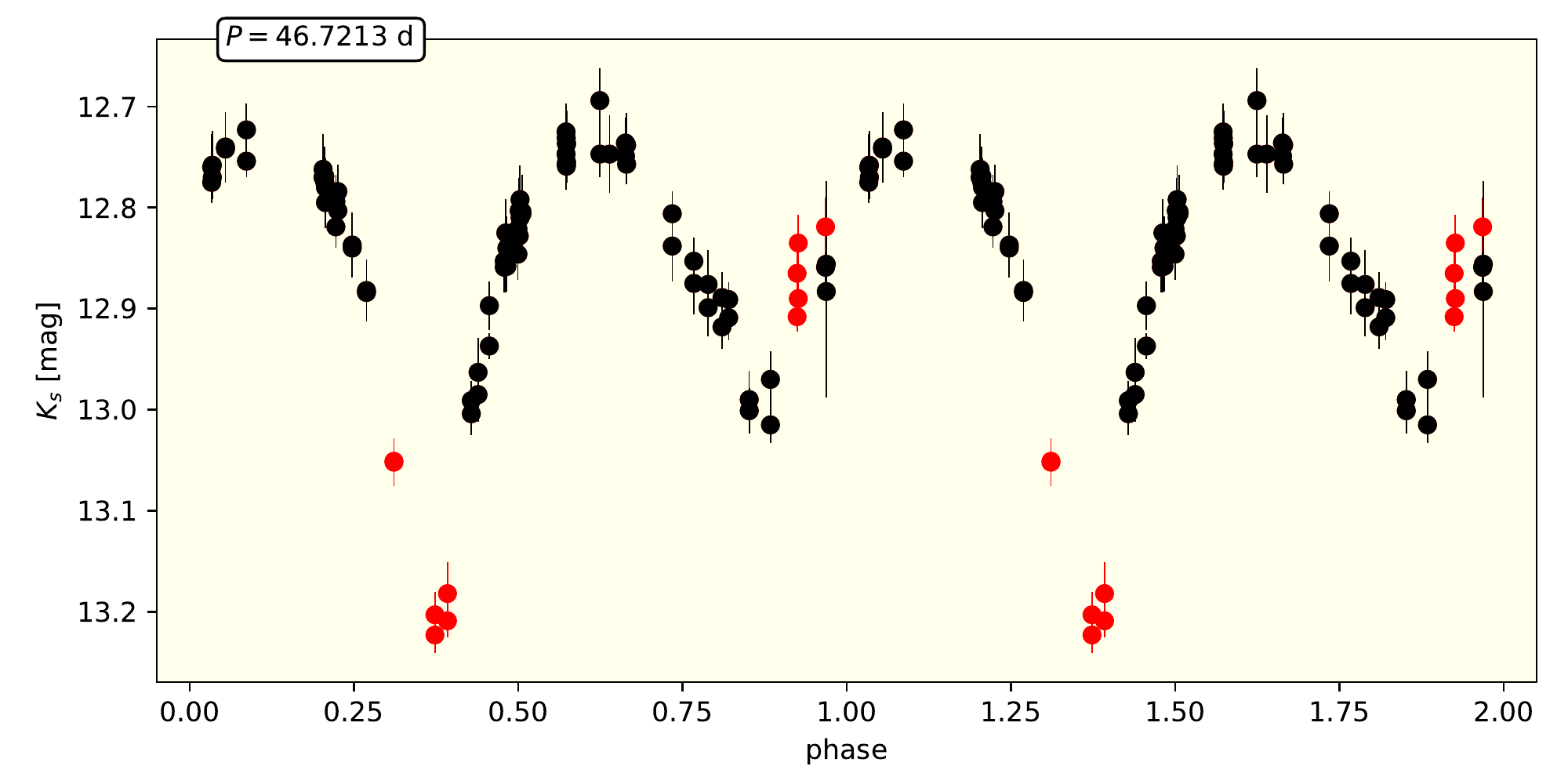}{0.5\textwidth}{}}
\caption{
Light curve of an eclipsing binary star misclassified as a classical Cepheid by our CNN classifier. The photometric measurements are phase-folded with the (incorrect) period found by our algorithm (left panel) and its doubled value (right panel). The red curve shows the fitted light curve model. Red points show data rejected by our regression algorithm.\label{fig:binary}}
\end{figure*}

Figure~\ref{fig:binary} illustrates the above problem with a concrete example. The primary minimum is under-sampled, which causes our regression algorithm to converge to an incorrect solution that mimics a Cepheid light curve. The misclassification is easily revealed by the visual inspection of the phase diagram created with twice the period fitted by our algorithm. As a result of the visual inspection, approximately $15\%$ of the light curves classified as Cepheid by the CNN were omitted from further analysis.

\subsection{Comparison with the OGLE-IV catalog of Galactic disk Cepheids}

The OGLE-IV survey collected time-series photometry in the optical $V$ and $I$ bands over the entire area of our study, and conducted a census of classical and type II Cepheids along the Galactic mid-plane. The resulting catalog of Cepheids \citep{2018AcA....68..315U} was publicly released shortly before the submission of our present paper. We investigated the overlap between these OGLE Cepheids and the catalog resulting from our analysis. The OGLE-IV disk catalog contains 1529 objects, among which 170 are type II, and 854 are fundamental-mode classical Cepheids. First, we cross-matched the coordinates of the OGLE Cepheids in the latter two subtypes with our broad selection of point-sources showing putative light variations, i.e., a sample of $\sim$660,000 objects that entered our period analysis (see Sect.~\ref{subsec:varsearch}). This resulted in 144 matching objects, $\sim$50$\%$ of which are brighter than 12 mag in the $K_s$ band. We note that the precision of the  positional cross-matching between OGLE and VVV is very high, with a residual rms of $\sim0.1''$ therefore astrometry is not a limiting factor in the number of cross-matched objects. The rest of the OGLE Cepheids are simply beyond the saturation limit of the VVV survey.

Subsequently, we cross-matched the positions of the same OGLE Cepheid sample with our stellar subsample that fulfilled the criteria of entering our classification procedure, which resulted in only 41 matches. Most of the missing objects have too many (or all) of their photometric measurements marked as saturated and were therefore omitted from the analysis, and the rest had too small $S/N$ and/or too sparse phase coverage in VVV. Based on their visual inspection, we concluded that only less than half of them have sufficiently high-quality $K_s$ light curves to be used as training data for classification. Therefore, we opted to use the 41 objects in common as a test set, in oder to give an independent estimate of our classification performance (assuming that the OGLE classifications represent the ground truth). According to OGLE, 38 of the objects in common are classical and 3 of them are type II Cepheids. Our CNN classification differs for 3 classical Cepheids, which we classified as type II, and for one type II Cepheid, which we classified as non-Cepheid. 

The common sample does not allow us to estimate the classification precision because we cannot measure the number of false positives, i.e., we do not know whether the objects that are classified as Cepheids by our CNN model and are missing from the OGLE catalog because they are to faint or because they are not Cepheids. On the other hand, the common sample yields an estimated recall of 0.92 for classical Cepheids with respect to the OGLE sample, with the caveat that it is based on small number statistics.

In the following analysis, for all objects present in both the OGLE Cepheid catalog and our sample, we use classifications provided by OGLE, overriding the results from our CNN classifier.

\subsection{The final sample of Cepheids}\label{subsec:finalsample}

\begin{figure*}
\centering
\gridline{\fig{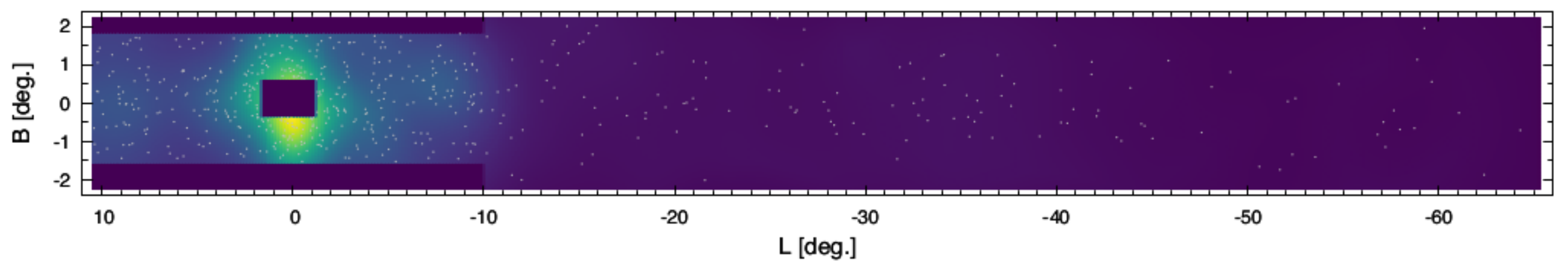}{\textwidth}{}
             }
\gridline{\fig{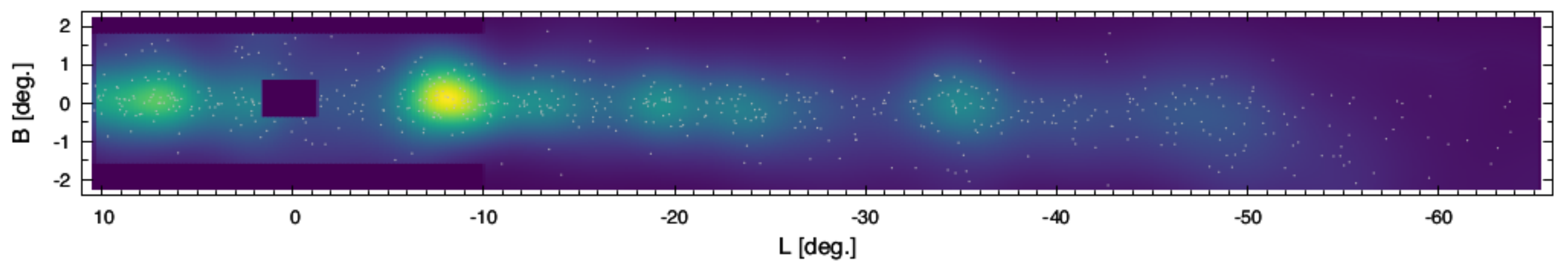}{\textwidth}{}
          }
\caption{
Distribution of objects classified as type II Cepheids (top) and classical Cepheids (bottom) in our final sample, shown in the Galactic coordinate system. Individual objects are represented by gray dots, the color scale represents their kernel density estimate.
\label{fig:lb_cep}
}
\end{figure*}

\noindent After the visual inspection step, our final sample contains 608 objects classified as type II Cepheids. Their celestial distribution is shown in the upper panel of Fig.~\ref{fig:lb_cep}. Their majority, 433 objects, are located in the bulge section of the VVV survey, strongly concentrated around the sight-line of the Galactic center. Among these objects, 82 had been previously catalogued by the OGLE-IV survey \citep{2017AcA....67..297S}, and 4 of them were earlier discovered and classified as classical Cepheids by us \citep{2015ApJ...812L..29D}. Two of the latter objects were also included in the samples of \citet{2016MNRAS.462..414M} and \citet{2018ApJ...859..137C}. The rest of the bulge type II Cepheid sample, 347 objects, are new discoveries. Among the 175 objects in the disk section of VVV, 6 stars appear in the catalog of \citet{2018ApJ...859..137C} with ambiguous classifications, and the remaining 169 stars are newly discovered. 

A total of 689 objects in our final sample are classified as classical Cepheids, among which 238 objects are located in the bulge, and 451 in the disk section of the survey. The distribution of their Galactic coordinates is displayed in the lower panel of Fig.~\ref{fig:lb_cep}. Three of the bulge objects were previously discovered by OGLE \citep{2017AcA....67..297S}, 29 had been reported in our former studies \citep{2015ApJ...799L..11D,2015ApJ...812L..29D}, 11 were found by \citet{2016MNRAS.462..414M}, and 7 objects of the latter two samples are common. 16 classical Cepheids from the disk area are listed in the catalog of \citet{2018ApJ...859..137C}, who gave this class to only 6 of them; and 38 classical Cepheids in our disk sample were previously found by the OGLE-IV survey. In total, 640 classical Cepheids in our sample cannot be found elsewhere in the literature, therefore we consider them as new discoveries. Table~\ref{tab:catalog} lists the names, coordinates, classes, periods, $K_s$ photometric properties  and cross-identifications of all Cepheids in our final sample. The photometric time-series in the $J$, $H$, and $K_s$ bands of all objects classified as classical or type II Cepheids are provided in Table~\ref{tab:tseries}.

\begin{figure}
%\plotone{period_hist_log.pdf}
\includegraphics[width=0.47\textwidth]{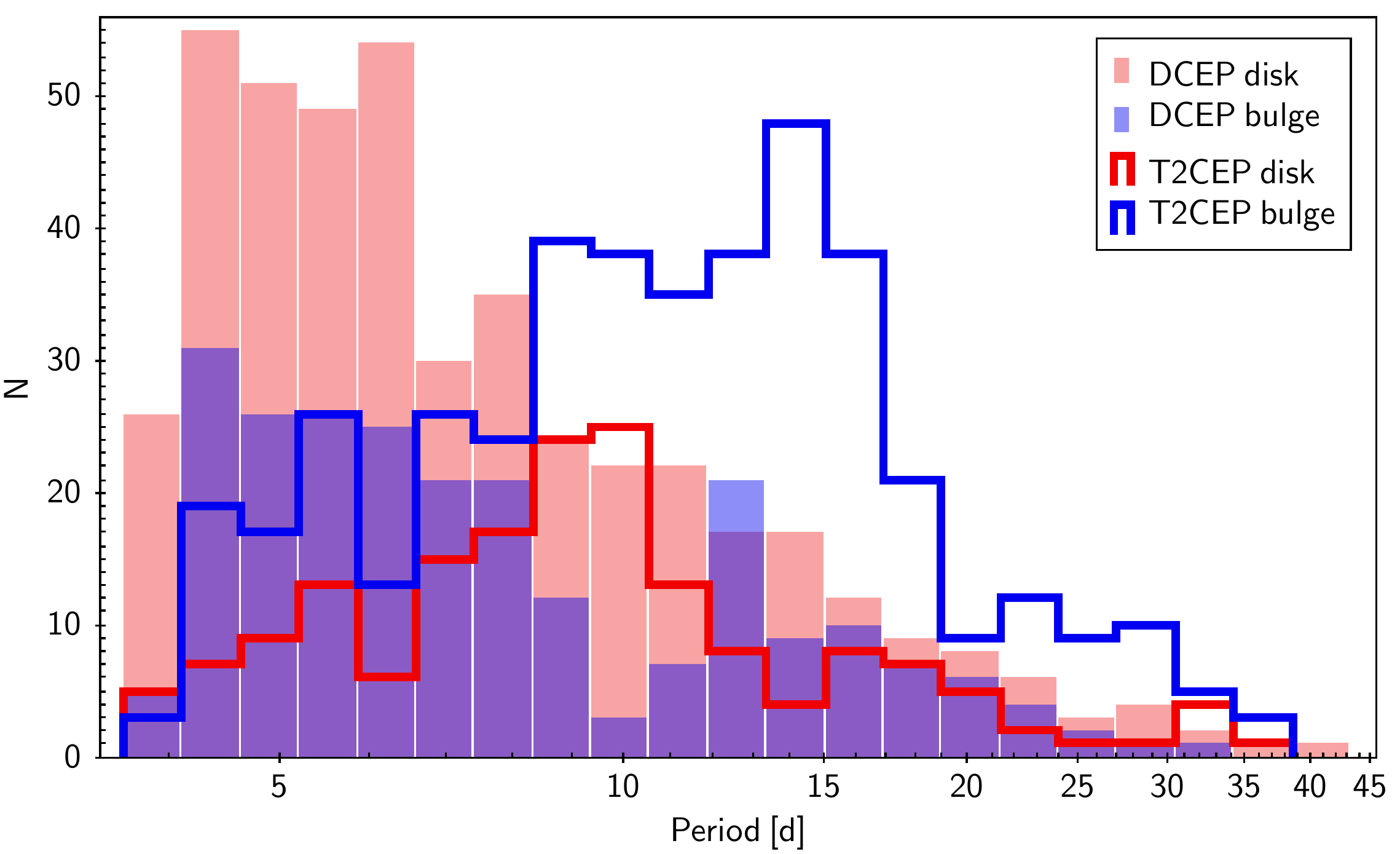}
\caption{
Histogram of the periods of the objects classified as classical Cepheids (DCEP, filled bars) or type II Cepheids (T2CEP, step curves) shown on a logarithmic scale. The blue and the red colors represent subsamples from the VVV survey's bulge and disk footprints, respectively.
\label{fig:period_hist}}
\end{figure}

Figure~\ref{fig:period_hist} shows the period distributions of the Cepheids identified in the VVV survey's bulge and disk footprints. There is an apparent deficiency of classical Cepheids with $P\simeq10$~d in the bulge footprint. This is due to the higher confusion rate between the two types of Cepheids, arising from the rapid change of light curve shape due to the Hertzsprung progression \citep[][]{1926BAN.....3..115H}. Around this period, the Hertzsprung bump in classical Cepheids coincides with the brightness maxima of the light curves, and the blending of the two features in noisy data makes it more difficult to distinguish between the two Cepheid types. The effect of the resulting confusion on the classical Cepheids' period distribution is more pronounced in the bulge subsample due to the intrinsically high concentration of type II Cepheids toward these sight-lines.

\begin{deluxetable*}{ccccDDcDDc}[]
\tablecaption{Coordinates, classes, basic photometric properties, and cross-identifications of the Cepheids. \label{tab:catalog}}
\tablehead{
\colhead{Name} & \colhead{RA~[hms]} & \colhead{DEC~[dms]} & \colhead{Class} & \multicolumn2c{Period~[d]} & \multicolumn2c{$\langle K_s \rangle$} & \colhead{Aperture\tablenotemark{a}} &  \multicolumn2c{$S/N$} &  \multicolumn2c{$\mathcal{A}_{\rm tot.}$} &\colhead{Cross-identifications}}
\decimals
\startdata
  1  &    11:40:12.93  & -62:08:29.0  & DCEP   &    4.87971    &  11.099  & 5   &   120.5  &  0.20  & J114012.8-620829\,(WISE) \\
  2  &    11:47:32.59  & -62:32:45.7  & DCEP   &    9.34493      &  15.161  & 2   &   60.7    &  0.25  & \dots  \\
  3  &    11:48:04.58  & -62:41:21.9  & T2CEP   &    21.09985  &  11.431  & 4   &   88.8      &  0.58  & \dots  \\
  4  &    11:52:15.57  & -62:51:11.6  & DCEP   &    4.50209    &  11.204     & 5  &    200.5  &   0.20 & OGLE-GD-CEP-0714  \\
  5  &    11:54:05.12  & -62:04:58.0  & T2CEP   &    10.22196  &  13.781  & 4   &   126.1  &  0.32  & \dots  \\
\enddata
\tablenotetext{a}{Optimal aperture (see Sect.~\ref{subsec:lcfit}).}
\tablecomments{This table is available in its entirety in machine-readable form.}
\end{deluxetable*}

\begin{deluxetable}{ccDDDD}[]
\tablecaption{Photometric time-series of the Cepheids. \label{tab:tseries}}
\tablehead{
\colhead{Name} & \colhead{filter} & \multicolumn2c{HJD-2400000} & \multicolumn2c{mag.} & \multicolumn2c{mag. err.} & \multicolumn2c{ZP err.}}
\decimals
\startdata
  1  &    $J$       &    57114.51688    &  13.672 &  0.004  &  0.010  \\ 
  1  &    $J$       &    57114.51778    &  13.684 &  0.005  &  0.010  \\
  1  &    $K_s$   &    55702.490324 &  11.194  & 0.026   & 0.026   \\
  1  &    $K_s$   &    55702.490733 &  11.178  & 0.025   & 0.025   \\
  1  &    $K_s$   &    55764.511305  & 11.108  & 0.025   & 0.024   \\
\enddata
\tablecomments{This table is available in its entirety in machine-readable form.}
\end{deluxetable}

\section{Interstellar extinction}\label{sec:extinction}

\subsection{Estimation of the mean color indices}\label{sec:cipred}

\noindent Both classical and type II Cepheids are excellent tracers of the interstellar reddening, thanks to their precise PL relations. We estimate the color excess $E(X-Y)$ of a Cepheid as follows:
\begin{equation}
E(X-Y) = \langle X-Y \rangle - (M_X - M_Y)\,, \label{eq:ce}
\end{equation}

\noindent where $M_X$ and $M_Y$ are the absolute magnitudes of the object in the photometric bands $X$ and $Y$, predicted by the corresponding  PL relations, and $\langle X-Y \rangle$ is the object's apparent color index averaged over its pulsation cycle. For our sample, however, we cannot obtain an unbiased estimate of the latter from the star's mean magnitudes in the $J,H,K_s$ bands due to the small number of measurements in the $J$ and $H$ bands.

In order to get an unbiased estimate of the mean $J-K_s$ and $H-K_s$ color indices, we express them in the following way:
\begin{equation}
\langle X-K_s \rangle = \langle X(\varphi) - F_{K_s}(\varphi) - \Delta_{X-K_s}(\varphi) \rangle\,,  \label{eq:cipred}
\end{equation}
\noindent where $X(\varphi)$ is the measured apparent magnitude in the $X\in\{J,H\}$ bands at pulsation phase $\varphi$, $F_{K_s}(\varphi)$ is the apparent $K_s$ magnitude at the same pulsation phase predicted by the light curve model (Eq.~\ref{eq:lcmodel}), and the last term of Eq.~\ref{eq:cipred} is the deviation of the $X-K_s$ color index in phase $\varphi$ from its mean value. 

Traditionally, the $\Delta_{X-K_s}$-like correction terms in similar problems are trivially estimated from light curve templates, i.e., using the mean color index variations computed from the observational data of an ensemble of objects \citep[e.g.,][]{2015A&A...576A..30I}. In order to avoid the information loss inherent to this method, in an earlier study we took a machine-learning approach for computing accurate mean color indices of RR~Lyrae stars based on VVV photometry \citep{2018ApJ...857...55H}, whereby the deviation of $J$ magnitudes from their mean was estimated for a grid of phases from the principal component amplitudes of the $K_s$ light curves via linear regression.

%whereby we estimated the deviation of the $H$ and $J$ magnitudes from their mean values for a grid of phases from the principal components of the $K_s$ light curves via linear regression.

For the current analysis, we opted to employ artificial neural networks to obtain a machine-learned model for predicting the $\Delta_{X-K_s}(\varphi,\mathbf{p})$ correction terms from the $\mathbf{p}$ parameter vector of the $K_s$ light curve of the same object. The advantage of this approach, in contrast with the aforementioned methods, is that it utilizes the full information content of the training set, and instead of data binning or dimensionality reduction, it mitigates observational noise and overfitting by regularization.

We assembled a training data set by compiling photometric time-series of Cepheids from the literature observed in the combination of the $J$, $H$, and $K_s$ bands. Similarly to the classification problem in Sect.~\ref{sec:classif}, at the time of this writing, only a modest amount of training data are available for solving this regression problem. For the classical Cepheids, we relied on the photometric data of 
LMC Cepheids by \citet{2004AJ....128.2239P} and Galactic field Cepheids by \citet{2011ApJS..193...12M}. The training data for type II Cepheids were collected from the studies of \citet{2006MNRAS.370.1979M,2013MNRAS.429..385M}. All of these datasets consist of simultaneous $JHK_s$ photometry, allowing us to obtain color curves by simply subtracting the measurements in the corresponding bands, and then fitting a light curve model as described in Sect.~\ref{subsec:lcfit}, using the literature values for the periods. The same procedure was applied to the $K_s$-band data, yielding the $\mathbf{p}=(P,a_i,\phi_i)$ parameter vectors, which were used as the descriptive variables (features) of the regression problem, together with $\varphi$. The fitted models of the color index variation were evaluated over a grid of 50 equidistant pulsation phases, which serve as the dependent variables in the regression. We enforce a periodic boundary condition by slightly extending the phase grid beyond the $[0,1]$ interval for the training data.

We used fully connected MLP architectures as implemented in the {\tt scikit-learn} programming framework. All models had output layers of a single neuron with linear activation and standard squared loss function and Tikhonov ($L_2$) regularization. The model parameters were fitted using the Adam optimization algorithm \citep{2014arXiv1412.6980K}. We varied the number of hidden layers, the number of neurons in each layer, and their activation functions, and also tried different subsets of light curve parameters in the $\mathbf{p}$ feature vector; and evaluated each model's performance via 10-fold cross-validation using the $R^2$ score as our metric. 

We found that a 4-layer MLP with 100 neurons in each layer using rectified linear unit (`relu') activations \citep{relu} gave the best performance in each regression problem. The optimal values of the regularization parameter and the cross-validation $R^2$ scores are summarized by Table~\ref{tab:lcpred} for each regression problem. We could not fit a reasonable model for predicting $\Delta_{H-K_s}$ for type II Cepheids due to the small amount and low $S/N$ of the available data, but concluded that the amplitude of this color variation probably does not exceed a few hundredths of a magnitude. Therefore, we assume $\Delta_{H-K_s}=0$ for these objects throughout this study.

\begin{deluxetable}{c|cc|c}[t]
\tablecaption{Optimal regularization parameters and cross-validation $R^2$ scores of the color index estimators\label{tab:lcpred}}
\tablehead{
& \multicolumn2c{DCEP} & \colhead{T2CEP}
}
%\decimals
\startdata
& $\Delta_{J-K_s}$ & $\Delta_{H-K_s}$ & $\Delta_{J-K_s}$ \\
\hline
$\alpha$ & 0.023 & 0.005 & 0.028 \\
$R^2$ & 0.907 & 0.371 & 0.701 \\
\enddata
\end{deluxetable}

\begin{figure*}
\gridline{\fig{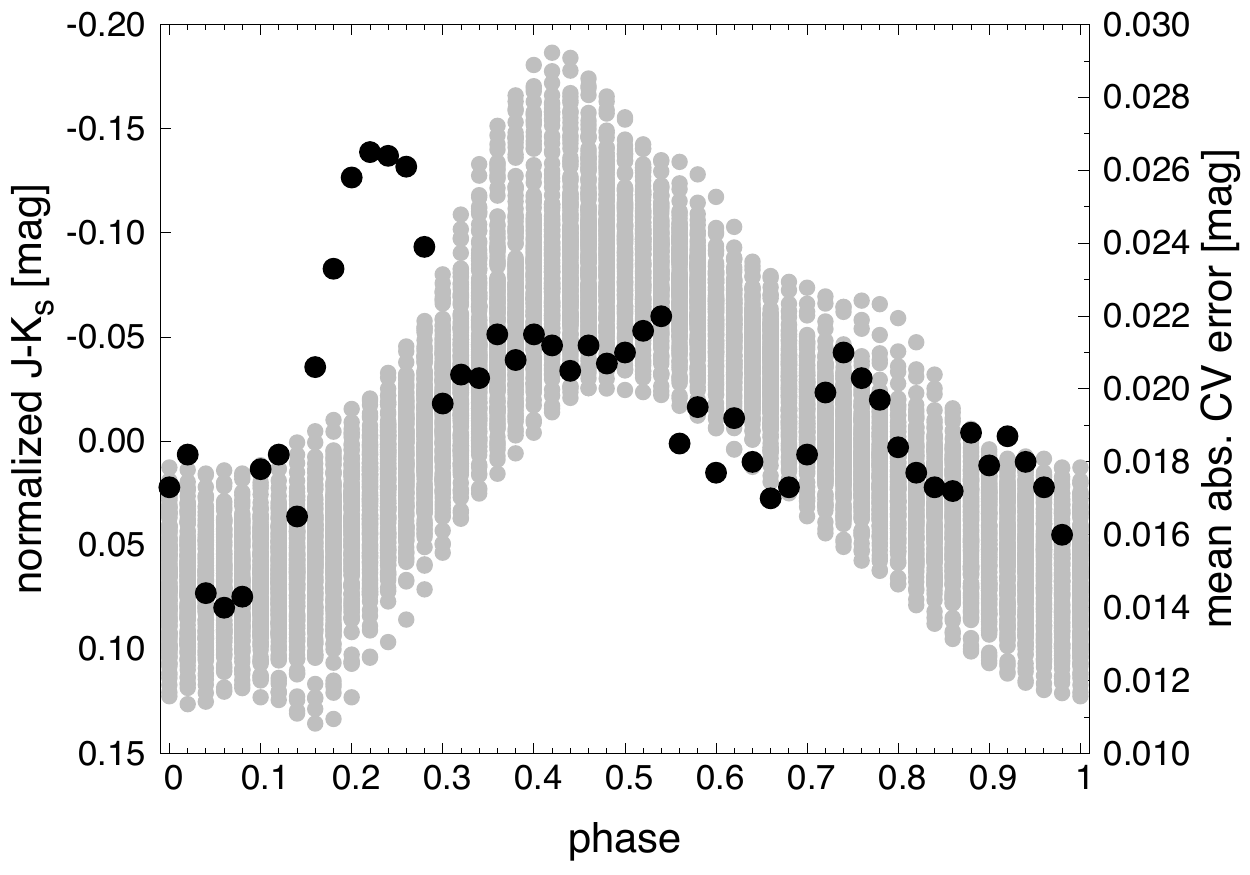}{0.33\textwidth}{}
          \fig{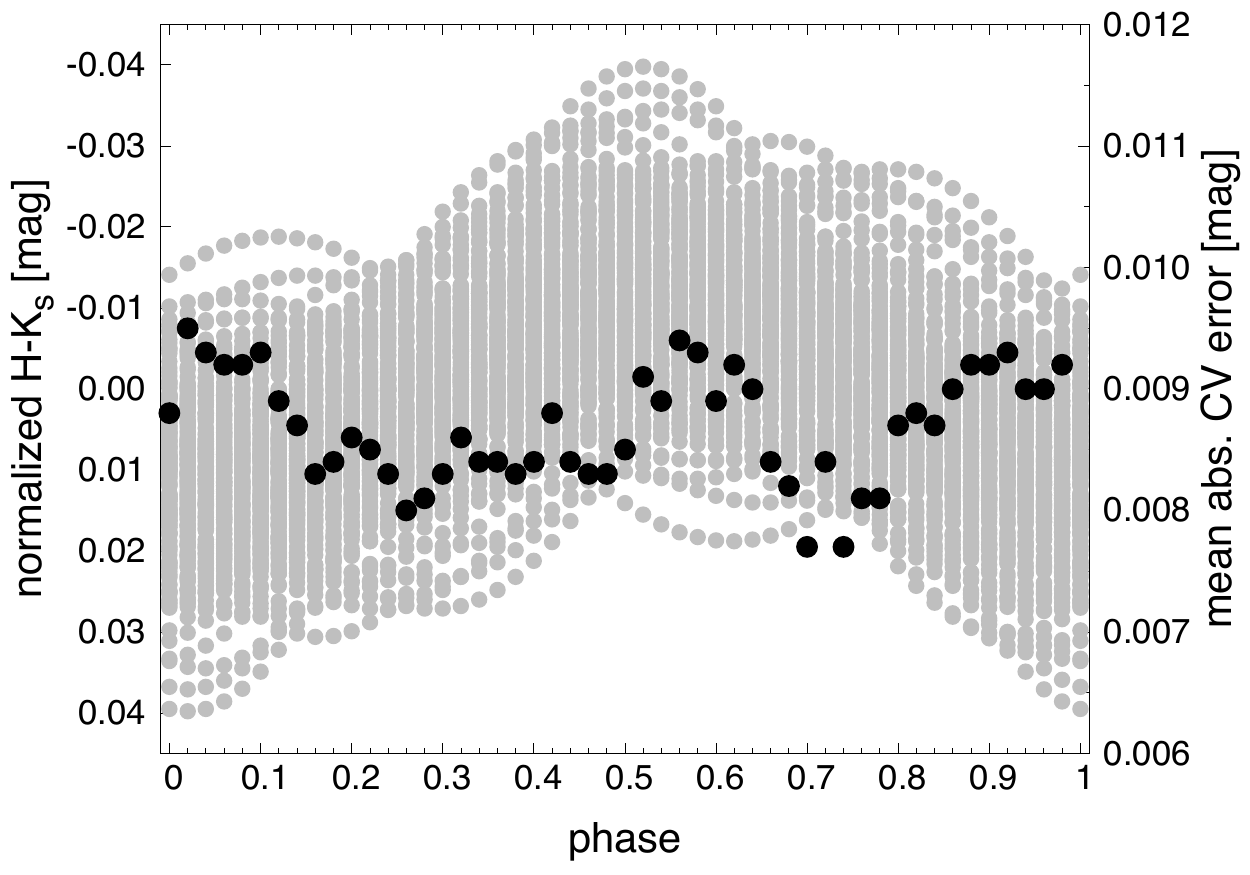}{0.33\textwidth}{}
          \fig{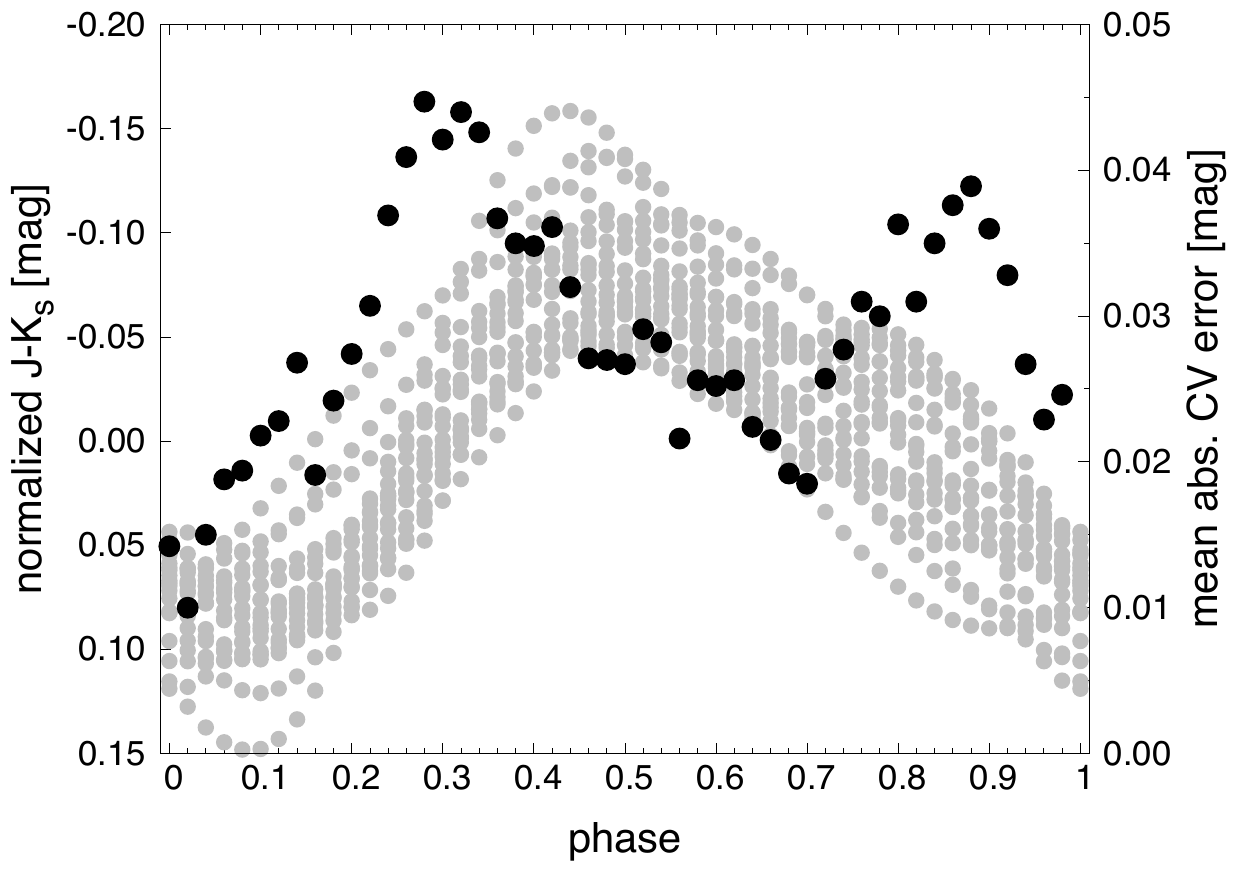}{0.33\textwidth}{}
          %\fig{jmk_mean_cv_error_t2cep.pdf}{0.33\textwidth}{}
          }
\caption{
Gray points: normalized and phase-aligned color curves of the classical Cepheids (left and middle) and type II Cepheids (right) in the training sets of our predictive models of the color index. Black points: mean absolute cross-validation error of our predictive models as a function of pulsation phase. The ordinate scales corresponding to the gray and black points are shown in the left and right sides of the panels, respectively.
\label{fig:jhkpred}
}
\end{figure*}

\begin{figure*}
\gridline{\fig{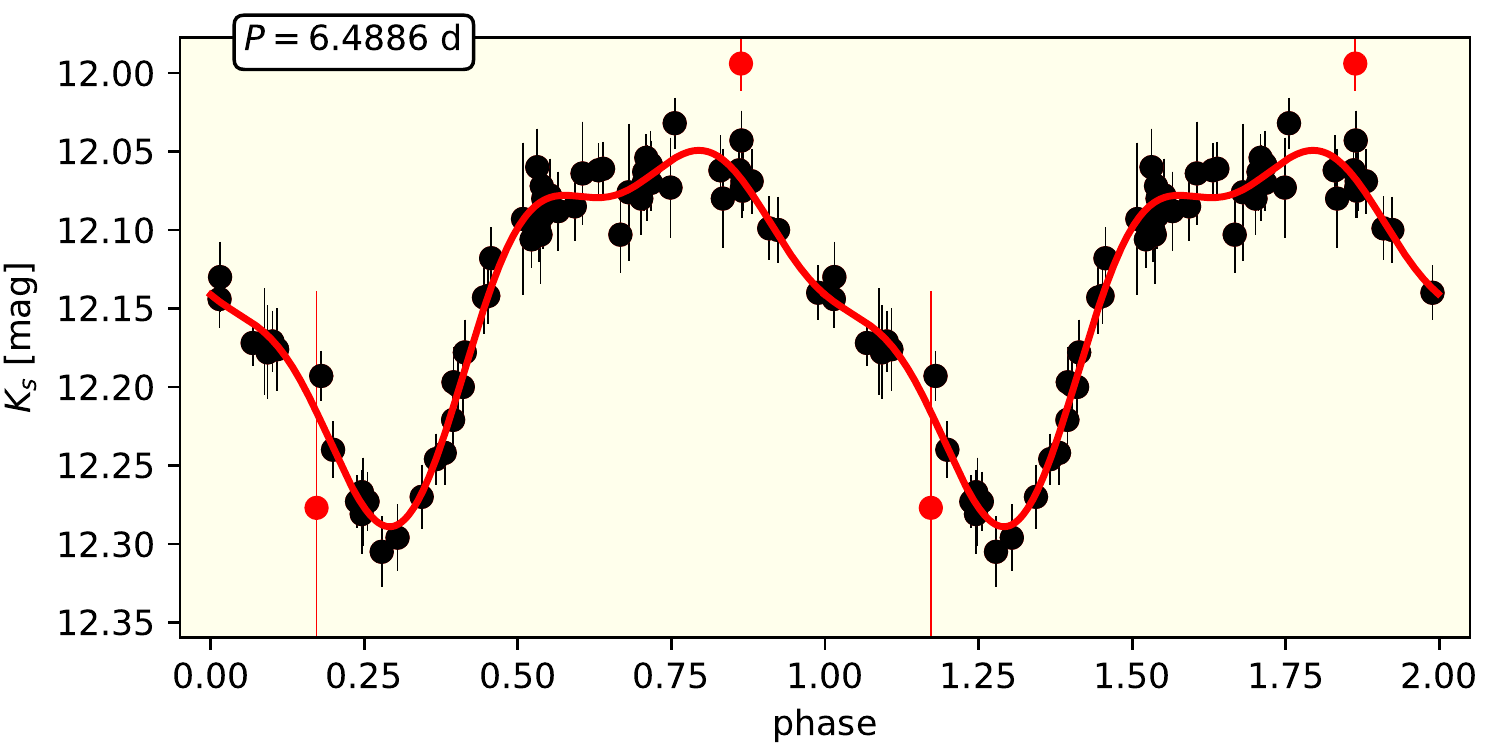}{0.5\textwidth}{}
          \fig{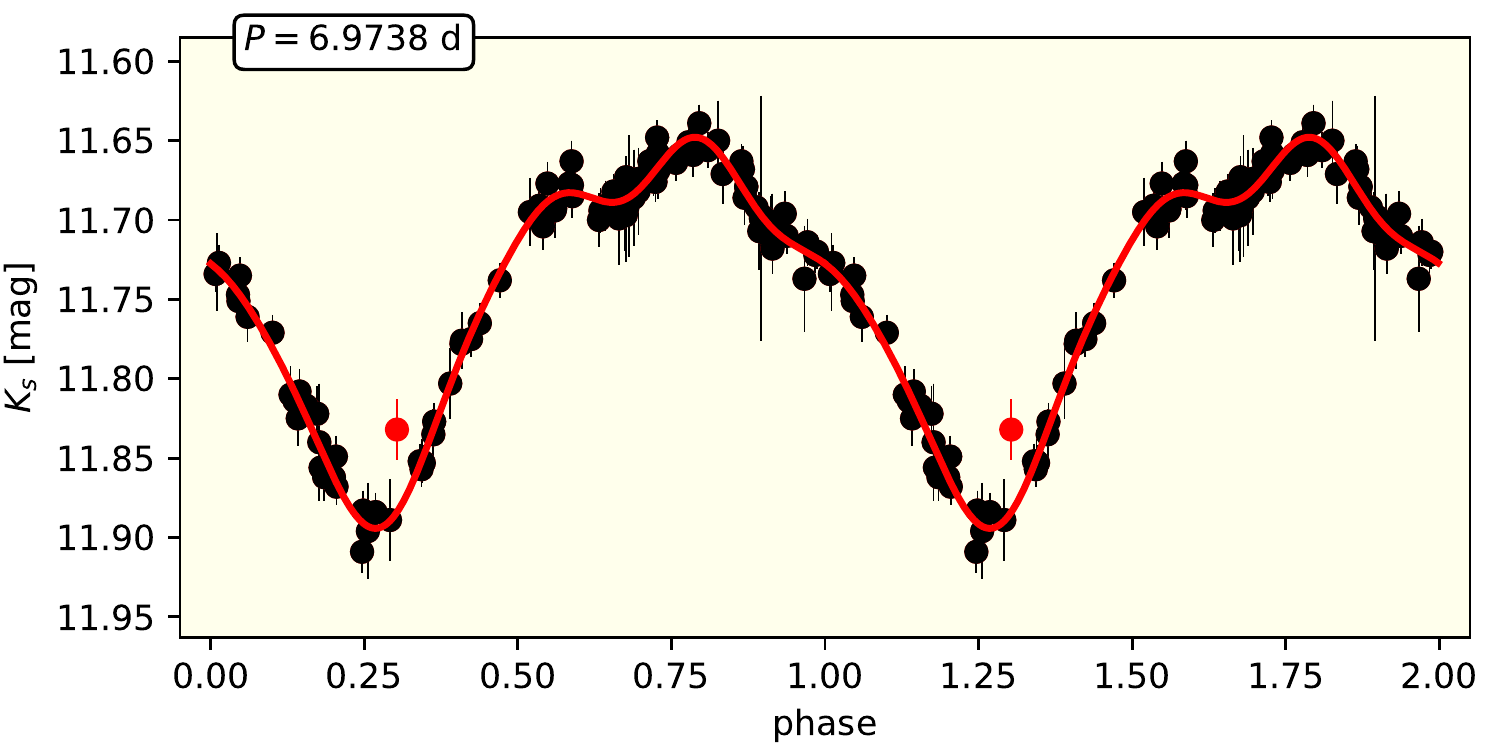}{0.5\textwidth}{}
          }
\gridline{\fig{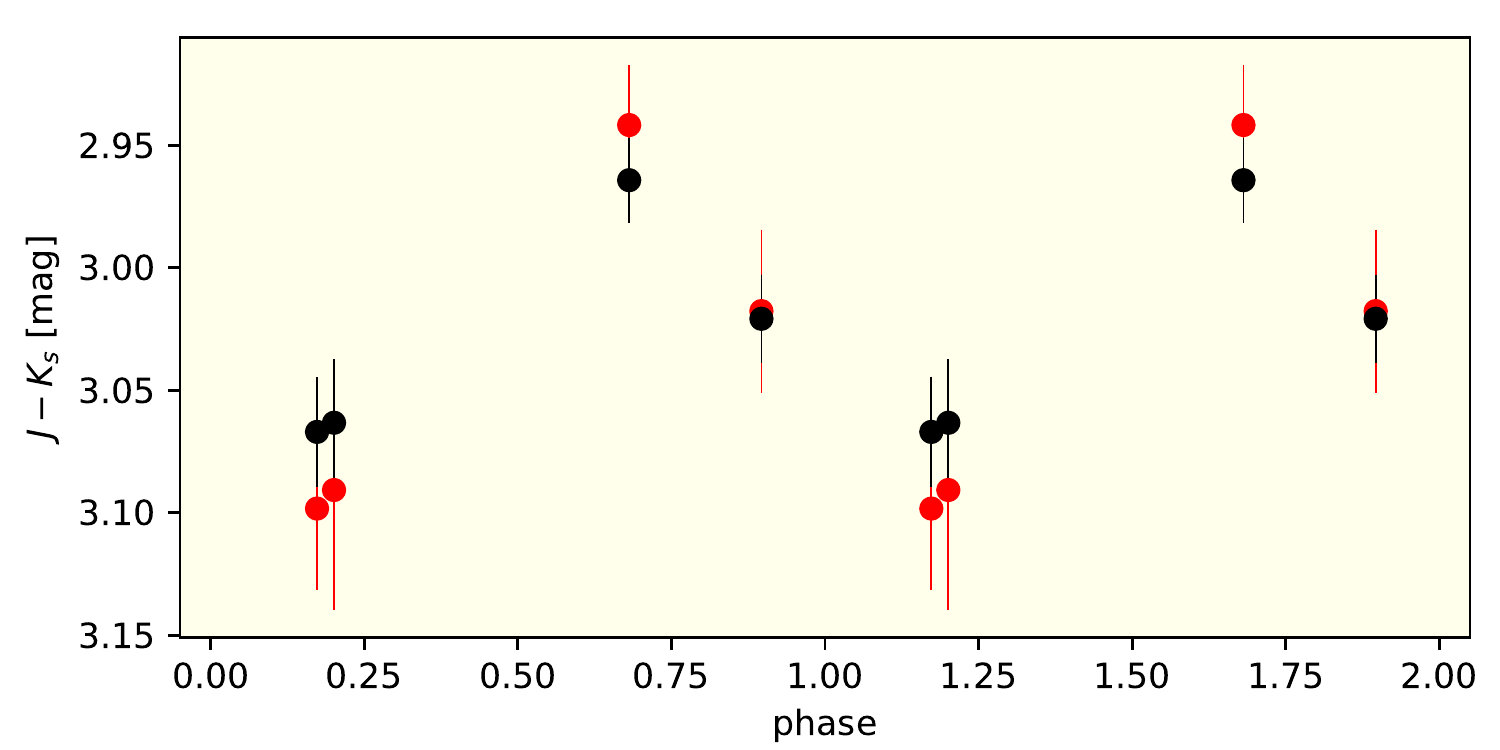}{0.5\textwidth}{}
          \fig{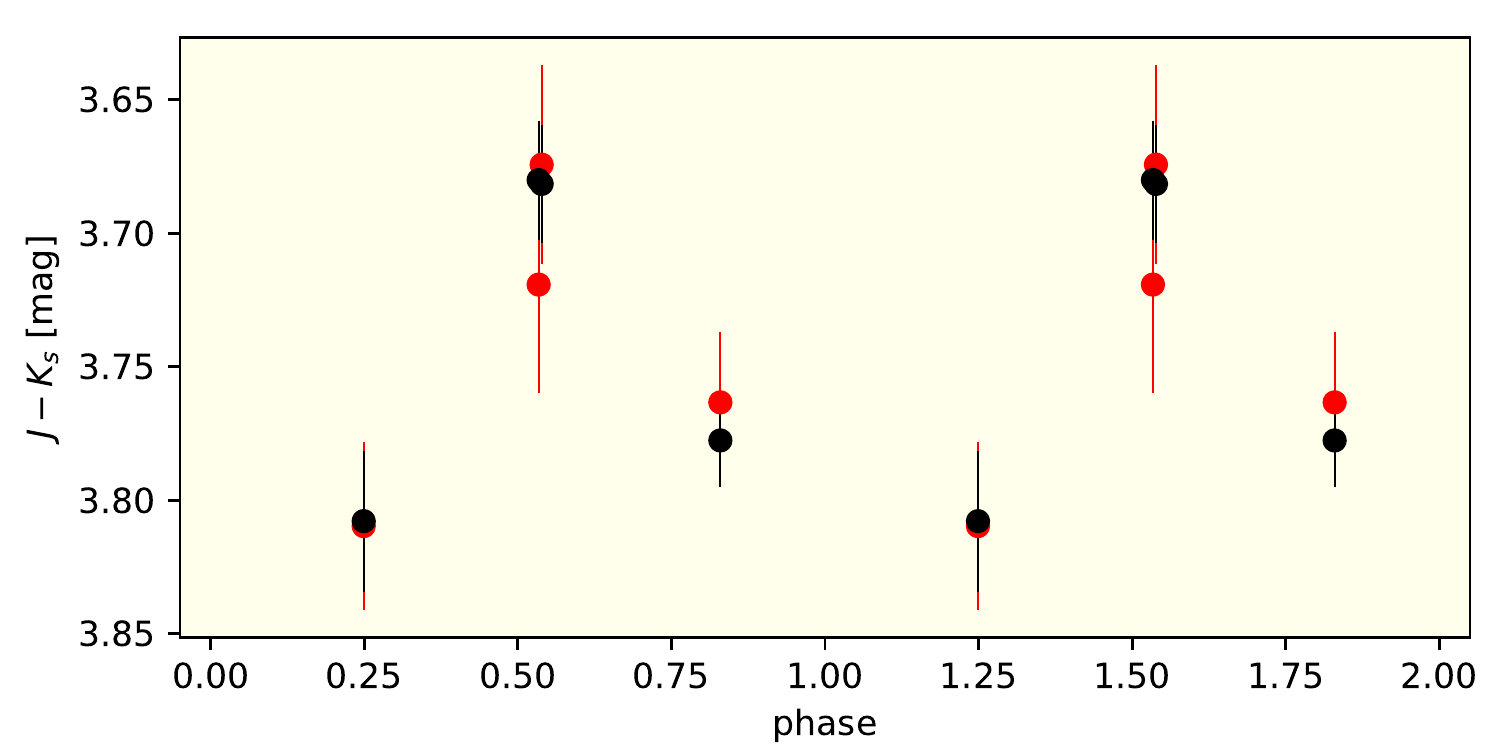}{0.5\textwidth}{}
          }
\caption{
Phase-folded $K_s$-band light curves (upper panels) and $J-K_s$ color index curves (lower panels) of two classical Cepheids from our disk sample. In the upper panels, black symbols mark the observational data and their errors, red curves denote the fitted light curve models and measurements rejected from the regression are shown in red. In the bottom panels, $J(\varphi) - F_{K_s}(\varphi)$ (see Eq.~\ref{eq:cipred}) values and their errors are denoted by red symbols, while black symbols show the $\Delta_{J-K_s}$ predictions and their errors as defined in the text, with their means shifted to the observations.
\label{fig:jmk_pred_example}
}
\end{figure*}

The mean absolute prediction errors measured by cross-validation at various pulsation phases are shown in Fig.~\ref{fig:jhkpred} with black points. As a comparison, the normalized and phase aligned data of the training sets are also shown by gray points (note the different ordinate scales), outlining the possible ranges of color index biases in the absence of a $\Delta_{X-K_s}$ correction for the stars' intrinsic color index variation over their pulsation cycles. Our predictive models decrease the bias in the mean color indices by approximately an order of magnitude compared to the case of using no color correction. To further illustrate the accuracy of the prediction, Fig.~\ref{fig:jmk_pred_example} compares the predicted and observed $J-K_s$ color indices for two Cepheids with multiple $J$-band measurements, showing a good agreement between the two.

\begin{deluxetable}{ccc@{$\pm$}ccc@{$\pm$}cc}[]
\tablecaption{Mean color indices of the Cepheids \label{tab:reddening}}
\tablehead{
\colhead{Name} & \colhead{class} & \multicolumn2c{$\langle J-K_s \rangle$} & \colhead{$N_J$} & \multicolumn2c{$\langle H-K_s \rangle$} & \colhead{$N_H$}
}
%\decimals
\startdata
  1   &   DCEP    &   2.670    &    0.031 &  1  &  \multicolumn2c{\dots}             &        0  \\
  2   &   DCEP    &   \multicolumn2c{\dots}        &             0  &  \multicolumn2c{\dots}      &               0  \\
    3   &   T2CEP  &     1.809   &    0.037  &  1  &  \multicolumn2c{\dots}       &              0   \\   
  4   &   DCEP    &   1.685   &    0.030  &  2  &  \multicolumn2c{\dots}           &          0   \\
  5   &   DCEP    &   1.234    &   0.040  &   3  &  0.322   &    0.027   &  3  
\enddata
\tablecomments{This table is available in its entirety in machine-readable form.}
\end{deluxetable}

\subsection{The near-IR reddening ratio}\label{subsec:reddening}

Estimates of the $J-K_s$ and $H-K_s$ color excesses were computed for each and every classical and type II Cepheid in our sample using the method described in Sect.~\ref{sec:cipred}.
%Eqs.~\ref{eq:ce} and \ref{eq:cipred}. 
First, the individual $J$ and $H$ measurements were binned for each epoch, then the $\Delta_{J-K_s}$ and $\Delta_{H-K_s}$ correction terms in Eq.~\ref{eq:cipred} were determined by the predictive models discussed in Sect.~\ref{sec:cipred}, and finally the mean color index was computed as the weighted mean of the corrected color indices at each epoch. In case of large ($3\sigma$) discrepancies in the color indices between different epochs, the corresponding images were visually inspected and the erroneous measurements were subsequently culled. Table~\ref{tab:reddening} lists the (predicted) mean color indices and the number of $J$ and $H$ measurements of the classical and type II Cepheids and the number of $J$ and $H$ measurements used to compute them.

The absolute magnitudes were obtained from the PL relations by \citet{2017AJ....153..154B} for type II Cepheids and those by \citet{2015AJ....149..117M,2016AJ....151...48M} for classical Cepheids, after converting them into the VISTA photometric system using the transformation formulae by \citet[][their Eqs.~7--9]{2018MNRAS.474.5459G}. Moreover, the zero-points of the PL relations were adjusted to match the highly accurate LMC distance modulus of $\mu=18.477\pm0.026$ determined by \citet{2019Natur.567..200P}. The resulting PL relations are as follows for type II Cepheids if $\log P < 1.3$:
\begin{eqnarray}
M_J &=& -2.066~[\pm .038] \times P_1 + 15.548~[\pm .017] - \mu  \label{eq:plc2j}\\
M_H &=& -2.199~[\pm .046] \times P_1 + 15.149~[\pm .017] - \mu   \label{eq:plc2h}\\
M_{K_s} &=& -2.233~[\pm .037] \times P_1 + 15.067~[\pm .015] - \mu   \label{eq:plc2k}
\end{eqnarray}

\noindent for type II Cepheids if $\log P > 1.3$:
\begin{eqnarray}
M_J &=& -2.247~[\pm .072] \times P_1 + 15.129~[\pm .039] - \mu  \label{eq:plj_rv}\\
M_H &=& -2.298~[\pm .071] \times P_1 + 14.759~[\pm .038] - \mu   \label{eq:plh_rv}\\
M_{K_s} &=& -2.173~[\pm .071] \times P_1 + 14.651~[\pm .036] - \mu   \label{eq:plk_rv}
\end{eqnarray}

\noindent and for classical Cepheids:
\begin{eqnarray}
M_J &=& -3.159~[\pm .004] \times P_1 + 13.214~[\pm .002] - \mu   \label{eq:plc1j}\\
M_H &=& -3.186~[\pm .004] \times P_1 + 12.854~[\pm .002] - \mu   \label{eq:plc1h}\\
M_{K_s} &=& -3.248~[\pm .004] \times P_1 + 12.773~[\pm .001] - \mu   \label{eq:plc1k}
\end{eqnarray}

\noindent where $P_1=\log P-1$.

We determined the mean $R_{JKHK}=E(J-K_s)/E(H-K_s)$ reddening ratio by fitting a linear function to the $E(J-K_s)$ {\em vs} $E(H-K_s)$ distribution of the Cepheids that have measurements in all three bands, by the following procedure. First, we assumed the selective-to-absolute extinction ratio of $A(K_s)/E(J-K_s)=0.49$ found by \citet{2016A&A...593A.124M}, and computed the distances of the objects using the relations in Eq.~\ref{eq:plc2j}--\ref{eq:plc1k}. Then we omitted all objects with heliocentric distances larger than 25~kpc from the fit of the reddening ratio, as misclassification can easily result in anomalously large distances (see Sect.~\ref{sec:distribution} for more details). We emphasize that the precision of the \citet{2016A&A...593A.124M} extinction coefficient is not critical in performing this rough distance cut. We also omitted all type II Cepheids with $\log P > 1.3$ from this part of the analysis, because these long-periodic stars of the RV~Tau subtype follow slightly different, more uncertain PL-relations than the rest of the sample \citep[see][]{2017AJ....153..154B}. The remaining sample was divided into bulge and disk subsamples (see Sect.~\ref{subsec:data}) and they were fitted separately.

\begin{figure}
%\plotone{Rjkhk_fit.pdf}
\includegraphics[width=0.48\textwidth]{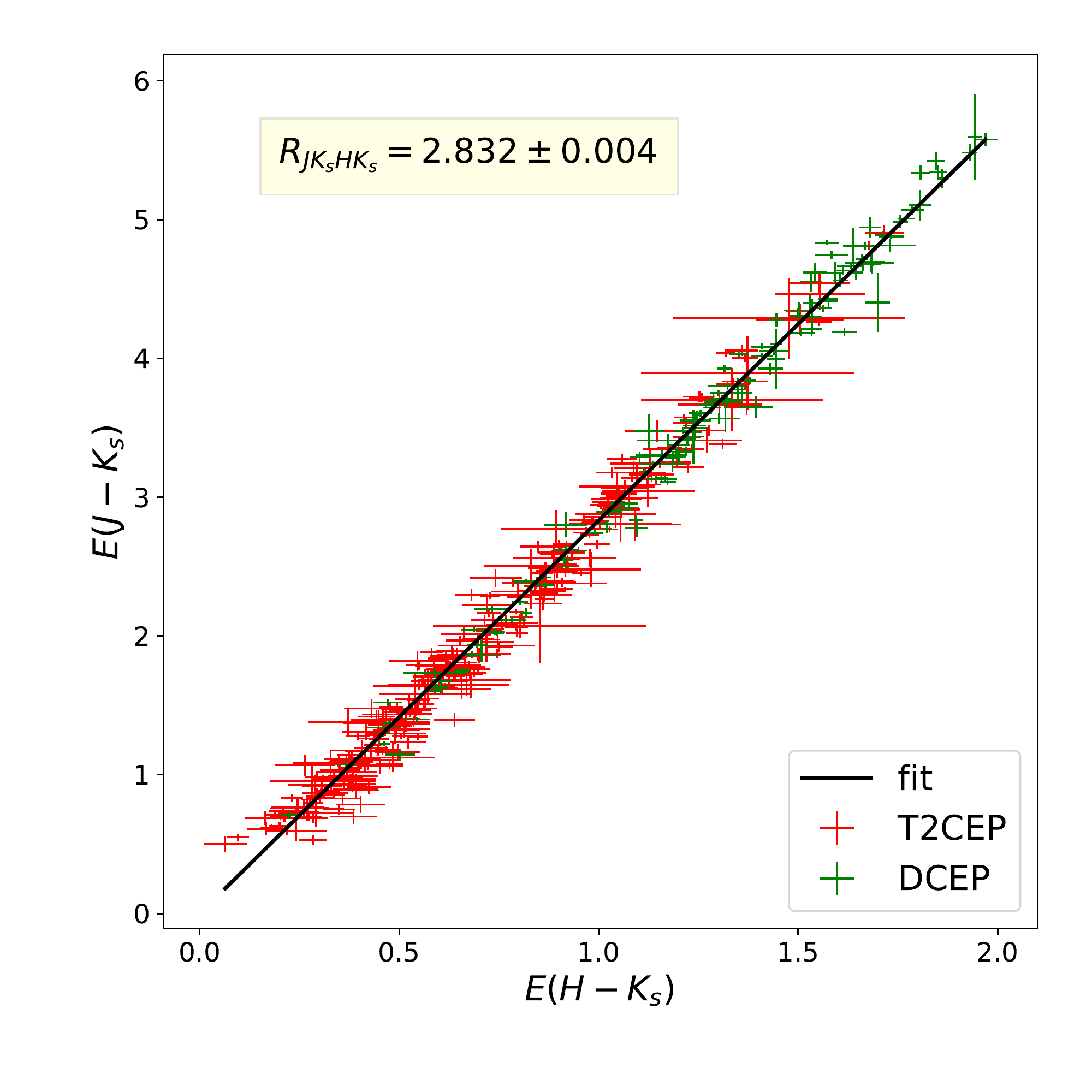}
\caption{
Regression of the mean color excess ratio for Cepheids toward the bulge area. Red and green symbols show the values measured for individual type II and classical Cepheids and their total errors, respectively. The black line denotes the result of the regression, with its slope shown in the inset. 
\label{fig:Rjkhk_fit}
}
\end{figure}

First, we performed iterative linear fits combined with $3$--$\sigma$ threshold rejection in order to omit strong outliers. This was followed by Bayesian fits of a linear function on the remaining data with Markov-chain Monte Carlo (MCMC) error sampling using the PyMC package. We used uniform priors on the coefficients around their values obtained the initial linear fit, and computed the means and the errors of the reddening ratios from their posterior distributions. The MCMC approach enabled us to properly take into account the correlated statistical and systematic errors in both color excesses in the fitting procedure. Figure~\ref{fig:Rjkhk_fit} shows the reddening values of the bulge Cepheid subsample, together with the best fit. The resulting $R_{JKHK}$ reddening ratios are $2.832\pm0.004$ and $2.833\pm0.004$ for the bulge and the disk subsamples, respectively. The errors correspond to the interquartile ranges of the posterior distributions and include all statistical and systematic uncertainties. The excellent agreement between the two values suggests that there are no changes in the mean near-IR reddening law over very large $\geq 10^{\circ}$ angular scales along the Galactic equator.

The bulge Cepheid subsample has a sufficiently large number density to detect possible spatial variations in the near-IR reddening ratio on smaller angular scales. We divided this subsample into longitudinal and latitudinal bins and repeated the fitting procedure discussed above, incorporating only the statistical errors in the Bayesian likelihoods, since the systematic errors originating from the PL relations would affect all binned subsamples in the same way, and would not contribute to spatial variations in $R_{JKHK}$.

\begin{figure}
%\plotone{Rjkhk_lb.pdf}
\includegraphics[width=0.48\textwidth]{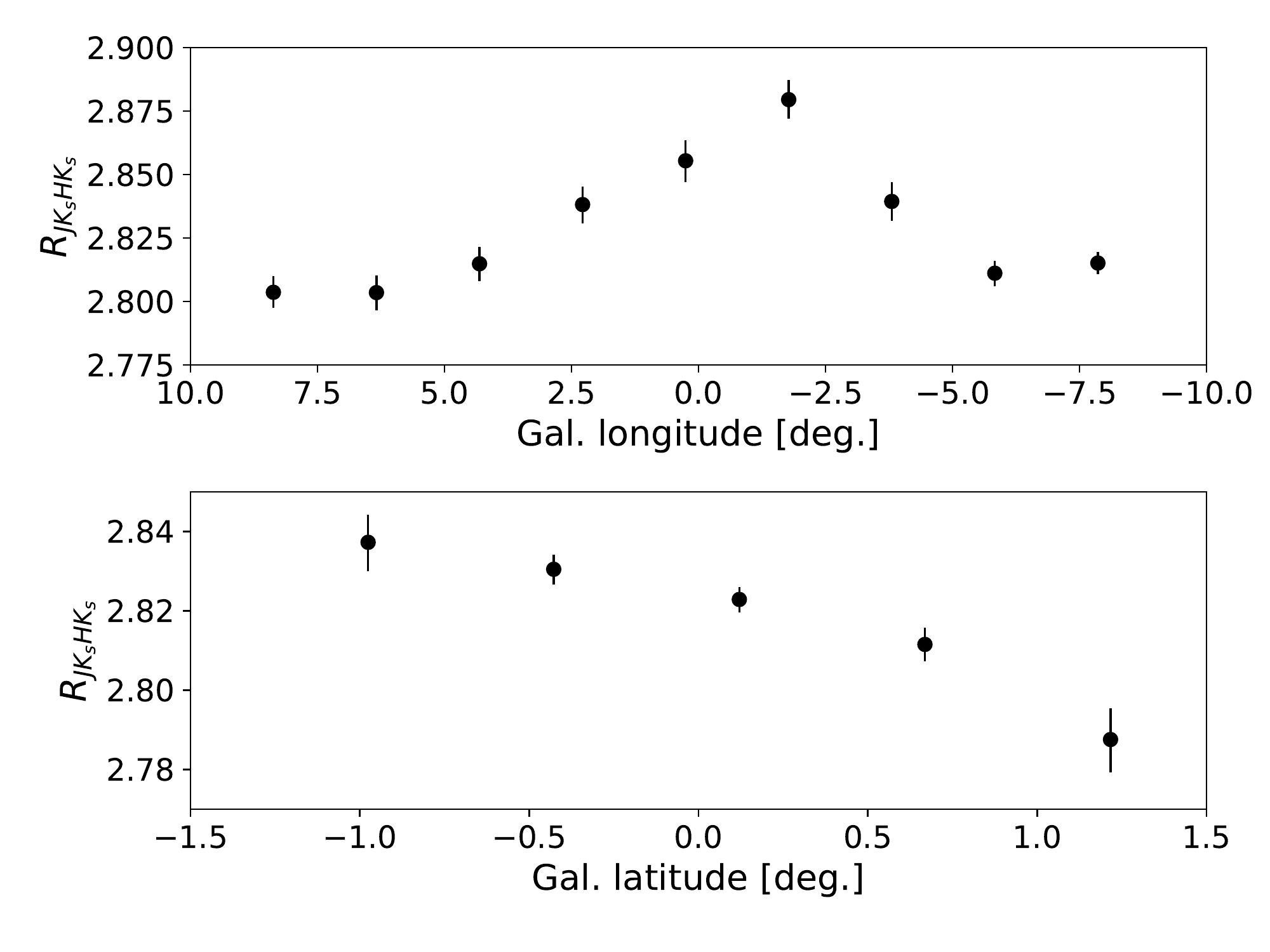}
\caption{
Spatial variations in the $R_{JKHK}$ reddening ratio as a function of Galactic longitude (top) and latitude (bottom) over the bulge area of our study, using bin sizes of $5^\circ$ and $0.5^\circ$, respectively.
\label{fig:jhkvar}
}
\end{figure}

Figure~\ref{fig:jhkvar} shows the fitted values of $R_{JKHK}$ and its error as a function of Galactic longitude, measured in 10 overlapping bins of $5^\circ$ width, and in latitude, using 6 overlapping bins of $0.5^\circ$ width. These bin configurations provide a good tradeoff between spatial sensitivity and sample size, but other arrangements yield to similar results: the near-IR reddening ratio has very significant spatial variations over relatively small ($\sim$5$^\circ$) angular scales. We emphasize that due to the limited spatial density, the use of overlapping bins was necessary, which blurs the underlying true spatial variation. Therefore, the distributions in Fig.~\ref{fig:jhkvar} can only be used to {\em detect} the variation, and should not be considered as a low resolution ``map'' of the latter, i.e., they are unsuitable for obtaining a $R_{JKHK}$ at a given sight-line by interpolation.

\subsection{The selective-to-absolute extinction ratios}\label{subsec:extinction}

Earlier evidence shows that the type II Cepheids observed toward our bulge footprint are concentrated around the Galactic center \citep{2017AcA....67..297S}, and their spatial distribution within the bulge \citep{2018A&A...619A..51B} is similar to that of the RR~Lyrae stars \citep{2013ApJ...776L..19D,2015ApJ...811..113P}, suggesting that they belong to an old spheroid or inner halo. We can take advantage of this important property to determine the mean selective-to-absolute extinction ratios toward the bulge subsample. We assume that type II Cepheids have a centrally symmetric distribution around the central supermassive black hole. Note that this assumption includes the possibility of an elongated core as observed in the case of the RR~Lyrae stars \citep{2013ApJ...776L..19D,2015ApJ...811..113P}. The distance to the central black hole, i.e., the Galactic center is known to an extremely high accuracy, thanks to the recent observations by the GRAVITY experiment, i.e., $R_0= 8178 \pm 25$~pc \citep{2019A&A...625L..10G}. We use this distance as the reference point in the determination of the extinction coefficients.

\begin{figure}
%\plotone{Rkjk_KDE.pdf}
\includegraphics[width=0.48\textwidth]{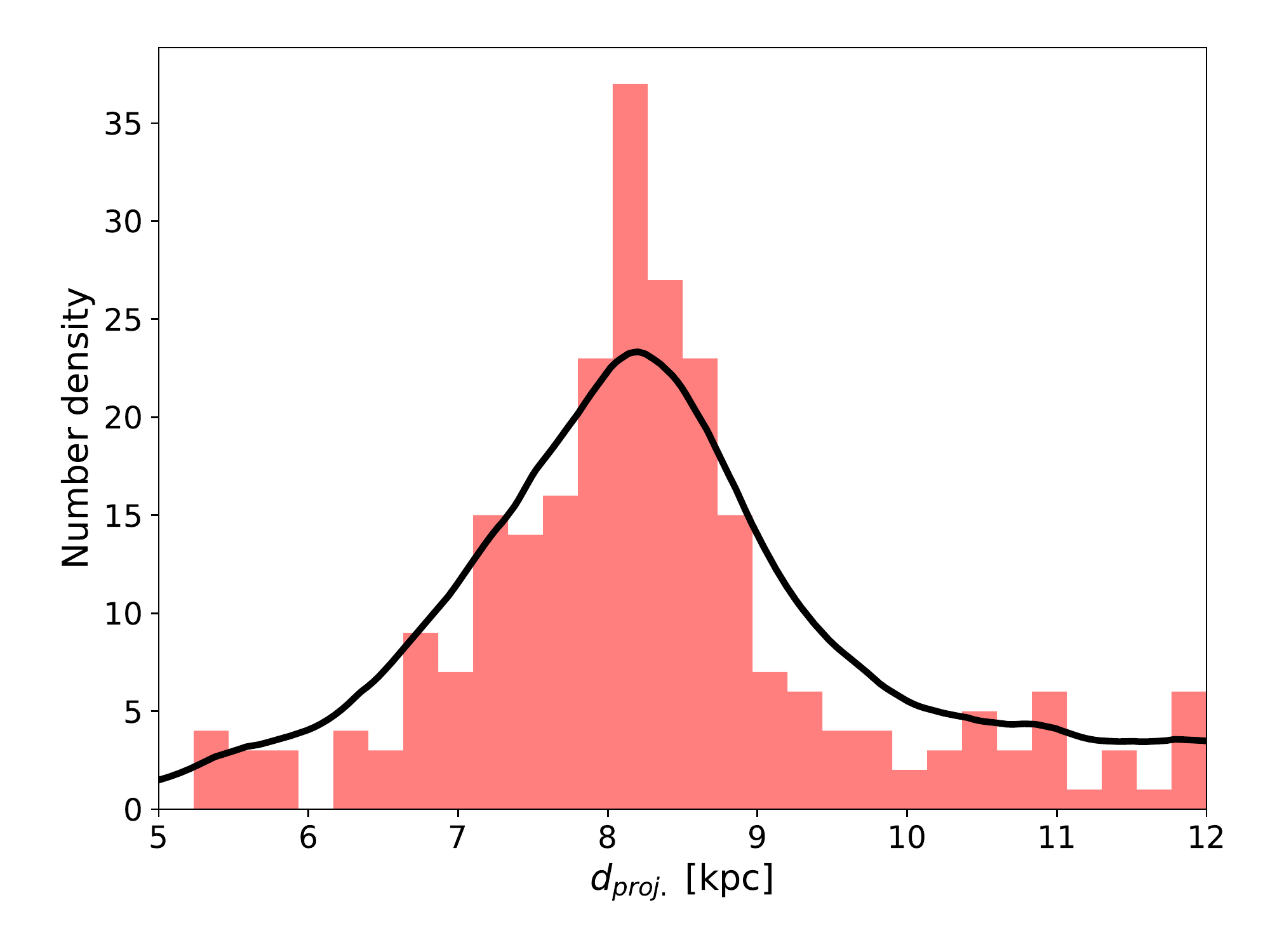}
\caption{
Histogram of the $d_Y$ distance components (see text.) of the bulge type II Cepheid sample with $\log P < 1.3$ and the corresponding kernel density estimate (black curve, arbitrarily scaled up to match the histogram) for a trial value of $R_{KJK}$.
\label{fig:Rkjk_kde}
}
\end{figure}

\begin{figure}[]
%\plotone{Rkjk_fit_staterr.pdf}
\includegraphics[width=0.48\textwidth]{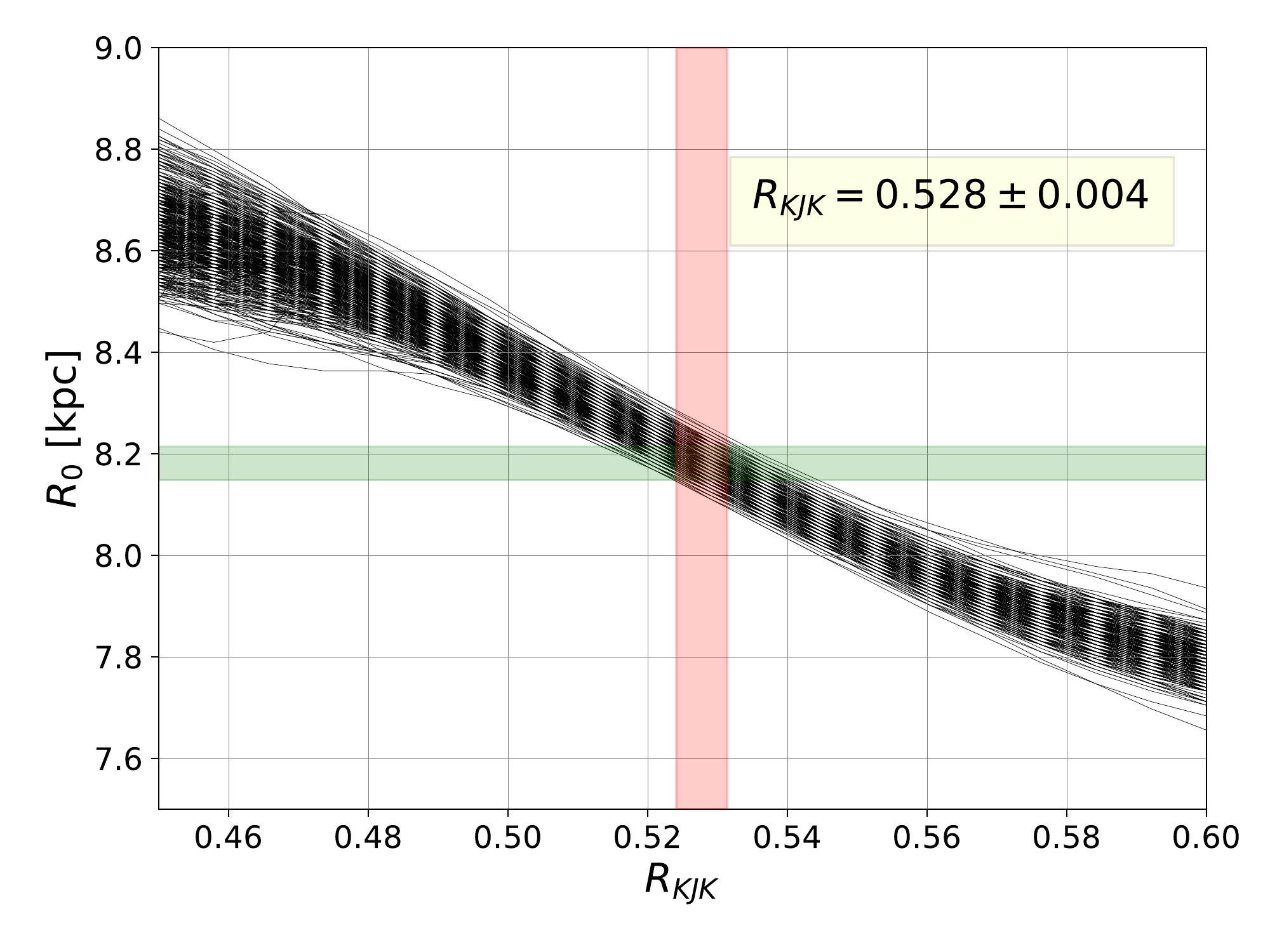}
\caption{
Variation of the $\hat{R}_0$ as a function of the $R_{KJK}$ (see text) are shown by black curves, each curve representing one realization of the Monte Carlo simulation of statistical errors. The green shaded area shows the total uncertainty around the value of $R_0$ by \citet{2019A&A...625L..10G}, the red shaded area shows the statistical error range of the optimal $R_{KJK}$ value indicated in the inset.
\label{fig:Rkjk_fit}
}
\end{figure}

The mean extinction coefficients $R_{KJK}=A(K_s)/E(J-K_s)$ and $R_{KHK}=A(K_s)/E(H-K_s)$ were determined separately as follows. For a fixed trial value $R_{KJK}$ or  $R_{KHK}$, we computed individual distances of the type II Cepheid sample with $\log P < 1.3$ using Eqs.~\ref{eq:plc2j}--\ref{eq:plc2k}. For each resulting distance distribution, we computed a kernel density estimate of the projected distance components parallel with the sight-line of the GC, i.e., $d_Y=d_H\cos b \cos l$, where $d_H$ denotes a Cepheid's heliocentric distance. The optimal kernel bandwidth was determined by 10-fold cross-validation. Figure~\ref{fig:Rkjk_kde} shows an example of the $d_Y$ distribution for a trial value of $R_{KJK}$. Then, we determined the $\hat{R}_0$ distance corresponding to the peak kernel density. The optimal values of the extinction coefficients were computed by minimizing the difference between $R_0$ and $\hat{R}_0$ for $R_{KJK}$ and $R_{KHK}$ separately. The statistical and systematic errors of the coefficients were computed by Monte Carlo simulations, repeating the above procedure in each realization. Figure~\ref{fig:Rkjk_fit} shows the $\hat{R}_0$ as a function of the $R_{KJK}$ coefficient for all realizations of the Monte Carlo simulation of the statistical errors, together with the optimal value of the former.

The optimal mean selective-to-absolute extinction coefficients found by our procedure are as follows:

\begin{eqnarray}
R_{KJK} = 0.528 \pm 0.004~{\rm (stat.)} \pm 0.019~{\rm (sys.)}  \label{eq:Rkjk} \\
R_{KHK} = 1.488 \pm 0.021~{\rm (stat.)} \pm 0.080~{\rm (sys.)}  \label{eq:Rkhk} 
\end{eqnarray}

The resulting $R_{JKHK} = R_{KHK}/R_{KJK} = 2.818 \pm 0.19$ reddening ratio is consistent with the one obtained in Sect.~\ref{subsec:reddening}, although much less accurate. Due to the large uncertainties in $R_{KHK}$ computed above, instead of relying on the value in Eq.~\ref{eq:Rkhk}, for the calculation of distances, we adopt an $R_{KHK}$ value by combining the mean $R_{JKHK}$ from Sect.~\ref{subsec:reddening} with Eq.~\ref{eq:Rkjk}, resulting in:

\begin{equation}
%R_{KHK}=A(K_s)/E(H-K_s)=R_{JKHK}\times R_{KJK} = 1.495 \pm 0.01~{\rm stat.} \pm 0.05~{\rm sys.}. \label{eq:Rkhk_final}
R_{KHK} = 1.50 \pm 0.01~{\rm (stat.)} \pm 0.05~{\rm (sys.)}. \label{eq:Rkhk_final}
\end{equation}

\section{Spatial distribution}\label{sec:distribution}

Using the mean selective-to-absolute extinction ratios determined in Sect.~\ref{subsec:extinction}, we computed the heliocentric distances of classical and type II Cepheids, using the PL relations in Eqs.~\ref{eq:plc1j}--\ref{eq:plk_rv} in conjunction with Eqs.~\ref{eq:Rkjk} and \ref{eq:Rkhk_final}, by:

\begin{equation}
\log d_H = 1+0.2(\langle K_s \rangle - M_{K_s} -A(K_s)),  \label{dist}
\end{equation}

\noindent where $\langle K_s \rangle$ is the magnitude of the intensity mean in the $K_s$-band. If both $J$ and $H$ measurements were available, we computed the $A(K_s)$ extinction from both the $\{J,K_s\}$ and $\{H,K_s\}$ filter pairs and calculated $d_H$ as the weighted mean of the corresponding two distances. Table~\ref{tab:distances} shows the absolute $K_s$-band extinctions and heliocentric distances of the Cepheids together with their statistical and systematic errors based on the $JK_s$ and $HK_s$ photometry. The errors are standard deviations computed from Monte Carlo error simulations of the input data and parameters.

\begin{deluxetable*}{ccD@{$\pm$}l@{$\pm$}lD@{$\pm$}l@{$\pm$}lD@{$\pm$}l@{$\pm$}lD@{$\pm$}l@{$\pm$}l}[]
\tablecaption{Heliocentric distances and $K_s$-band interstellar extinctions of the Cepheids\tablenotemark{a}. \label{tab:distances}}
\tablehead{
\colhead{Name} & \colhead{class} &
\multicolumn4c{$d_H(J,K_s$)~[kpc]}  & \multicolumn4c{$A_{K_s}(J,K_s$)~[mag]}  &
\multicolumn4c{$d_H(H,K_s$)~[kpc]} & \multicolumn4c{$A_{K_s}(H,K_s$)~[mag]} 
}
\decimals
\startdata
  1   &   DCEP   &    8.31  &  0.06   &    0.19   &   1.19  &  0.02   &   0.04   &   \multicolumn4c{\dots} & \multicolumn4c{\dots}      \\
  2   &   DCEP   &    \multicolumn4c{\dots} & \multicolumn4c{\dots} & \multicolumn4c{\dots} & \multicolumn4c{\dots}        \\
  3   &   T2CEP &    11.20  &  0.10   &   0.40  &    0.72  &  0.02   &   0.04   &   \multicolumn4c{\dots} & \multicolumn4c{\dots}         \\
  4   &   DCEP   &    10.52 &  0.08    &   0.17  &   0.67  &  0.02  &   0.02   &   \multicolumn4c{\dots} & \multicolumn4c{\dots}         \\
  5   &   DCEP   &    66.01 &  0.63   &   0.92  &   0.42  &  0.02   &   0.02  &    67.83  &  1.26  &   1.54  &   0.359  &  0.04    &   0.01    
\enddata
\tablecomments{This table is available in its entirety in machine-readable form.}
\tablenotetext{a}{Statistical and systematic errors are given separately, in this order.}
\end{deluxetable*}

\subsection{Type II Cepheids in the bulge}

\begin{figure}[]
%\plotone{xy_t2cep_W1_vgrid300_sm.pdf}
\includegraphics[width=0.48\textwidth]{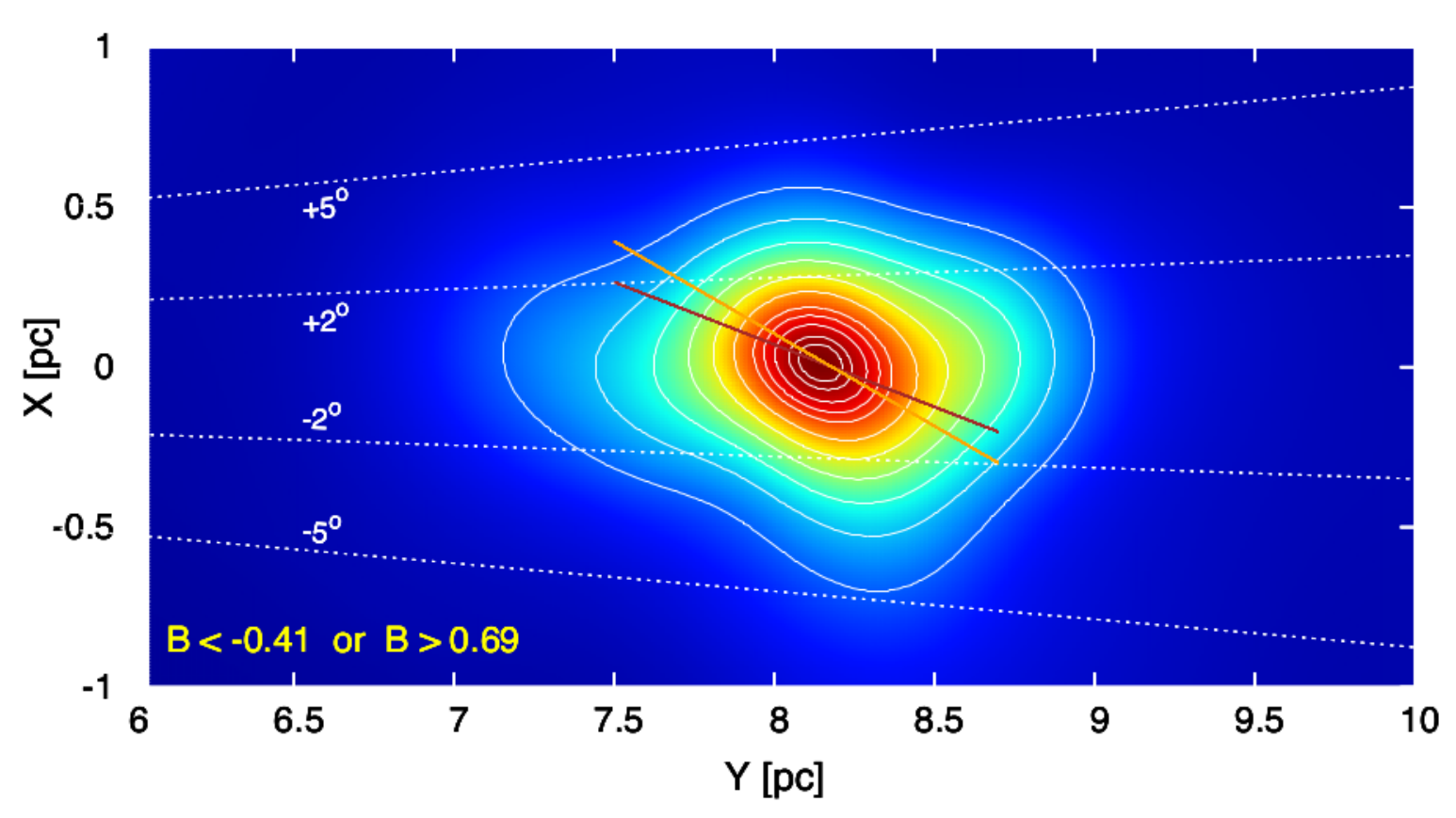}
\caption{
Kernel density estimate of the projected distance distribution of type II Cepheids in the Galactic bulge from latitudinal ranges shown in the figure, represented as a color scale and contour lines. The white dashed lines show the $l=\pm2^\circ$ and $l=\pm5^\circ$ sight-lines. The brown and orange lines illustrate the $21^\circ$ and $30^\circ$ inclination angles of the triaxial ellipsoid fitted to the outer and inner parts of the spatial distribution of bulge RR~Lyrae stars, by \citet{2015ApJ...811..113P}. 
\label{fig:c2_faceon}
}
\end{figure}

The on-sky distribution of type II Cepheids is highly concentrated around the Galactic center (Fig.~\ref{fig:lb_cep}), and nearly uniform over the studied volume outside the bulge. We further examined the spatial distribution of the type II Cepheids projected onto the Galactic plane in two parallel stripes with $b<-0.41^\circ$ and $b>0.69^\circ$, in order to avoid the gap in our celestial coverage toward the nuclear bulge, and thus attain a sample with contiguous longitudinal coverage. Figure~\ref{fig:c2_faceon} shows a kernel density estimate of the distance distribution projected onto the Galactic plane for the sum of the two aforementioned subsets, using a Gaussian kernel with a bandwidth of 300~pc. The slight elongation toward the Sun in the outskirts of the distribution is probably due to large photometric errors and/or misclassifications of a few objects. The core of the distribution is slightly inclined with respect to the Galactic center sight-line. A similar feature was detected in the distribution of bulge RR~Lyrae stars by \citet{2013ApJ...776L..19D} and \citet{2015ApJ...811..113P} based on the OGLE-III and OGLE-IV samples, respectively. The inclination angles of $21^\circ$ and $30^\circ$ computed by \citet{2015ApJ...811..113P} corresponding to the triaxial ellipsoid fits of the outer and inner parts of the RR~Lyrae distribution are indicated in Fig.~\ref{fig:c2_faceon}. The latter shows a remarkably good agreement with the density estimate of the type II Cepheids. 

\subsection{Classical Cepheids}

\begin{figure}[]
%\plotone{c1_dist_hist.pdf}
\includegraphics[width=0.48\textwidth]{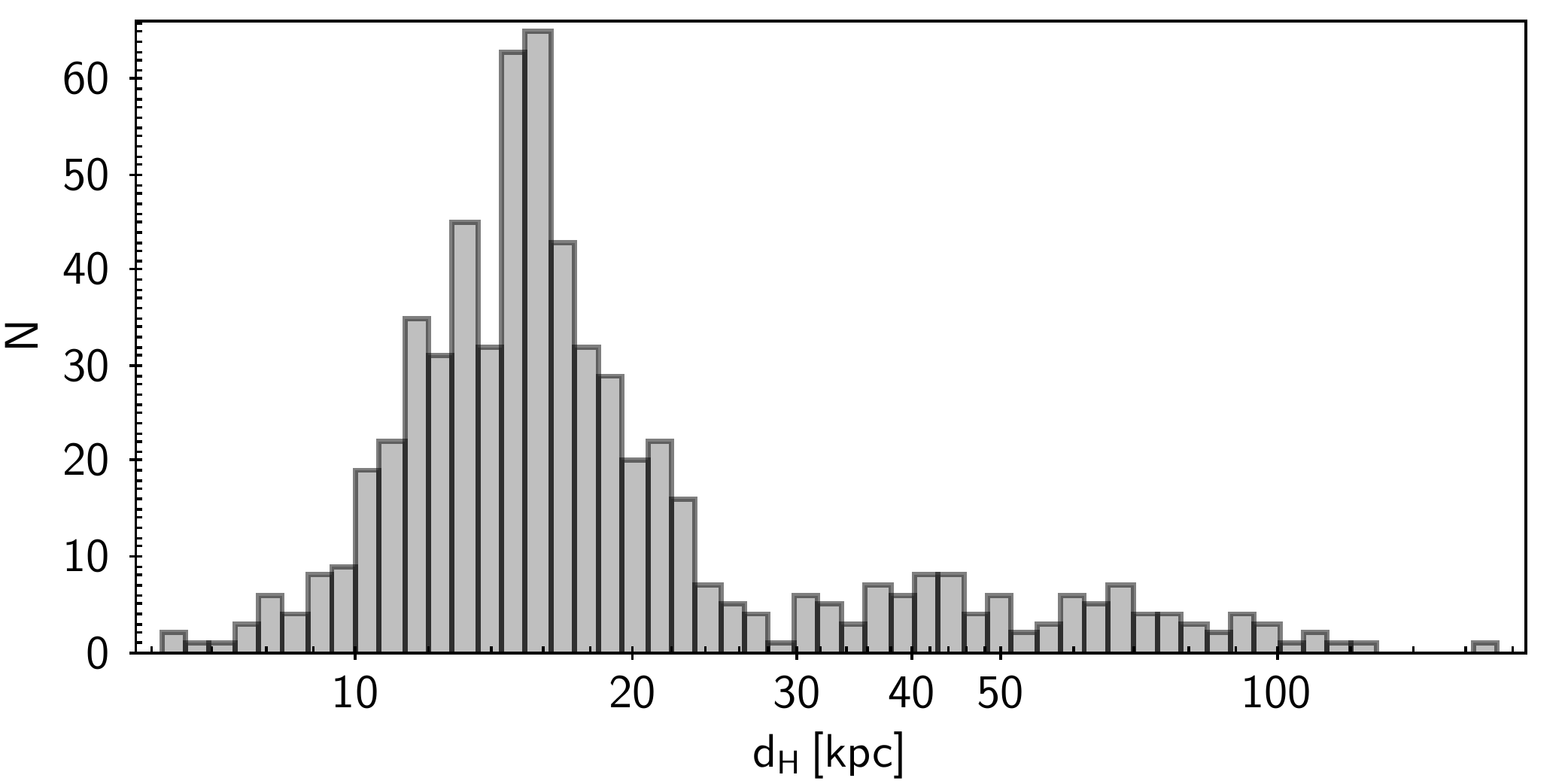}
\caption{
Histogram of the Heliocentric distances of classical Cepheids on a logarithmic scale.
\label{fig:c1disthist}
}
\end{figure}

Heliocentric distances for the classical Cepheids were computed in the same way as for type II Cepheids, using the PL relations in Eq.~\ref{eq:plc1j}--\ref{eq:plc1k}. In total, 624 objects had $J$ and/or $H$ measurements, allowing us to estimate their extinctions and distances, the rest of the sample had only $K_s$ photometry. Figure~\ref{fig:c1disthist} shows the resulting distance distribution. There is a sharp decrease in the objects' number density beyond $\sim$20~kpc, but a small fraction of objects have very large distances. Although the existence of classical Cepheids in the flared outer disk at Galactocentric distances of $\gtrsim20$~kpc has been reported earlier \citep{2014Natur.509..342F,2019NatAs...3..320C}, the presence of Cepheids beyond $\sim$40--50~kpc is unexpected. Only $13\%$ of our sample have distance estimates of $d_H>40$~kpc, and only $8\%$ of them have $d_H>50$~kpc. Most of these objects with anomalously large distances have small $S/N$ ($\lesssim 100$), and are most probably misclassified. The latter is also indicated by their relatively small $A(K_s)$ estimates even at very low latitudes. Their number fraction is also consistent with the classification precision estimated in Sect.~\ref{subsect:performance}. Nevertheless, we cannot rule out that our sample contains real Cepheids at very large Galactocentric distances, therefore we encourage the follow-up observations of these objects.

\begin{figure*}[]
%\plotone{dists_marg.pdf}
\includegraphics[width=\textwidth]{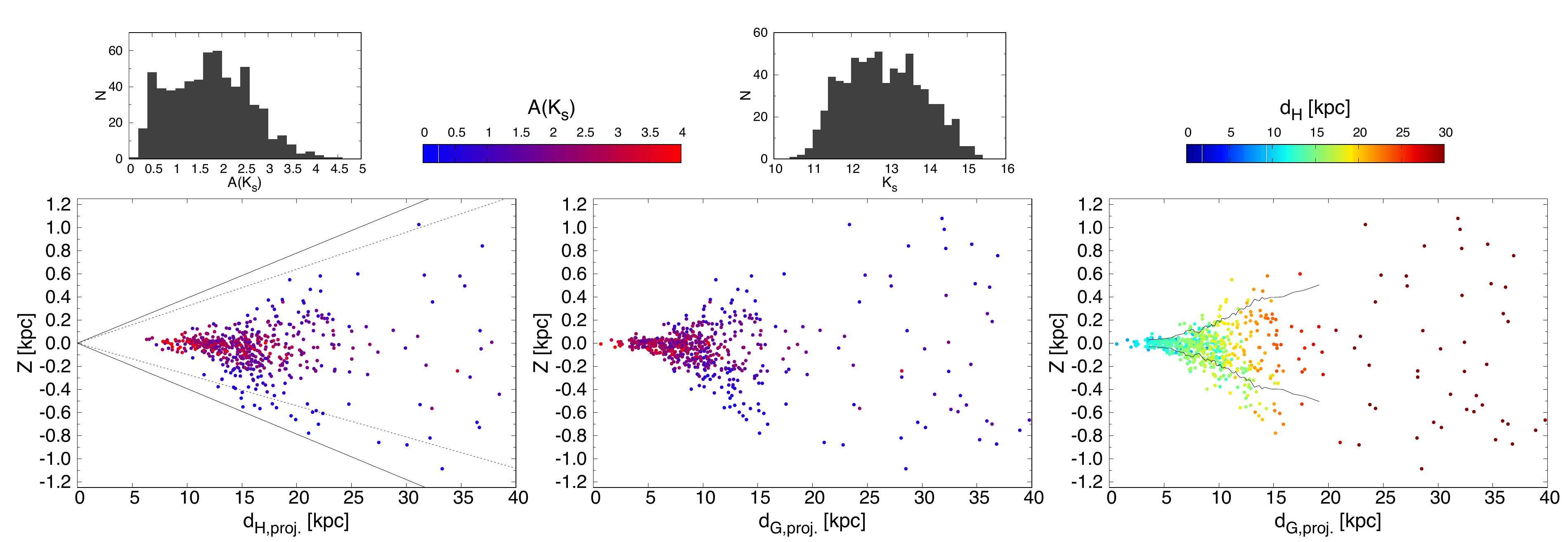}
\caption{
Top panels: histograms of the $K_s$-band extinction (left) and the apparent mean $K_s$ magnitude of the classical Cepheids in our sample. Bottom panels: distributions of the $Z$ distance from the Galactic plane marginalized over the projected Heliocentric ($d_{\rm H, proj.}$, left panel) and the Galactocentric cylindrical distance ($d_{\rm G,proj.}$), with the extinction or Heliocentric distance color coded. The latitudinal ranges of the bulge and disk subsamples are shown in the left panel by dotted and solid black lines, respectively. The black curves in the right panel denote HWHM values of $Z$ as a function of $d_{\rm G,proj.}$ (see text for details).
\label{fig:dists_marg}
}
\end{figure*}

Figure~\ref{fig:dists_marg} shows marginalized $d_H$ heliocentric and $d_G$ Galactocentric distance distributions of the classical Cepheid sample up to 40~kpc, along with the distributions of their mean apparent $\langle K_s \rangle$ magnitudes and extinctions. Most of the objects have very high extinction, rendering their detection impossible for even the deepest optical surveys. The latitudinal range of our surveyed area (marked in the left panel of Fig.~\ref{fig:dists_marg}) allows us to probe the flaring of the outer disk with the Cepheids. The flaring is immediately apparent in the distribution of the $Z$ distance from the Galactic plane marginalized over $d_G$ (lower middle and left panels of Fig.~\ref{fig:dists_marg}). In order to quantify the thickness of the disk, we determined the half width at half maximum (HWHM) of $Z$ as a function of $d_G$ using a boxcar with a width of 50 points and a step size of 5 points in $d_G$, computed as ${\rm HWHM}(Z)=\sqrt{2\ln 2}\sigma$, where $\sigma$ is the standard deviation of $Z$ in each bin. The resulting symmetric disk thickness estimate up to 20~kpc is shown in the right panel of Fig.~\ref{fig:dists_marg}. Although the Cepheids' number density drops dramatically at $d_G\gtrsim12$~kpc, we can still observe an apparent break in the flaring at around 15~kpc.

\begin{figure*}
\gridline{\fig{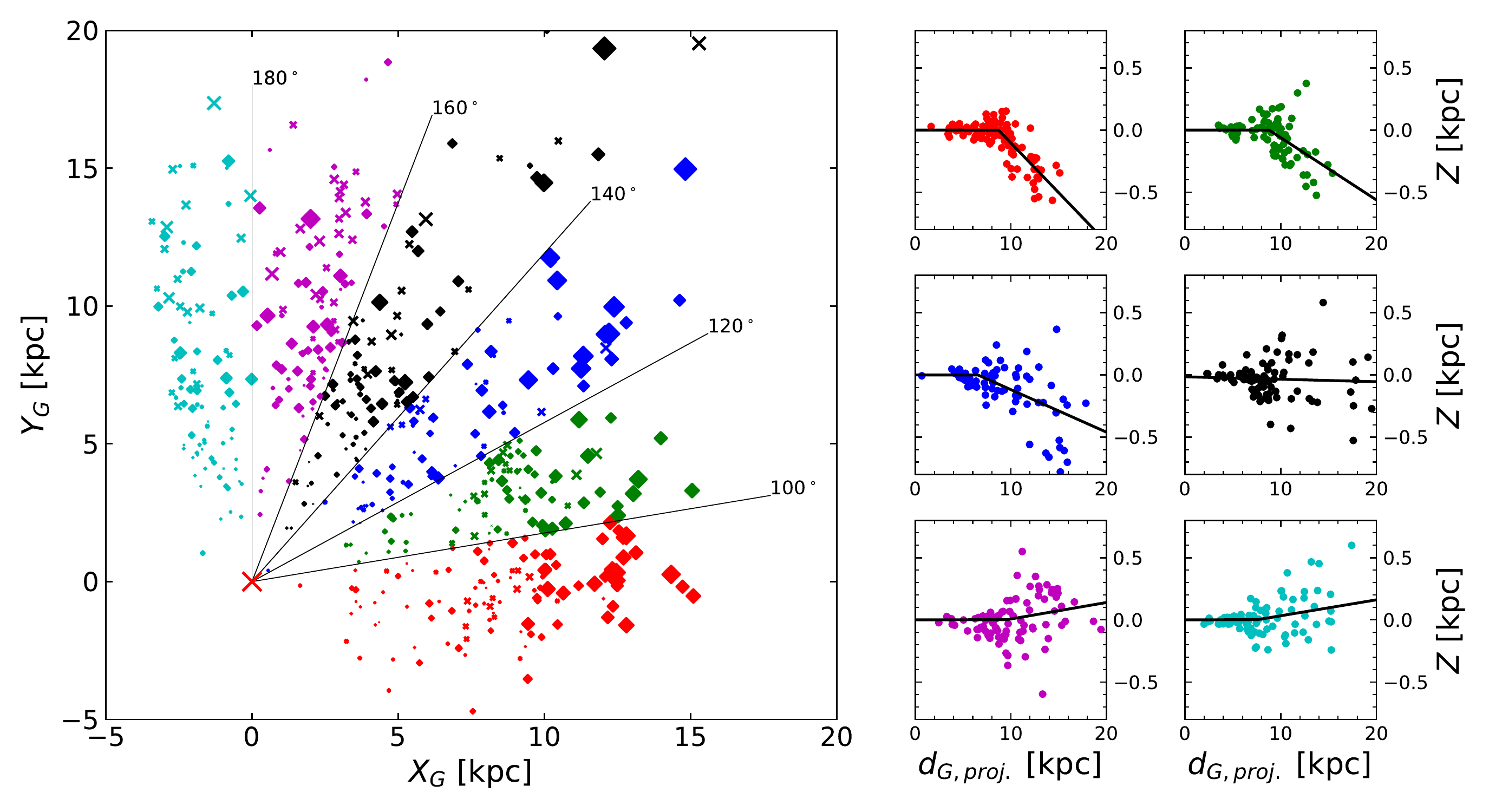}{0.65\textwidth}{}
          \fig{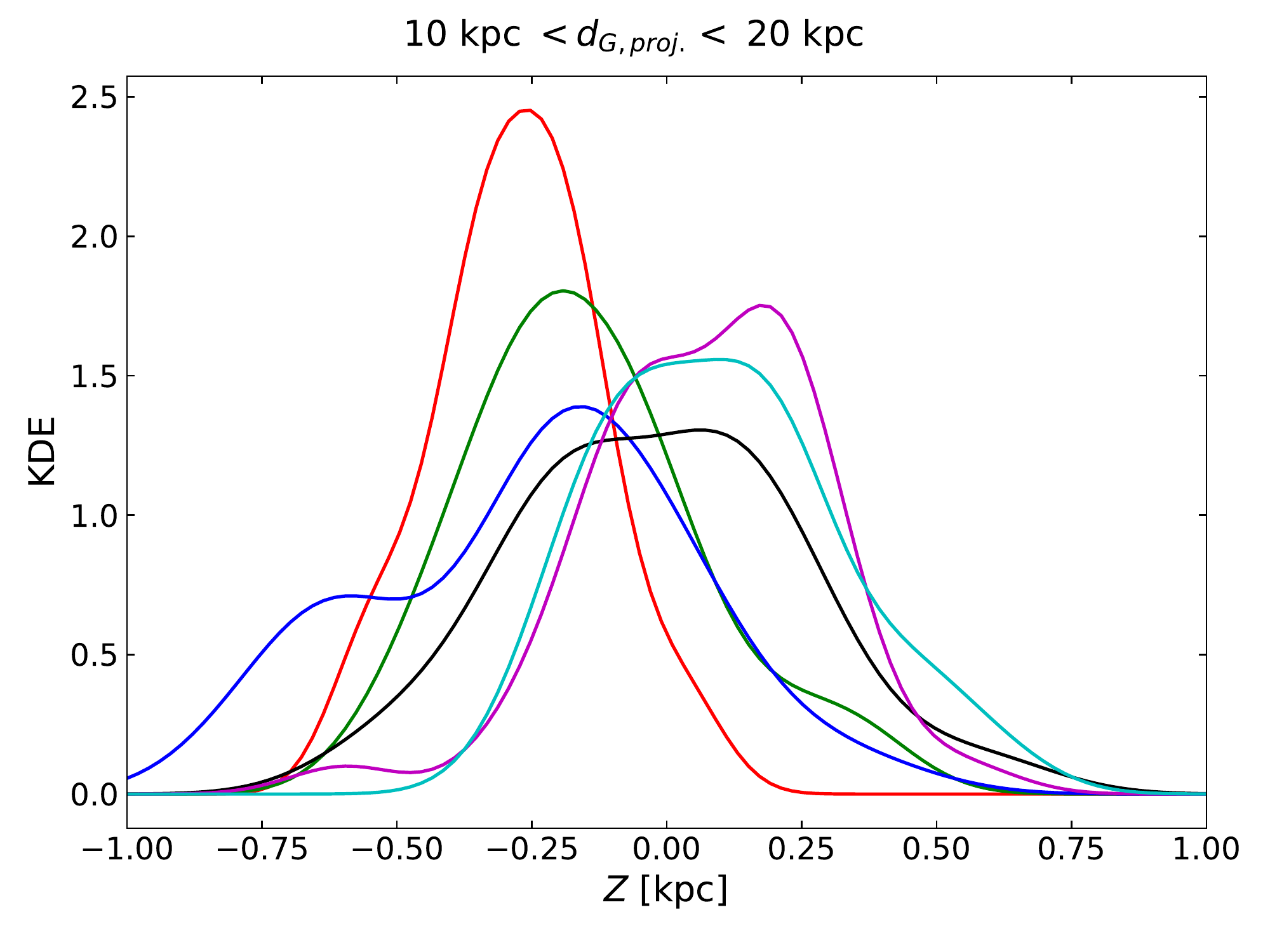}{0.35\textwidth}{}
          }
\caption{
Left: the spatial distributions of the classical Cepheids in our sample projected onto the Galactic plane. The red `x' symbol marks the position of the Galactic center. 
Crosses and diamonds of different sizes indicate how far above or below, respectively, the objects are located from the Galactic plane.
Various ranges of Galactocentric azimuth are color-coded. Middle panels: the distributions of the $Z$ distance from the Galactic plane {\em vs} the Galactocentric cylindrical distance $d_{G,proj.}$ in azimuthal ranges color coded as in the left panel. Black lines denote linear warp models fitted to the data. Right: kernel density estimates (KDEs) of the $Z$ distance component in various azimuthal ranges color coded as in the left panel.\label{fig:warp}
}
\end{figure*}

In order to spatially resolve the flaring, we divided the classical Cepheids into 6 sets of disjunct ranges of Galactocentric azimuth, and plotted the distribution of $Z$ {\em vs} $d_{G,proj.}$ for each set in Fig.~\ref{fig:warp}. It is immediately evident that the outer disk is warped within the azimuthal range of our Cepheid sample. We also computed kernel density estimates of $Z$ in each azimuthal range for stars in the $10~{\rm kpc}<d_{G,proj.}<20$~kpc range, shown in the right panel of Fig.~\ref{fig:warp}. The location of peak density changes in a range of 0.5~kpc between azimuths of $\sim$260$^\circ$ and $\sim$10$^\circ$, indicating the warp.

The limited azimuthal range of our Cepheid sample does not allow us to model the warp in detail. We fitted simple linear warp models to the binned distributions to guide the eye (Fig.~\ref{fig:warp}). The line of nodes, where the warp emerges from negative to positive $Z$ distances, falls within the azimuthal range of $(140^\circ,160^\circ)$. The mean onset radius of the warp is $8.6$~kpc, computed from the linear models fitted to the 4 azimuthal bins lying farthest from the line of node. These approximate values are in good agreement with the detailed warp model by \citet{2019NatAs...3..320C} fitted to Cepheids within a $>$180$^\circ$ azimuthal range at the near side of the Galactic disk. 

The classical Cepheids allow us to trace the radial distribution of stellar ages in the Galactic disk. We used the theoretical period-age relations derived by \citet{2016A&A...591A...8A} to compute individual age estimates for each star. In addition to the pulsation period and metallicity, a Cepheid's age estimate also depends on which instability strip crossing it is going through, as well as its rotational history. Although in an ideal case, these could be constrained by measurements of the period change rate, effective temperature and surface CNO abundances, in the present case these quantities remain unknown, therefore we relied on period-age relations averaged over the full range of possible stellar rotation and instability strip crossings, derived for metallicities $Z=\{0.002,0.006,0.014\}$ \citep[Table~4 of][]{2016A&A...591A...8A}.

In the absence of individual metallicities, we estimated [Fe/H] values from the radial metallicity gradient found by \citet{2014A&A...566A..37G} and \citet{2018AJ....156..171L} using Cepheids at the near side of the Galactic disk. These were converted into estimates of absolute heavy element content by $\log Z = {\rm [Fe/H]}-1.77$, assuming a helium content of $Y = 0.245$, no $\alpha$-element enhancement, and the Solar heavy element mixture measured by \citet{1998SSRv...85..161G}, and we considered only objects with ${\rm [Fe/H]>-0.4}$. Age estimates were obtained by quadratically interpolating between the relations for the 3 tabulated $Z$ values.

\begin{figure*}[]
\gridline{
          \fig{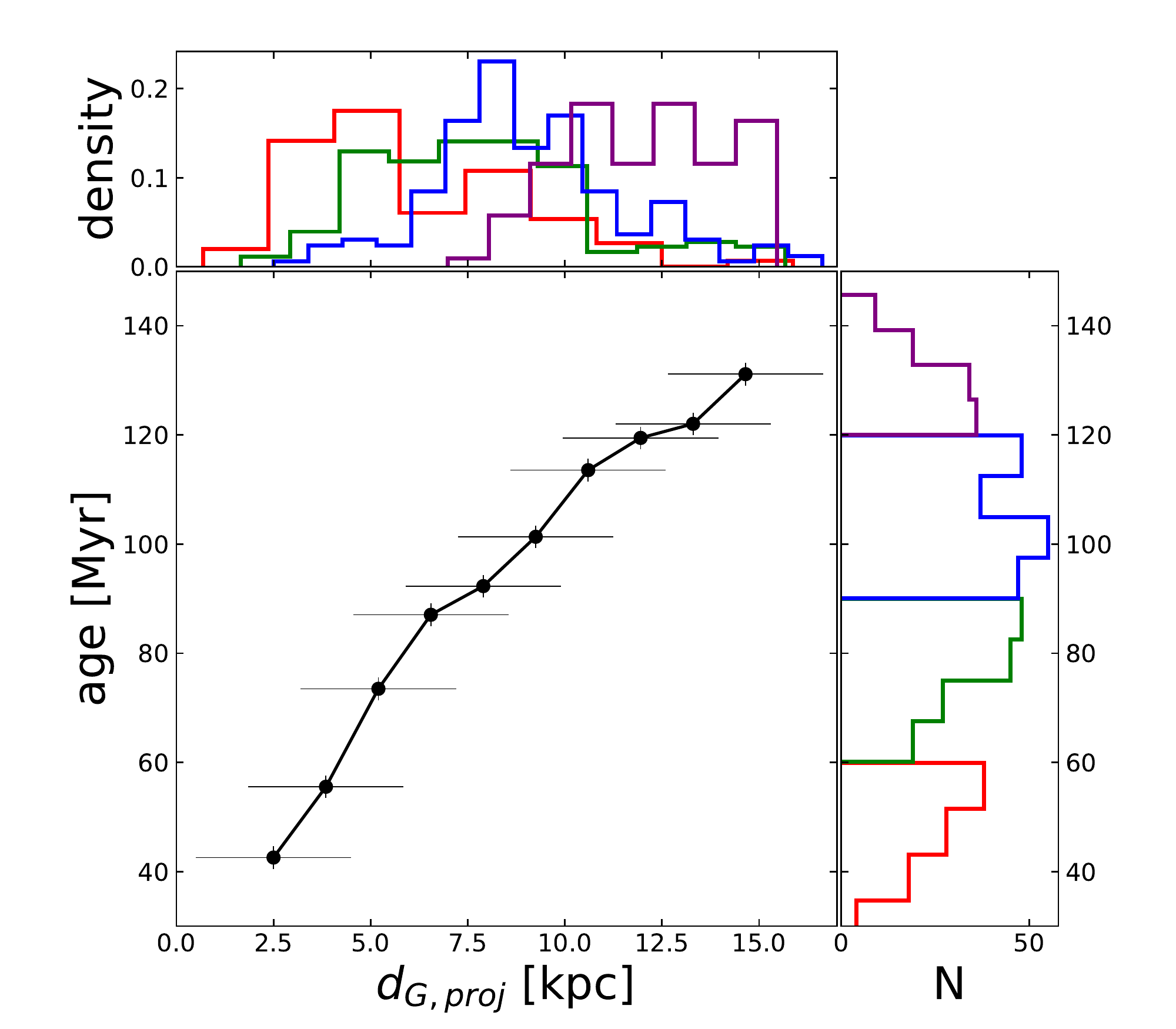}{0.33\textwidth}{}
          \fig{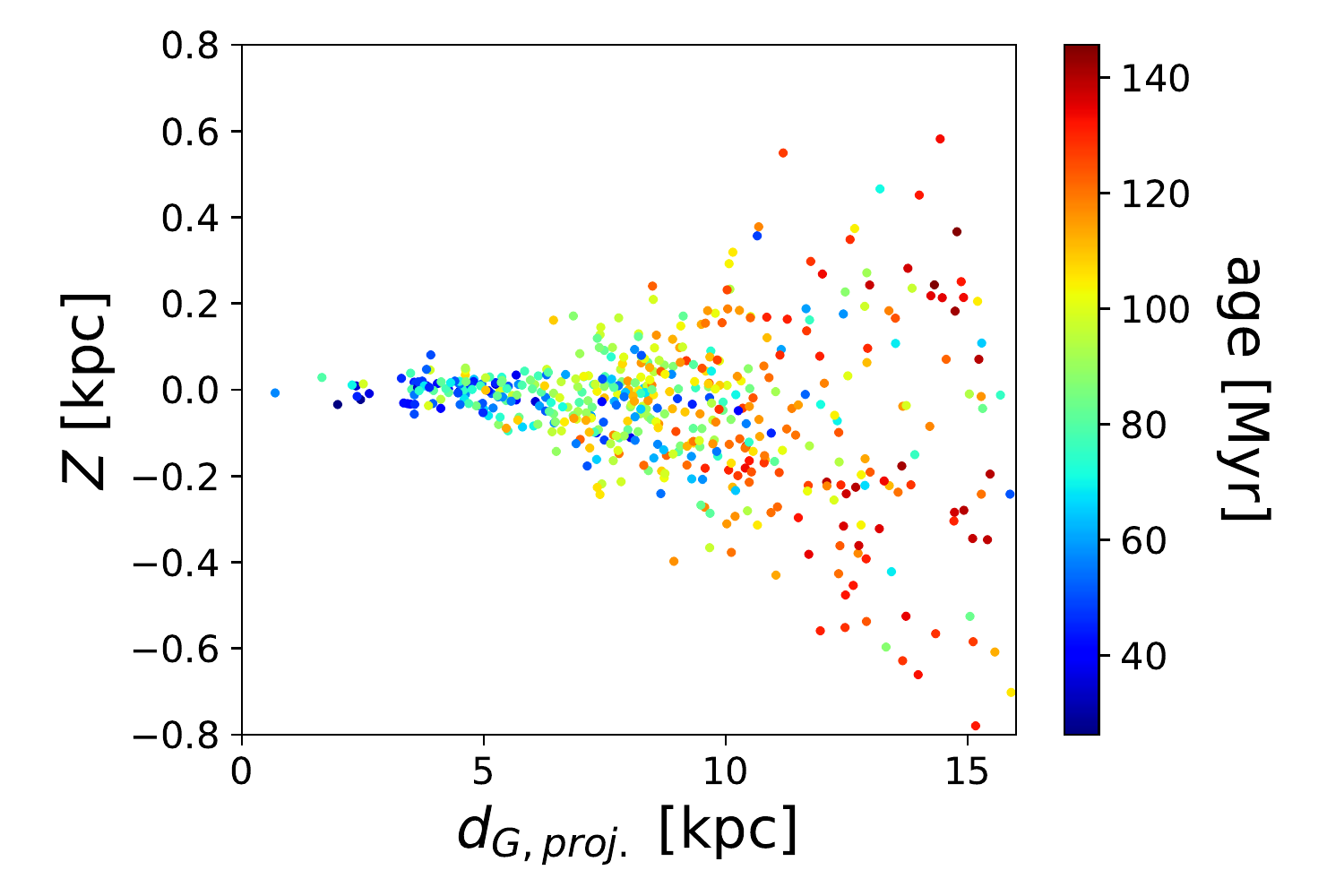}{0.33\textwidth}{}
          \fig{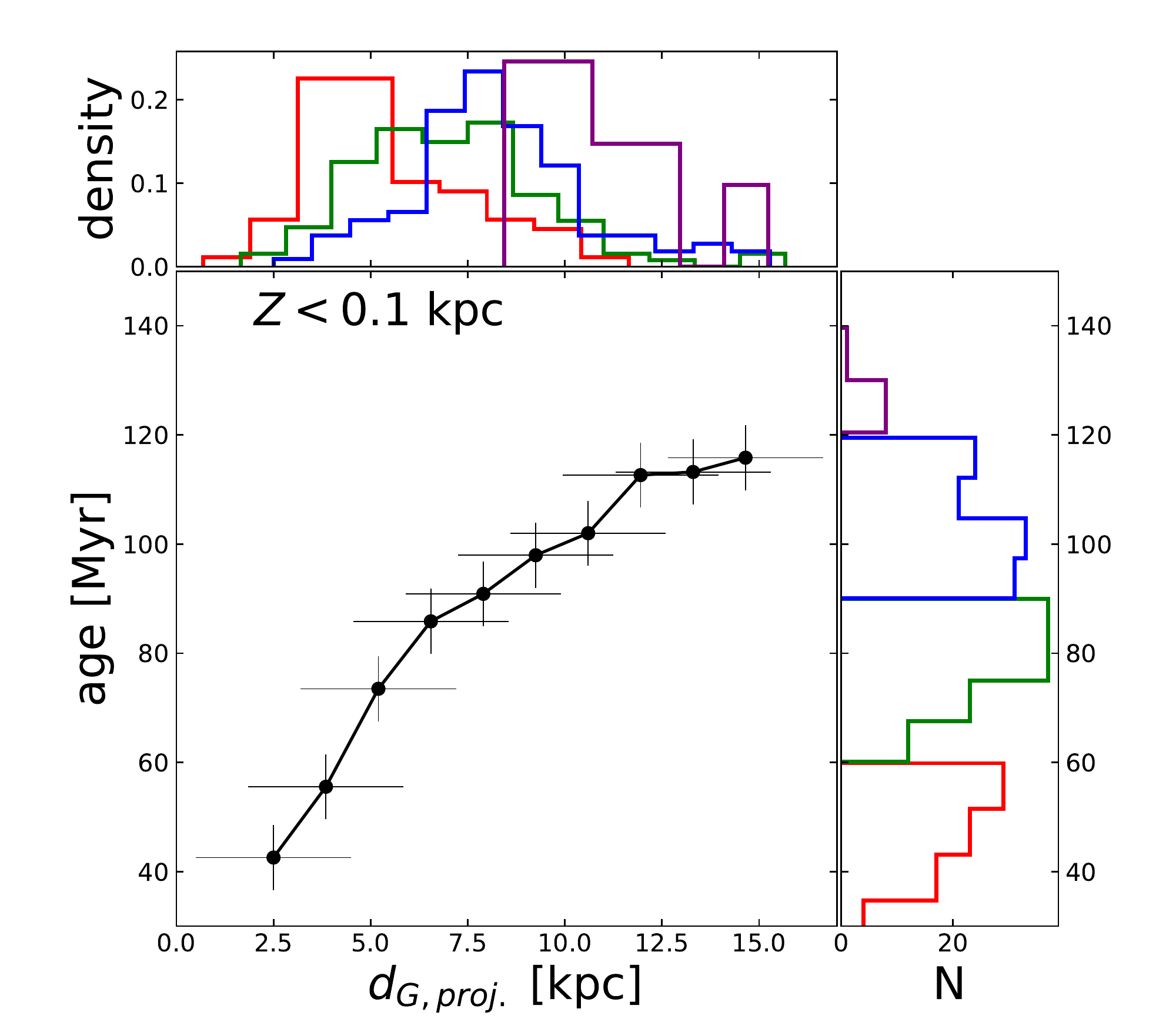}{0.33\textwidth}{}
          }
\caption{
Left panel: black points denote median ages of classical Cepheids in the Galactic disk computed for 10 subsamples in 2~kpc-wide overlapping bins of Galactocentric cylindrical distance. The vertical bars show error estimates of the median, the horizontal bars mark the width of each bin. The adjacent sub-panels show histograms of the stellar age estimates of the Cepheids (see text for details) with 4 non-overlapping age ranges color coded (right), and Galactocentric radial distributions of the Cepheids in the 4 age bins denoted with the same color coding (top).\\
Middle panel: distance from the Galactic plane ($Z$) as a function of Galactocentric cylindrical distance of the classical Cepheids, with their age estimates color coded.\\
Right panel: same as in the left panel, but for a subsample with $Z<0.1$~kpc.
\label{fig:agegrad}
}
\end{figure*}

We mapped the radial age distribution in the Galactic disk by dividing the Cepheid sample into 10 overlapping 2~kpc wide bins according to their Galactocentric cylindrical distances $d_G$, and computed the median age and its error for each bin. The errors were estimated as $1.4826\times MAD/\sqrt{N}$, where $N$ is the number of stars in a bin and $MAD$ is the median absolute deviation. The results are summarized in Fig.~\ref{fig:agegrad}. It is important to emphasize that the age limits of our sample are not intrinsic, but are set by the limits imposed on our period search (see Sect.~\ref{subsec:varsearch}). The corresponding age limits are approximately $[30,140]$~Myr for Solar metallicity, and $[35,170]$~Myr for $[$Fe/H$]=-0.4$.

The Galactocentric radial distribution of median ages clearly indicates that the most recent star formation took place in the inner part of the disk. The majority of Cepheids younger than 70~Myr are located inside the Solar circle, while Cepheids older than $\sim$120~Myr are almost exclusively found outside of it. The middle panel in Fig.~\ref{fig:agegrad} shows the ages color coded in the $Z$ {\em vs} $d_G$ distribution. Most of the older Cepheids are located further off the Galactic plane, particularly its flared outer part. At the same time, the Cepheids lying in close proximity of the plane also exhibit an age gradient. The right panel shows the age gradient for the $Z<0.1$ subsample, where the same age trend can be seen as for the entire sample, except for the truncation at the oldest ages. 

The observed age trend cannot be due to an observational selection effect. First, our $Z$ range is large enough at all values of $d_G$ ({\em cf.}~Fig.~\ref{fig:dists_marg}), thus the lack of old Cepheids further off plane inside the Solar circle cannot be due to a truncated sample. Furthermore, a selection effect due to limiting magnitudes cannot lead to the radial age distribution of the sample because it acts in the exact opposite way: young Cepheids tend to be more luminous (having longer periods, {\em cf.} Eqs.~\ref{eq:plc1j}--\ref{eq:plc1k}), thus due to the saturation limit of the VVV survey, they are preferentially detected at larger Heliocentric radii in the absence of ``sufficient'' extinction; and the opposite applies to old Cepheids. We note that the increased confusion between classical and type II Cepheids at $P\simeq10$~d discussed in Sect.~\ref{subsec:finalsample} likely introduces some upward bias in the age gradient.

\begin{figure*}[]
\plotone{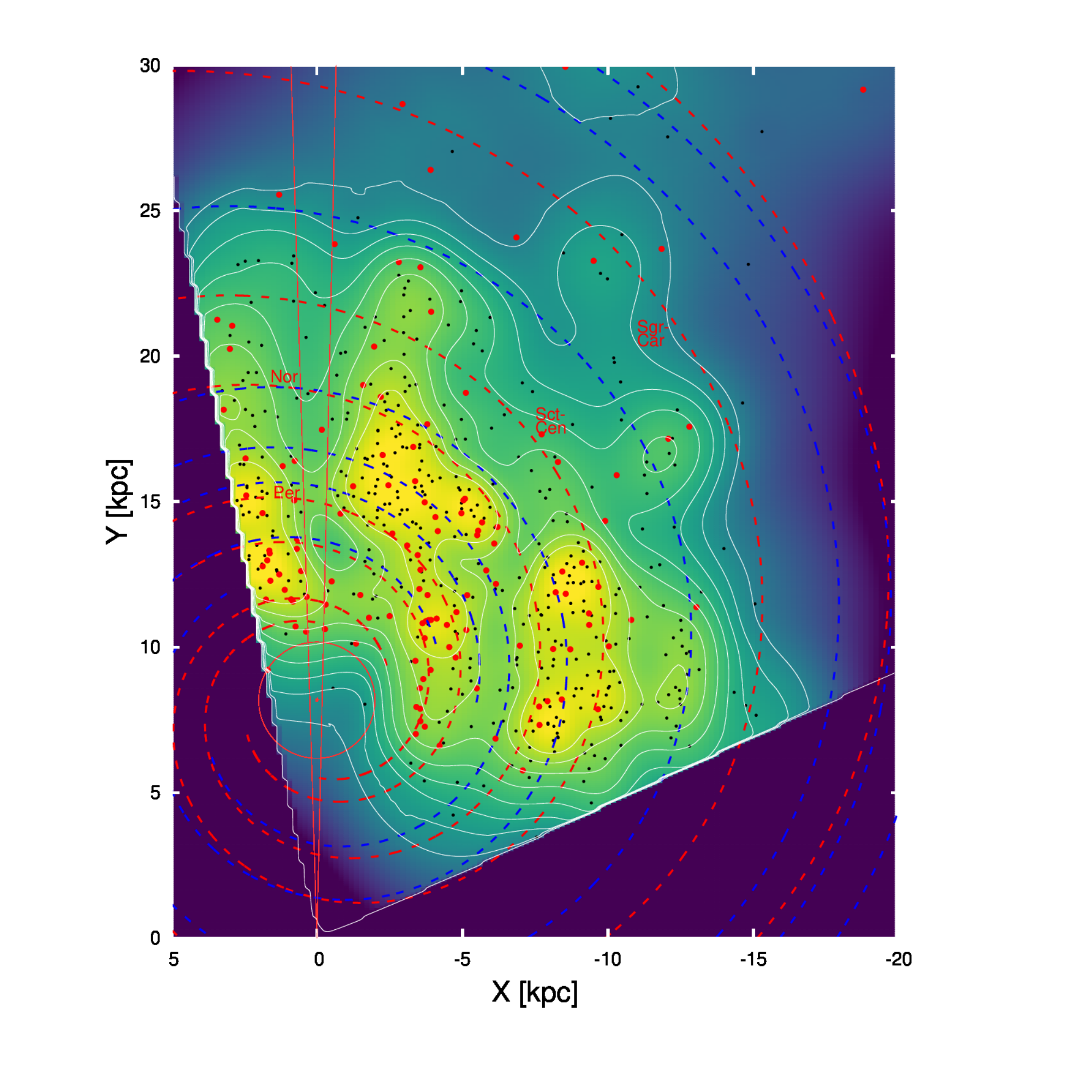}
\caption{
Spatial distribution of the classical Cepheids projected onto the Galactic plane. Black and red points show the positions of Cepheids with $P \leq 10$~d and $P > 10$~d, respectively. The color scale and the white contours denote a kernel density estimate of the distribution (see text for details). The longitudinal range of the gap in the surveyed area toward the nuclear bulge is marked with red lines. The red dot and circle show the position of the Galactic center and its 2~kpc radius. The blue and red dashed curves denote four-arm logarithmic models of the Milky Way's spiral arms by \citet{2015MNRAS.454..626H} and \citet{2017PASP..129i4102K}, respectively.
\label{fig:c1_faceon_arms}
}
\end{figure*}

The spatial distribution of the classical Cepheids projected onto the Galactic plane is shown in Fig.~\ref{fig:c1_faceon_arms}. In order to map overdensities, we computed edge-corrected kernel density estimates using adaptive Gaussian kernels with a pilot kernel size of 800~pc using the {\tt sparr} routine \citep{sparr}. The resulting face-on number density distribution shows a lot of substructure, in part due to varying detection completeness. The latter is most apparent toward the nuclear bulge, where VVV fields b333 and b334 were omitted from the analysis, causing a gap in our spatial coverage between the sight-lines marked by red lines in Fig.~\ref{fig:c1_faceon_arms}. On the other hand, features not oriented toward the Sun's position are likely real, such as the area almost devoid of Cepheids at around $5~{\rm kpc} \lesssim X \lesssim7~{\rm kpc},~6~{\rm kpc}\lesssim Y \lesssim 12~{\rm kpc}$, which is probably an interarm region.

Four-arm logarithmic models based on the HI surface density \citep{2017PASP..129i4102K}, and a compilation of HII regions, giant molecular clouds and methanol masers \citep{2014A&A...569A.125H} are shown as red and blue dashed curves, respectively, in Fig.~\ref{fig:c1_faceon_arms} for context, together with the names of the various arms corresponding to their extrapolated near-side segments.

The classical Cepheid sample was searched for tight spatial groupings, similar to the Twin Cepheids found earlier behind the bulge \citep{2015ApJ...799L..11D}. Such groups might be indicative of their common origin, e.g., that they still reside within the tidal radius of an open cluster where they formed, or that their parent cluster dissolved only recently. In contrast to the Magellanic Clouds, Galactic open clusters typically contain only one (if any) Cepheids, thus finding two or more Cepheids in surviving stellar agglomerations is rather rare \citep[see, e.g.,][]{2013MNRAS.434.2238A}.
We selected candidates based on the following simple criteria: (i) an angular separation not exceeding 5', and (ii) heliocentric distance estimates deviating by less than $2\sigma$ statistical uncertainty. 

We found 2 new classical Cepheid pairs in the bulge (b2 \& b3), and 5 pairs in the disk footprint (d1--d5) fulfilling the above selection criteria. Their positions, periods and distance estimates are shown in Table~\ref{tab:cepheidpairs}. The first pair (b1) toward the bulge is the Twin Cepheid with updated extinction and distance estimates, shown for comparison. The other two pairs in the VVV bulge fields contain Cepheids with similar periods. Employing the period-metallicity-age relations by \citet{2016A&A...591A...8A}, assuming a metallicity of $Z = 0.014$ and that they are in the second crossing of the instability strip, result in mean age estimates of $\sim$82~Myr and $\sim$113~Myr for pairs b2 and b3, respectively.

\begin{deluxetable*}{l|cccc|cccc}[]
\tablecaption{Periods, coordinates and distances\tablenotemark{a} of the candidate classical Cepheid pairs
\label{tab:cepheidpairs}
}
\tablehead{
  & \multicolumn4c{Cepheid~1} & \multicolumn4c{Cepheid~2} \\
%}
%\tablehead{
\colhead{Pair} & \colhead{Period} & \colhead{R.A.} & \colhead{DEC.} & \colhead{$d_H$\tablenotemark{a}} & 
\colhead{Period} & \colhead{R.A.} & \colhead{DEC.} & \colhead{$d_H$\tablenotemark{a}} \\
 & [d] & [hms] & [dms] & [kpc] & [d] & [hms] & [dms] & [kpc]
}
\decimals
\startdata
b1\tablenotemark{b} & 11.23424 & 18:01:24.49 & -22:54:44.6 & 13.3$\pm$0.4 & 11.21709 & 18:01:25.08 & -22:54:28.3 & 13.4$\pm$0.3 \\
b2 & ~7.46199 & 18:04:49.69 & -21:19:21.3 & 16.5$\pm$0.2 & ~7.80276 & 18:04:53.37 & -21:16:12.0 & 15.7$\pm$0.2 \\
b3 & ~4.40730 & 17:23:17.55 & -35:02:43.2 & 22.5$\pm$0.4 & ~4.36703 & 17:23:15.93 & -35:01:25.4 & 22.3$\pm$0.3 \\
\hline
d1 & ~6.93387 & 15:46:43.29 & -54:28:00.8 & 13.3$\pm$0.3 & ~8.01823 & 15:46:48.13 & -54:28:23.9 & 12.9$\pm$0.3 \\
d2 & 14.56058 & 16:37:28.51 & -46:56:58.8 & ~9.3$\pm$0.2 & 13.03596 & 16:37:33.31 & -46:54:29.8 & ~9.6$\pm$0.2 \\
d3 & 10.47008 & 16:40:30.82 & -46:27:53.9 & 15.4$\pm$0.4 & 14.58135 & 16:40:35.24 & -46:30:46.3 & 14.9$\pm$0.3 \\
d4 & 12.88066 & 16:48:04.94 & -44:48:59.7 & 10.1$\pm$0.2 & ~5.43532 & 16:48:19.64 & -44:48:36.4 & 10.1$\pm$0.2 \\
d5 & 10.82043 & 16:50:47.36 & -44:27:15.5 & 11.5$\pm$0.3 & ~8.33871 & 16:51:10.85 & -44:27:00.9 & 11.5$\pm$0.2 \\
%d6 & 11.18755 & 17:12:59.40 & -38:59:59.5 & 18.1$\pm$0.3 & ~5.60199 & 17:13:15.08 & -39:01:02.8 & 21.5$\pm$0.4
\enddata
\tablenotetext{a}{Weighted means and standard deviations of distances obtained by using the \{$J,K_s$\} and \{$H,K_s$\} filter pairs for the estimation of the extinction are given (when available).}
\tablenotetext{b}{The Twin Cepheids discovered by \citet{2015ApJ...799L..11D}.}
%\tablecomments{}
\end{deluxetable*}

The classical Cepheid pairs d3 and d4 have large period differences that are incompatible with them having similar ages, even assuming that the shorter-period Cepheid is in the second, while its longer-period pair is in the third crossing of the instability strip. At the same time, the above assumption yields comparable ages of $\sim$88~Myr and $\sim$94~Myr for d1, $\sim$63~Myr and $\sim$64~Myr for d2, and $\sim$77~Myr and $\sim$80~Myr for d5; making these pairs' associations to common birthplaces a plausible scenario. We searched the VVV Infrared Astrometric Catalogue \citep[VIRAC,][]{2018MNRAS.474.1826S} for the proper motions of the remaining Cepheid pairs. Unfortunately, due to the large distances of the stars coupled with large proper motion errors, the VIRAC measurements are insufficient to confirm or refute the physical association of these Cepheid pairs.

\section{Discussion and conclusions}\label{sec:conclusions}

We leveraged the near-IR photometric database of the VVV survey to extend the census of Cepheids toward highly attenuated Galactic regions lying in close proximity of the southern Galactic mid-plane. Periodic variable stars were classified as classical or type II Cepheid or neither, using a small convolutional neural network. We detected over 600 type II Cepheids, over $80\%$ of which are new discoveries. Likewise, 689 objects were identified as classical Cepheids, among which 640 do not appear in any previous literature. We estimated that both samples suffer from $\sim 10\%$ contamination.

The observed color indices of the Cepheids were corrected for biases due to their sparse photometric sampling by predicting their $J-K_s$ and $H-K_s$ color variations due to pulsation from their $K_s$ light-curves using neural networks. This enabled us to use the Cepheids as unbiased proxies of the interstellar reddening, thanks to their tight near-IR period-luminosity relations. The Cepheids were employed to investigate the mean properties of the interstellar extinction and its spatial variations in the $JHK_s$ bands of the VISTA photometric system. We found that our bulge and disk footprints share a mean near-IR reddening ratio of $R_{JKHK}=E(J-K_s)/E(H-K_s)\simeq2.83$. The Cepheids lying toward the bulge have a sufficient number density for probing the variations of the reddening ratio between different lines of sight, which is a proxy of the extinction curve's shape (derivative) in the near-IR regime. We find small, but highly significant spatial variations of $R_{JKHK}$ at the $\sim$2$\%$ level. Our results indicate coherent variations in the near-IR extinction law with angular distance from the Galactic Center, as well as a north--south latitudinal asymmetry. Further investigations are required using tracers with much higher number density to properly map the changes in the extinction curve.

\begin{figure}[]
\plotone{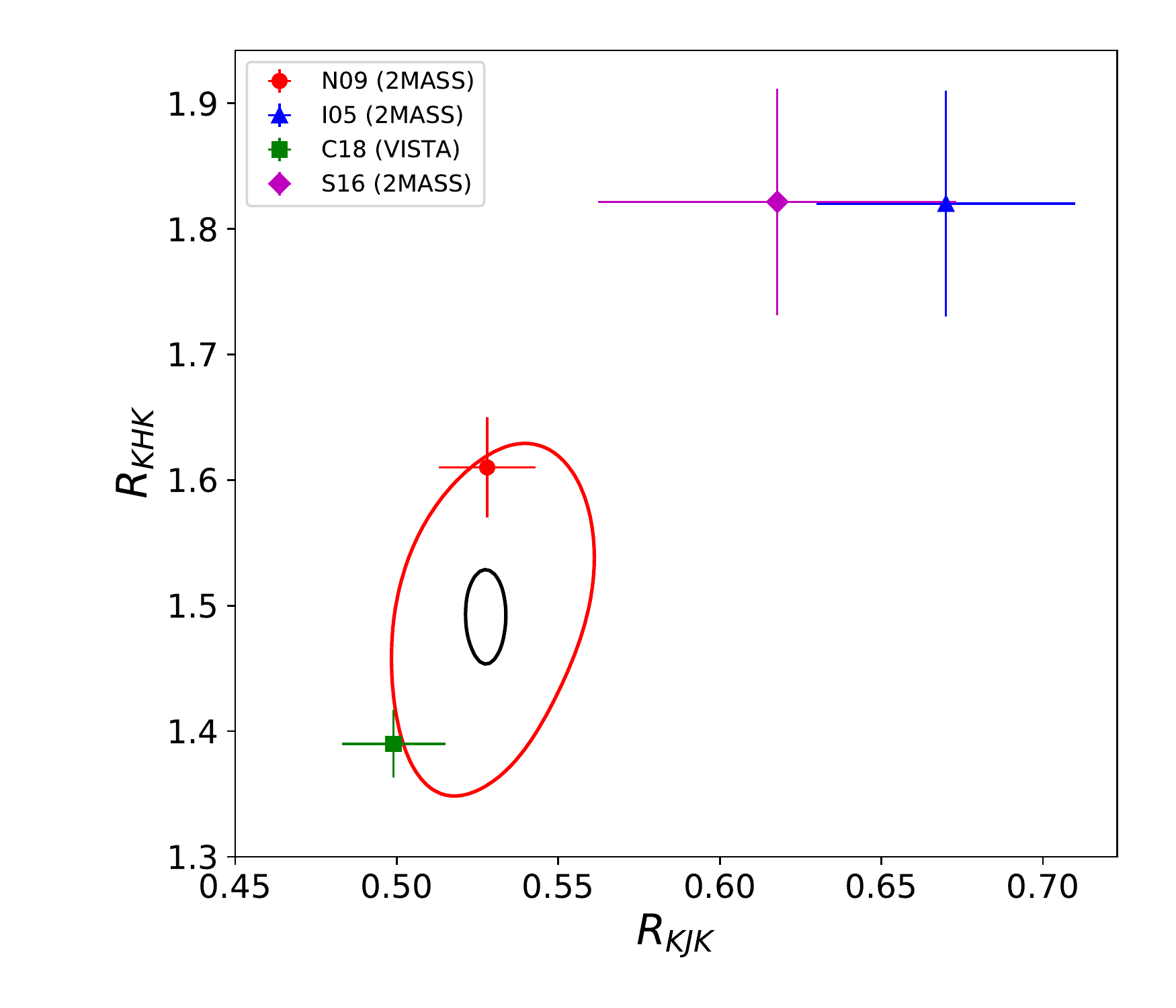}
\caption{
The mean selective-to-absolute extinction ratio derived from the Cepheid sample in this paper, in comparison with literature values from \citet[][N09]{2009ApJ...696.1407N}, \citet[][I05]{2005ApJ...619..931I}, \citet[][C18]{2018ApJ...859..137C} and \citet[][S16, in conjunction with $A(H)/A(K_s)=1.55$ from I05]{2016ApJ...821...78S}. The black and the red curves show statistical and systematic error ranges representing interquartile ranges of the posterior distributions. The error bars on literature values show statistical uncertainties. \label{fig:Rkjk_Rkhk}
}
\end{figure}

We employed the type II Cepheids endemic to the Galactic bulge to constrain the near-IR mean selective-to-absolute extinction ratios, using the recently published, extremely accurate measurement of the Galactic center's distance as our main reference point. The results are summarized by Eqs.~\ref{eq:Rkjk} and \ref{eq:Rkhk_final}, and are equivalent to $A(J)/A(K_s)=2.89\pm0.02\,{\rm(stat.)}\pm0.1\,{\rm(sys.)}$ and $A(H)/A(K_s)=1.67\pm0.01\,{\rm(stat.)}\pm0.06\,{\rm(sys.)}$. Figure~\ref{fig:Rkjk_Rkhk} shows our results in the context of selected contemporary values of the near-IR selective-to-absolute extinction ratios in the literature. Our results are consistent with a relatively steep extinction curve in the near-IR, but are in tension with the very small $R_{KJK}$ and $R_{KHK}$ values found toward the inner Galaxy by, e.g., \citet{2018ApJ...859..137C} and \citet{2017ApJ...849L..13A}. These discrepancies can arise from various problems related to the determination of the gray components of the extinction law. Traditionally, the selective-to-absolute extinction ratios are obtained from linear fits to the locus of tracer objects in color--magnitude space within an observational cone, known as the ``common-distance method'' due to its requirement that the tracers should share similar distances. Since the latter is usually a rough approximation, the intrinsic distance spread, in conjunction with correlated errors in the colors and magnitudes can easily lead to biases and underestimated errors in the resulting coefficients. Moreover, in the case of a spatially varying extinction law, extinction ratios obtained by the common distance method become biased with respect to their mean values over the observational cone due to the effect of variations along diverging lines of sight, called the composite extinction bias \citep{2013ApJ...769...88N}. A possible contribution of the Central Molecular Zone (CMZ, \citealt{1996ARA&A..34..645M} omitted from our surveyed area) to the spatial variations in the extinction law might also be accounted for the discrepancies between various observations. Finally, we note that biases in the VVV survey's photometric zero-point calibration revealed by \citet[][see also Sect.~\ref{subsec:zpcal}]{zpcalib} might also affect studies of the extinction that are based on VVV data. 

The type II Cepheids were applied as tracers to investigate the spatial distribution of the old stellar population in the inner bulge. They show a spheroidal distribution with central symmetry and high concentration around the Galactic center, and with a slight elongation that is inclined to the Galactic center's line of sight. The spatial structure traced by bulge RR~Lyrae stars shows similar features \citep{2013ApJ...776L..19D,2015ApJ...811..113P}. We find that the inclination angle of the type II Cepheids' distribution is in agreement with that found for the inner part of the RR~Lyrae distribution, in qualitative agreement with a radius-dependent inclination of the old population suggested by \citet{2015ApJ...811..113P}. At the same time, the bar traced by intermediate-age stars (such as red clump stars) has a slightly higher ($\sim$40$^\circ$) inclination angle with respect to the Solar azimuth \citep[see][and references therein]{2019arXiv190411302A}. The limited number density of the type II Cepheids in the bulge together with the non- contiguous coverage of our study prevent us from a detailed 3-dimensional modeling of their spatial distribution. The latter is left for future studies that can exploit various complementary catalogs of these objects over the entire bulge volume from VVV, OGLE, and other surveys.

In an earlier study, we reported the discovery of 35 classical Cepheids in the bulge volume using VVV data, and concluded that they trace a young, thin stellar disk spanning across the inner Galaxy \citep{2015ApJ...812L..29D}. Our results were debated by \citet{2016MNRAS.462..414M}, who found no evidence for classical Cepheids in the bulge volume outside the nuclear bulge, based their photometry of a partly overlapping sample of classical Cepheids, acquired with the Infrared Survey Facility. Their conclusion was corroborated by \citet{2018ApJ...859..137C}. \citet{2016MNRAS.462..414M} argued that the tension between our and their distance estimates was mostly due to a difference in the adopted (mean) reddening law, and also partly due to systematic differences in photometric zero-points and PL relations. Since the Cepheid sample of \citet{2015ApJ...812L..29D} is fully incorporated in our current study, it is necessary to revisit the distance distribution of these objects.

\begin{figure*}[]
\gridline{
          \fig{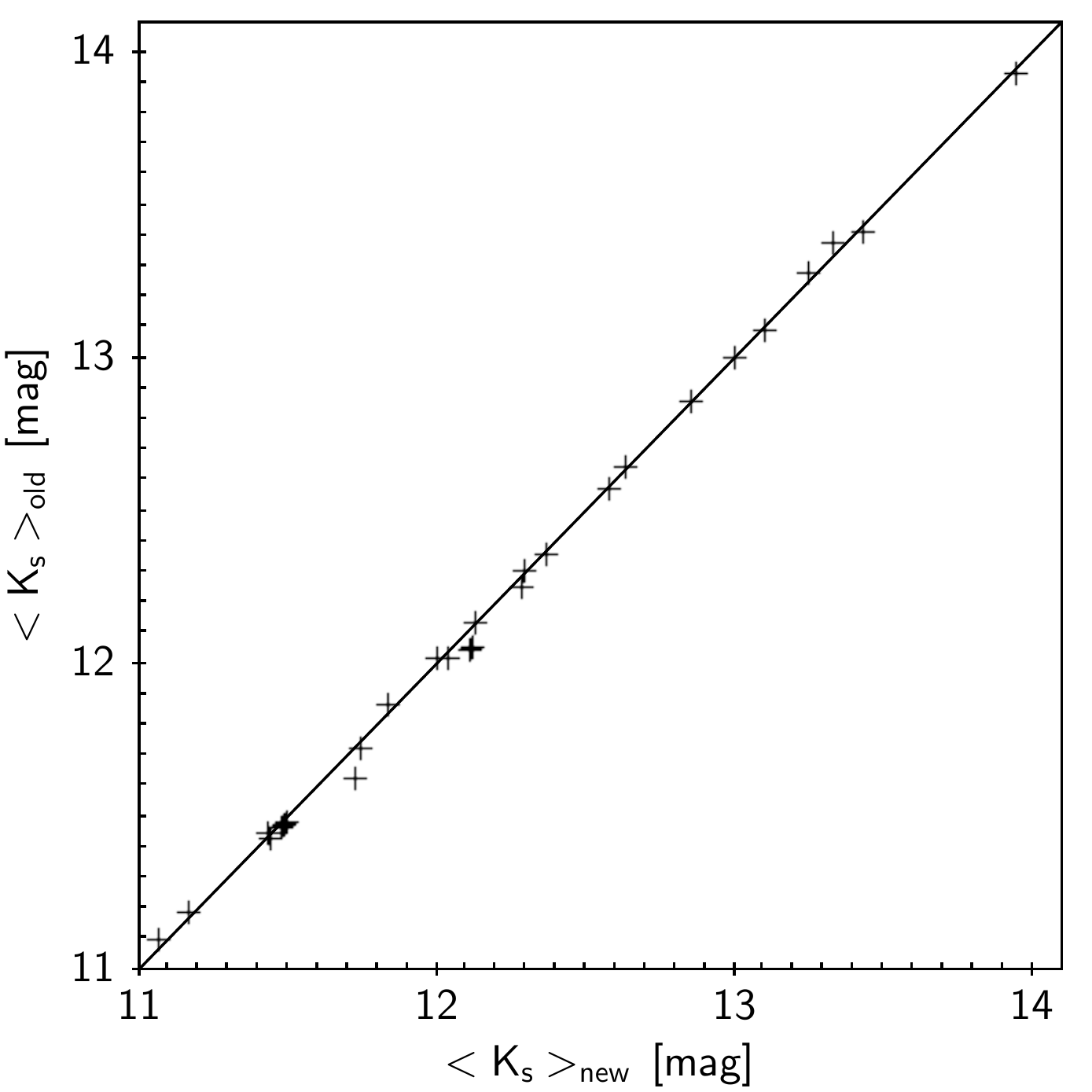}{0.3\textwidth}{}
          \fig{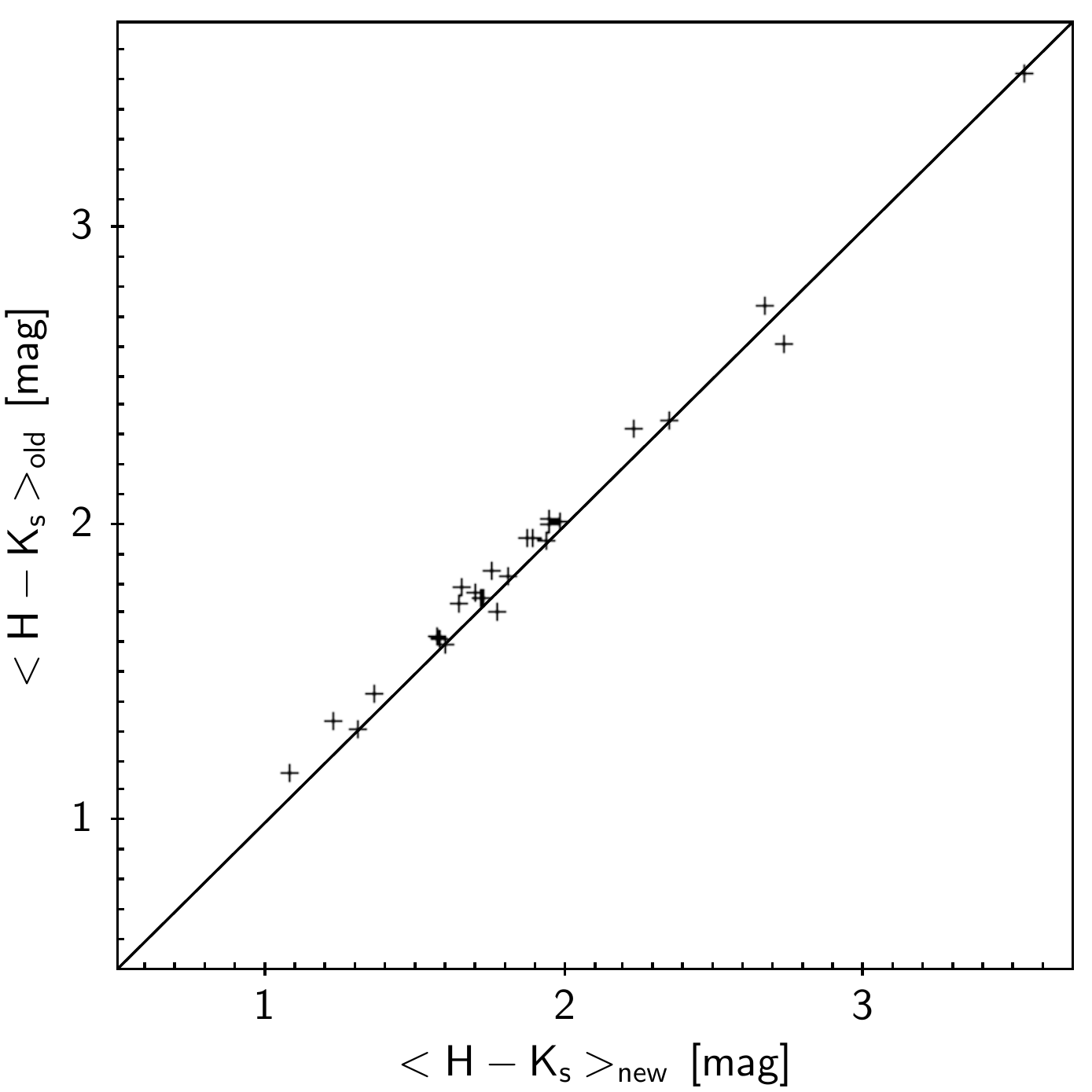}{0.3\textwidth}{}
          \fig{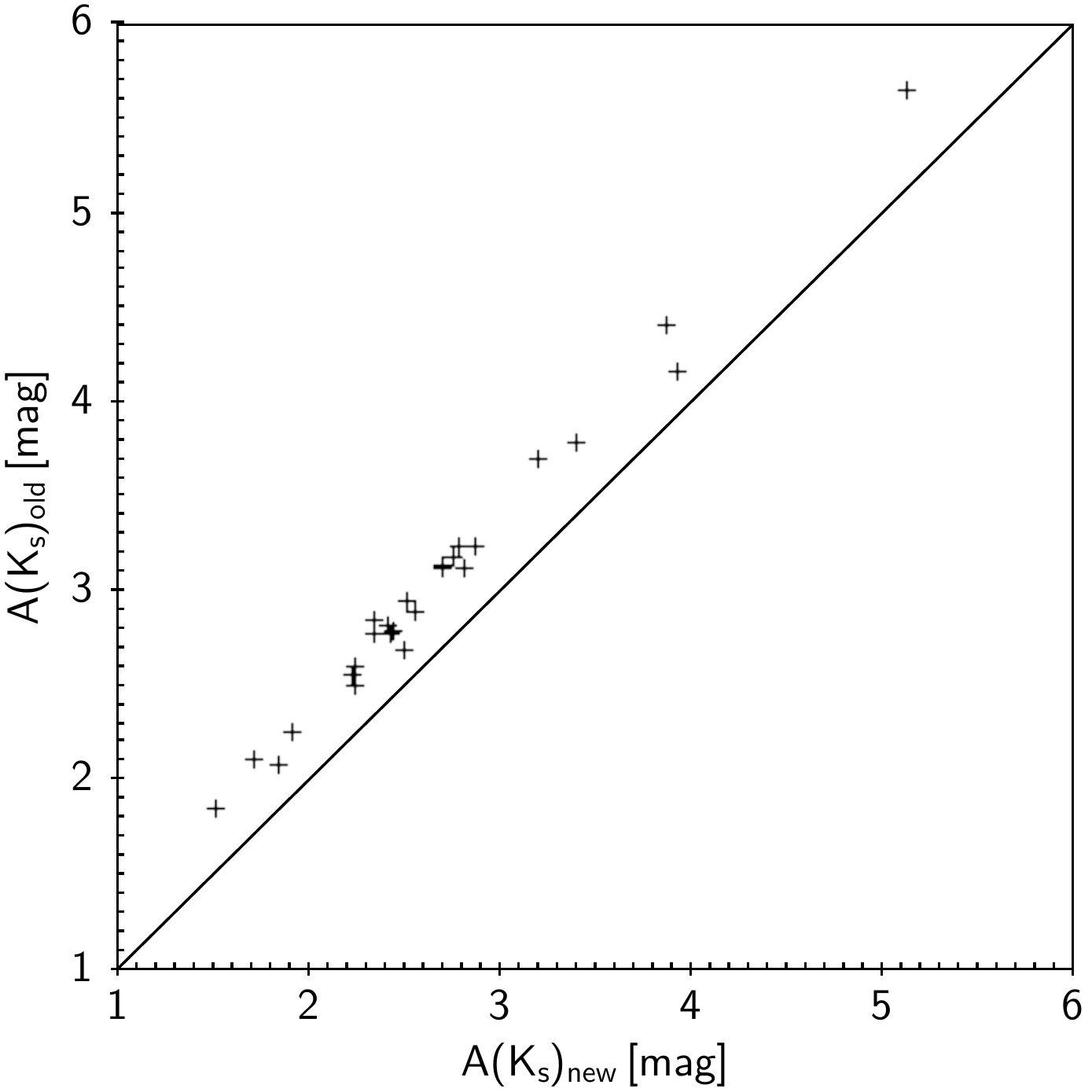}{0.3\textwidth}{}
          }
\caption{
Comparison of the mean apparent $K_s$ magnitudes (left), $H-K_s$ color indices (middle), and $K_s$-band absolute extinctions (right) of the \citet{2015ApJ...812L..29D} Cepheid sample as presented in that study (denoted as `old') {\em vs} our present analysis (denoted as `new').
\label{fig:id15cep}
}
\end{figure*}

As already mentioned in Sect.~\ref{subsec:finalsample}, only 27 of the 35 objects in \citet{2015ApJ...812L..29D} were classified as classical Cepheids by our CNN. Figure~\ref{fig:id15cep} compares the mean apparent $K_s$ magnitudes and $\langle H-K_s \rangle$ color indices, and the estimated $A(K_s)$ total extinction in \ref{subsec:finalsample} and this study. Importantly, in our earlier study, we used an earlier version of the standard VVV photometry as provided by CASU, while our present study is based on the version 1.5 CASU VVV photometry, with a custom ZP calibration based on the findings of \citet[][see Sect.~\ref{subsec:zpcal}]{zpcalib}. While the version update and subsequent recalibration did not cause systematic changes in the $\langle K_s \rangle$ mean magnitudes, the $\langle H-K_s \rangle$ measurements became systematically lower by $\sim$0.1~mag, in consistence with the findings of \citet{2016MNRAS.462..414M}. The latter, together with a steeper near-IR extinction curve (i.e., smaller $R_{KHK}$) compared to \citet{2009ApJ...696.1407N}, resulted in systematically lower $A(K_s)$ estimates, hence larger distances compared to those obtained by \citet{2015ApJ...812L..29D}. As a result, we found only 9 objects with classical Cepheid classification at $d_G<3$~kpc, and only 3 objects at $d_G<2$~kpc, but the presence of classical Cepheids in the inner Milky Way cannot be ruled out. 
We urge the photometric and spectroscopic follow-up observations of these objects, in order to further clarify their nature. In fact, there is plenty of other evidence for the presence of young stellar populations in the inner Milky Way (apart from the ongoing star formation in the CMZ), e.g., in the form of young massive star clusters such as Arches or Quintuplet, massive field stars (e.g., \citealt{2015MNRAS.446..842D}, \citealt{2015A&A...584A...2F}) and Miras \citep[e.g.,][]{2019MNRAS.482.5567M}.

% and reach a consensus regarding the presence of a young stellar population in the inner Milky Way.

The classical Cepheids identified in this study can be employed to trace spatial structures at the far side of the Galactic disk. The Cepheids clearly trace a flared outer disk in agreement with the findings of \citet{2007A&A...469..511K} based on interstellar gas. They also trace the Galactic warp with its onset radius and nodal line being in good agreement with the findings of \citet{2019NatAs...3..320C} and \citet{2019Sci...365..478S}. In particular, our sample closes the earlier gap in the Cepheids' census at the far side of the southern disk, and in combination with datasets from complementary surveys, they allow a more detailed 360$^\circ$ future modeling of the Galactic warp. 

By taking advantage of the relationship between the classical Cepheids' periods and ages, we investigated the radial age distribution of the young Galactic disk. The radial age gradient shown by the Cepheids is qualitatively consistent with the findings by \citet{2019Sci...365..478S}, who used OGLE Cepheids concentrated at the near side of the disk. The age distributions found by the two studies indicate that the most recent star formation episodes of the Milky Way were concentrated inside the Solar circle over the III-rd and IV-th Galactic quadrants, while the inner disk was mostly devoid of star formation within a range of $\sim$100--150~Myr lookback time.

The classical Cepheids can also contribute to constrain the face-on map of the Milky Way's disk, e.g. by tracing its spiral arms and other substructures. Some overdensities in the face-on distribution of the classical Cepheids are in tentative agreement with the extrapolated locations of the spiral arms from the models of \citet{2017PASP..129i4102K} and \citet{2014A&A...569A.125H} (Fig.~\ref{fig:c1_faceon_arms}), and the deviations between the KDE maxima and the spiral arm ridges of either model do not exceed the deviations between the different models. There are several factors that can contribute to the lack of a clearer spiral structure in the classical Cepheids' distribution, apart from the non-uniform incompleteness of the sample. Most importantly, the Cepheids show a wide age range, and the older they are, the more they drift away from the the spiral arms due to the difference between the standard Galactic rotation and the spiral arms' pattern speed expected from density wave theory \citep[e.g.,][]{2014PASA...31...35D,2018MNRAS.478.3590S}, and observed by \citet{2015MNRAS.454..626H}. On the other hand, young Cepheids are not present in our sample in sufficient numbers to trace spiral arms. In addition, the cumulative peculiar motions of older Cepheids can add further scatter. Although we found an agreement between the mean extinction coefficients toward the bulge and the disk footprints of this study, spatial variations in the extinction law over smaller angular scales, such as discovered toward the bulge footprint, can contribute to the scatter to the distance estimates of the Cepheids. Last but not least, it is still an open question whether the Milky Way is a grand-design spiral galaxy or shows flocculent substructures \citep[e.g.][]{2019MNRAS.487.1400C} and inter-arm segments connecting major arms such as the Local Arm \citep{2013ApJ...769...15X}. It is also unclear whether the global configuration of the spiral arms follows a logarithmic or a polynomial pattern (i.e., has a varying pitch angle, as suggested by \citealt{2014A&A...569A.125H}). The quantitative modeling of the spiral arm structure based on the classical Cepheids is beyond the scope of this paper, and is left for future studies on the basis of homogenized distance estimates of the joint Cepheid sample from VVV, OGLE, WISE, ASAS-SN, Gaia, etc.

The newly discovered Cepheids presented in this study close the observational gap in the Zona Galactica Incognita at the far side of the disk, complementing the census from other surveys with objects from the most inaccessible region of the Milky Way. The known classical Cepheids of the Galactic disk now populate the full azimuthal range out to Galactocentric distances of $\sim$20~kpc, supplementing the multitude of gaseous tracers known to date. With the abundant emergence of stellar tracers, the obscured disk of our home Galaxy is now ripe for detailed exploration.

%%%%%%%%%%%%%%%%%%%%%%%%%%%%%%%%%%%%%%%%%%%%%%%%%%%%%%%%%%%%%%%%%

\acknowledgments
%We warmly thank M\'arcio Catelan all the fruitful discussions about Cepheid variable stars.
Our results were based on observations collected at the European Southern Observatory under ESO programme 179.B-2002.
I.D. and E.K.G. were supported by Sonderforschungsbereich SFB 881 ``The Milky Way System'' (subproject A03) of the German Research Foundation (DFG).
G.H. acknowledges support from the Graduate Student Exchange Fellowship Program between the Institute of Astrophysics of the Pontificia Universidad Cat\'olica de Chile and the Zentrum f\"ur Astronomie der Universit\"at Heidelberg, funded by the Heidelberg Center in Santiago de Chile and the Deutscher Akademischer Austauschdienst (DAAD), and support by CONICYT-PCHA/Doctorado Nacional grant 2014-63140099.
G.H. and M.C. acknowledge the support provided by the Ministry for the Economy, Development, and Tourism's Millennium Science Initiative through grant IC\,120009, awarded to the Millennium Institute of Astrophysics (MAS); by Fondecyt through grant \#1171273; and by Proyecto Basal AFB-170002.
%G.H. and M.C. acknowledge support by the Chilean Ministry for the Economy, Development, and Tourism's Programa Iniciativa Milenio through grant IC120009.
%G.H. and M.C. acknowledge support by the Ministry for the Economy, Development, and Tourism's Programa Iniciativa Milenio through grant IC120009 and by Proyecto Basal PFB-06/2007, and by FONDECYT through grant \#1171273, and by CONICYT's PCI program through grant DPI20140066.
Post-processing and analysis of data were performed on the Milky Way supercomputer, which is funded by the Deutsche Forschungsgemeinschaft (DFG) through the Collaborative Research Center (SFB 881) ``The Milky Way System'' (subproject Z2), and on the {\mbox BWFor} supercomputer, which is supported by the state of Baden-W\"urttemberg through bwHPC and the German Research Foundation (DFG) through grant INST 35/1134-1 FUGG.

%%%%%%%%%%%%%%%%%%%%%%%%%%%%%%%%%%%%%%%%%%%%%%%%%%%%%%%%%%%%%%%%%


\vspace{5mm}
\facilities{ESO:VISTA}


\software{STIL \citep{2006ASPC..351..666T},  
                numpy \citep{numpy},  
                PyMC \citep{pymc},  
                scikit-learn \citep{2011JMLR....12.2825},  
                scipy \citep{scipy},  
                sparr \citep{sparr},  
                TensorFlow \citep{2016arXiv160304467A},  
                Keras \citep{keras} 
          }

%%%%%%%%%%%%%%%%%%%%%%%%%%%%%%%%%%%%%%%%%%%%%%%%%%%%%%%%%%%%%%%%%



\end{document}